\documentclass[onecolumn]{article}%
\usepackage{amsfonts}
\usepackage{amsmath}
\usepackage{amssymb}
\usepackage{color}
\usepackage{colortbl}
\usepackage{multicol}
\usepackage{graphicx}%
\begin{document}

\title{\hfill{\normalsize UMTG - 3}\\\ \\Quasi-hermitian Quantum Mechanics in Phase Space}
\author{Thomas Curtright$^{\text{\S ,\dag,}}${\small *} and Andrzej
Veitia$^{\text{\S ,}}${\small *}\medskip\\$^{\text{\S }}${\small Department of Physics, University of Miami, Coral
Gables, Florida 33124}\\$^{\text{\dag}}${\small School of Natural Sciences, Institute for Advanced
Study, Princeton, New Jersey 08540}}
\maketitle

\begin{abstract}
We investigate quasi-hermitian quantum mechanics in phase space using standard
deformation quantization methods: \ Groenewold star products and Wigner
transforms. \ We focus on imaginary Liouville theory as a representative
example where exact results are easily obtained. \ We emphasize spatially
periodic solutions, compute various distribution functions and phase-space
metrics, and explore the relationships between them. \ 

\vfill

\vfill

\noindent\underline
{\ \ \ \ \ \ \ \ \ \ \ \ \ \ \ \ \ \ \ \ \ \ \ \ \ \ \ \ \ \ \ \ \ \ \ \ \ \ \ \ \ \ \ \ \ \ \ \ \ \ \ \ \ \ \ }%

{\small *}curtright@physics.miami.edu \& aveitia@physics.miami.edu\newpage

\end{abstract}
\tableofcontents

\newpage

\section{Introduction}

Superficially non-hermitian Hamiltonian quantum systems are of considerable
current interest, especially in the context of PT symmetric models
\cite{BenderReview,MostafazadehReview}. \ For such systems the Hilbert space
structure is at first sight very different than that for hermitian Hamiltonian
systems inasmuch as the dual wave functions are not just the complex
conjugates of the wave functions, or equivalently, the Hilbert space metric is
not the usual one. \ While it is possible to keep most of the compact Dirac
notation\ in analyzing such systems (see Appendix E), in the main body of this
paper we will work with explicit functions and avoid abstract notation. \ Our
goal is to expose the underlying mechanisms (as in \cite{CM,CMS}) rather than
to hide them.

Our discussion is focussed on a system with potential $\exp\left(  2ix\right)
$. \ This model, as well as its field theory extension, is of interest for
applications to table-top physical systems \cite{Affleck,Berry} and to deeper
problems in string theory \cite{Schomerus,Strominger}. \ We will not discuss
those applications here, but rather we will simply develop the phase-space
formalism for the point particle model. \ We believe this formalism will be
helpful in understanding the applications cited, as well as others. \ Other
recent work along these same lines can be found in
\cite{Scholtz2005,Scholtz2006}.

\section{Imaginary Liouville quantum mechanics}

Consider \textquotedblleft imaginary\textquotedblright\ or \textquotedblleft
periodic\textquotedblright\ Liouville quantum mechanics as governed by the
\emph{apparently} non-hermitian Hamiltonian%
\begin{equation}
H=p^{2}+m^{2}e^{2ix} \label{ImaginaryLiouvilleHamiltonian}%
\end{equation}
Without essential loss of generality, we take $m=1$ in most of the following.
\ Obviously, this is a \textquotedblleft PT symmetric\textquotedblright%
\ model. \ But more to the point, this is actually a \textquotedblleft
quasi-hermitian\textquotedblright\ theory \cite{Scholtz1992} with a
\emph{real} energy spectrum, as explained in \cite{CM} and as we shall clarify
further here. \ We will analyze this system in phase-space using the methods
of deformation quantization.

\subsection{Eigenfunctions}

But first, let us briefly review the position representation Schr\"{o}dinger
eigenvalue problem for this system (see e.g. \cite{CM}). \ With $x$ on the
real line and with the condition that the wave functions\ remain bounded, the
corresponding Schr\"{o}dinger equation has energy eigenvalues given by all
real $E\geq0$. \ The eigenfunctions are just Bessel functions, $J_{\pm\sqrt
{E}}\left(  e^{ix}\right)  $. \ These are doubly degenerate when $\sqrt{E}\neq
n\in\mathbb{N}$, but they merge into a single, nondegenerate eigenfunction
when $E=n^{2}$. \ For these nondegenerate cases the eigenfunctions are $2\pi
$-periodic in $x$, and the solutions are the analytic Bessel functions $J_{n}$
\cite{Watson}.%
\begin{align}
n^{2}J_{n}\left(  e^{ix}\right)   &  =\left(  -\frac{d^{2}}{dx^{2}}%
+e^{2ix}\right)  J_{n}\left(  e^{ix}\right) \label{BesselEquation}\\
J_{n}\left(  e^{ix}\right)   &  =\frac{1}{2^{n}}e^{inx}\sum_{k=0}^{\infty
}\frac{\left(  -1\right)  ^{k}}{k!\left(  k+n\right)  !}\frac{e^{2ikx}}{4^{k}%
}\ ,\ \ \ \ n=0,1,2,\cdots\label{JSeries}%
\end{align}
We will consider only such periodic solutions here, since in our opinion this
is the most interesting situation. \ We note that these periodic
eigenfunctions are superpositions of only \emph{right-moving} plane waves.
\ In fact, in this periodic situation the discrete energy spectrum is
precisely the same as would be found for particles moving freely on a circle
\emph{but} restricted to non-negative momentum.

\subsection{Dual polynomials}

The periodic Bessel functions and their complex conjugates do \emph{not} form
an orthonormal set on the circle. To obtain an orthonormal set of functions it
is necessary to combine $\left\{  J_{n}\left(  z\right)  \right\}  $ with an
associated set of \emph{polynomials in }$z^{-1}$, $\left\{  A_{n}\left(
z\right)  \right\}  $, the so-called Neumann polynomials. \ These are dual to
$\left\{  J_{n}\left(  z\right)  \right\}  $ on any contour enclosing the
origin $z=0$, in the following sense: $\frac{1}{2\pi i}%
%TCIMACRO{\doint }%
%BeginExpansion
{\displaystyle\oint}
%EndExpansion
\frac{dz}{z}\ A_{j}\left(  z\right)  J_{k}\left(  z\right)  =\delta_{j,k}$.
\ So on the circle, we have%
\begin{equation}
\frac{1}{2\pi}\int_{0}^{2\pi}dx\ A_{j}\left(  e^{ix}\right)  J_{k}\left(
e^{ix}\right)  =\delta_{j,k} \label{OrthoAJ}%
\end{equation}
The Neumann polynomials on the circle are given explicitly by\footnote{For
convenience we have modified the usual notation of the associated polynomials
as given in \cite{Abram,Neumann,Watson}, namely $O_{n}$, and have defined
$A_{n}\left(  z\right)  =\varepsilon_{n}\ z\ O_{n}\left(  z\right)  $ where
$\varepsilon_{0}=1$ and $\varepsilon_{n}=2$ for $n\neq0$. \ }%
\begin{align}
A_{0}\left(  e^{ix}\right)   &  =1\ ,\ \ \ A_{1}\left(  e^{ix}\right)
=2e^{-ix}\ ,\ \ \ A_{n}\left(  e^{ix}\right)  =2^{n}ne^{-inx}\sum
_{k=0}^{\left\lfloor n/2\right\rfloor }\frac{\left(  n-k-1\right)  !}{k!}%
\frac{e^{2ikx}}{4^{k}}\label{ASeries}\\
\mathcal{I}_{n}\left(  e^{ix}\right)   &  =\left(  -\frac{d^{2}}{dx^{2}%
}+e^{2ix}-n^{2}\right)  A_{n}\left(  e^{ix}\right)  \label{InhomoBesselEqn}%
\end{align}
As indicated, the $\left\{  A_{n}\right\}  $ obey\emph{\ inhomogeneous}
modifications of Bessel's equation, where the inhomogeneity is either
$\mathcal{I}_{n}\left(  z\right)  \propto z^{2}$ for even $n$ or
$\mathcal{I}_{n}\left(  z\right)  \propto z$ for odd $n$, according to
\begin{equation}
\mathcal{I}_{n}\left(  e^{ix}\right)  =\left\{
\begin{array}
[c]{cc}%
\varepsilon_{n}e^{2ix} & \text{for even }n\geq0\\
2ne^{ix} & \text{for odd }n>0
\end{array}
\right.
\end{equation}
While (\ref{InhomoBesselEqn}) is inhomogeneous, nevertheless the usual proof
of orthogonality between pairs of non-degenerate $H$ eigenfunctions and their
duals goes through because the inhomogeneities are orthogonal to the
eigenfunctions.%
\begin{equation}
\int_{0}^{2\pi}dx\ \mathcal{I}_{j}\left(  e^{ix}\right)  J_{k}\left(
e^{ix}\right)  =0\ ,\ \ \ \text{for all}\ \ \ j,k\in\mathbb{N}\text{\ }%
\end{equation}

\section{Phase space distributions}

We next compute Wigner transforms of various function bilinears for the $2\pi
$-periodic Bessel/Neumann system. \ For a system so-defined, on a circle,
momentum is quantized. \ Thus we would expect that the $\left(  x,p\right)
$\ phase space is not the usual $\mathbb{R}^{2}$ nor even $\mathbb{S}%
^{1}\times\mathbb{R}$, but rather that it is reduced to $\mathbb{S}^{1}%
\times\mathbb{Z}$. \ In fact, the periodic energy eigenfunctions of the
imaginary Liouville Hamiltonian consist of superpositions of positive momentum
plane waves, so we would also expect not to need the $\mathbb{Z}_{<0}$
momentum sector at all. \ Well, both expectations are almost true. \ But not
quite. \ We shall see below to what extent these expectations are born out.

\subsection{Eigenfunction WFs}

As a first step, we remind the reader about the structure of real (diagonal)
Wigner functions (WFs) made from $2\pi$-periodic plane waves. \ They are just
Kronecker deltas, with $p\in\mathbb{Z}$ as expected.%
\begin{align}
e_{n}\left(  x,p\right)   &  \equiv\frac{1}{2\pi}\int_{0}^{2\pi}\phi
_{n}\left(  x-y\right)  ~\overline{\phi_{n}}\left(  x+y\right)  ~e^{2iyp}%
dy\label{FreeWF}\\
&  =\delta_{n,p} \label{FreeChiralWF}%
\end{align}
upon choosing $\phi_{n}\left(  x\right)  =\exp\left(  inx\right)  $. \ By
analogy, for the Liouville eigenfunctions $\psi_{n}$ the WFs are again
manifestly real, and again have support for $p\in\mathbb{Z}$ as expected, as
given by%
\begin{align}
f_{n}\left(  x,p\right)   &  \equiv\frac{1}{2\pi}\int_{0}^{2\pi}\psi
_{n}\left(  x-y\right)  ~\overline{\psi_{n}}\left(  x+y\right)  ~e^{2iyp}%
dy\label{WF}\\
&  =\frac{\left(  -1\right)  ^{p-n}}{4^{p}}\sum_{k=0}^{p-n}\frac{e^{2ix\left(
n-p+2k\right)  }}{k!\left(  n+k\right)  !\left(  p-k\right)  !\left(
p-k-n\right)  !} \label{LiouvilleWF}%
\end{align}
where the sum results from taking $\psi_{n}\left(  x\right)  =J_{n}\left(
e^{ix}\right)  $ as given by the series in (\ref{JSeries}). \ Note for the
Liouville case, as opposed to the free particle case, the support in $p$ is
\emph{infinite}: \ This particular $f_{n}\left(  x,p\right)  $\ is non-zero
for all $p\geq n\geq0$. \ Another way to write these WFs makes use of the
associated Legendre functions.\footnote{A form which facilitates a
continuation to non-integer $p$, should anyone wish to do that.}%
\begin{equation}
f_{n}\left(  x,p\right)  =\frac{\left(  -1\right)  ^{n}}{p!\left(  p-n\right)
!}\left(  \frac{i\sin2x}{2}\right)  ^{p}\operatorname{LegendreP}\left(
p,-n,i\cot2x\right)
\end{equation}
In particular, for any point in the reduced phase space, $\left(  x,p\right)
\in\mathbb{S}^{1}\times\mathbb{Z}$, we have
\begin{subequations}
\begin{align}
f_{0}\left(  x,p\right)   &  =1\times\delta_{p,0}-\frac{1}{2}\left(
\cos2x\right)  \times\delta_{p,1}+\frac{1}{2^{5}}\left(  2+\cos4x\right)
\times\delta_{p,2}\ +\cdots\label{GroundStateWF}\\
f_{1}\left(  x,p\right)   &  =\frac{1}{2^{2}}\times\delta_{p,1}-\frac{1}%
{2^{4}}\left(  \cos2x\right)  \times\delta_{p,2}+\frac{1}{2^{8}3}\left(
3+2\cos4x\right)  \times\delta_{p,3}+\cdots\\
f_{2}\left(  x,p\right)   &  =\frac{1}{2^{6}}\times\delta_{p,2}-\frac{1}%
{2^{7}3}\left(  \cos2x\right)  \times\delta_{p,3}+\frac{1}{2^{12}3^{2}}\left(
4+3\cos4x\right)  \times\delta_{p,4}+\cdots\\
f_{3}\left(  x,p\right)   &  =\frac{1}{2^{8}3^{2}}\times\delta_{p,3}-\frac
{1}{2^{11}3^{2}}\left(  \cos2x\right)  \times\delta_{p,4}+\frac{1}{2^{16}%
3^{2}5}\left(  5+4\cos4x\right)  \times\delta_{p,5}+\cdots
\end{align}
But now we have more functions at our disposal, namely the Neumann
polynomials, so we may build another set of WFs for comparison to those in
(\ref{FreeChiralWF}) and (\ref{LiouvilleWF}).

\subsection{Dual WFs}

For dual functions $\chi_{n}$ the WFs are also manifestly real, with
$p\in\mathbb{Z}$, as given by\footnote{To dispel any confusion about our
conventions for $\widetilde{f_{n}}$, the complex conjugation in (\ref{DualWF})
is \emph{different} from that in (\ref{FreeWF})\ and (\ref{WF})\ just because
we chose in \cite{CM,CMS}\ to define the duals such that $\delta_{n,k}%
=\frac{1}{2\pi}\int_{0}^{2\pi}\chi_{n}\left(  x\right)  \psi_{k}\left(
x\right)  dx$ without any explicit conjugations.}
\end{subequations}
\begin{align}
\widetilde{f_{n}}\left(  x,p\right)   &  \equiv\frac{1}{2\pi}\int_{0}^{2\pi
}\overline{\chi_{n}}\left(  x-y\right)  ~\chi_{n}\left(  x+y\right)
~e^{2iyp}dy\label{DualWF}\\
&  =4^{p}n^{2}\sum_{k=0}^{\left\lfloor n/2\right\rfloor }\sum_{l=0}%
^{\left\lfloor n/2\right\rfloor }\frac{\left(  n-k-1\right)  !}{k!}%
\frac{\left(  n-l-1\right)  !}{l!}e^{2ix\left(  l-k\right)  }\delta_{n-p,k+l}
\label{LiouvilleDualWFDoubleSum}%
\end{align}
where the sums result from taking $\chi_{n}\left(  x\right)  =A_{n}\left(
e^{ix}\right)  $ as given by the series in (\ref{ASeries}). \ Note the support
in $p$ is now \emph{finite}: \ This particular $\widetilde{f_{n}}\left(
x,p\right)  $\ is non-zero for $0\leq n-2\left\lfloor n/2\right\rfloor \leq
p\leq n$. \ That is to say, $0\leq p\leq n$ for $n$ even, and $1\leq p\leq n$
for $n$ odd. \ 

Resolving the constraint set by the Kronecker delta in
(\ref{LiouvilleDualWFDoubleSum}) eliminates one sum and further restricts the
range of the other to yield\footnote{The limits on the sum may also be written
as $\min\left(  \left\lfloor n/2\right\rfloor ,n-p\right)  $ and
$n-p-\min\left(  \left\lfloor n/2\right\rfloor ,n-p\right)  $.}%
\begin{equation}
\widetilde{f_{n}}\left(  x,p\right)  =4^{p}n^{2}\sum_{k=\max\left(
0,n-p-\left\lfloor n/2\right\rfloor \right)  }^{\min\left(  \left\lfloor
n/2\right\rfloor ,n-p\right)  }\frac{\left(  n-k-1\right)  !\left(
k+p-1\right)  !}{k!\left(  n-p-k\right)  !}e^{2i\left(  n-p-2k\right)  x}
\label{LiouvilleDualWF}%
\end{equation}
In particular, again for any phase-space point $\left(  x,p\right)
\in\mathbb{S}^{1}\times\mathbb{Z}$,
\begin{subequations}
\begin{align}
\widetilde{f_{0}}\left(  x,p\right)   &  =\delta_{p,0}\\
\widetilde{f_{1}}\left(  x,p\right)   &  =4\delta_{p,1}\\
\widetilde{f_{2}}\left(  x,p\right)   &  =4\times\delta_{p,0}+32\left(
\cos2x\right)  \times\delta_{p,1}+64\times\delta_{p,2}\\
\widetilde{f_{3}}\left(  x,p\right)   &  =36\times\delta_{p,1}+576\left(
\cos2x\right)  \times\delta_{p,2}+2304\times\delta_{p,3}%
\end{align}
Thus, at any given momentum level, we find the same set of functions of $x$
(i.e. cos$\left(  2kx\right)  $) no matter whether we consider $\left\{
f_{n}\right\}  $\ or $\left\{  \widetilde{f_{n}}\right\}  $.

There is a basic orthogonality relation for WFs and dual WFs.
\end{subequations}
\begin{equation}
\frac{1}{2\pi}\int_{x,p}\hspace{-0.25in}%
%TCIMACRO{\tsum }%
%BeginExpansion
{\textstyle\sum}
%EndExpansion
~f_{k}\left(  x,p\right)  \widetilde{f_{n}}\left(  x,p\right)  =\delta_{k,n}
\label{WFDualWFOrthonormality}%
\end{equation}
This follows in a straightforward way from the bilinear structure of the
Wigner transform and from the orthogonality of the wave functions and their
duals. \ There is also a corresponding pseudo-local\ relation on the phase
space that involves the Groenewold star product (see (\ref{WF*DualWF}) and
(\ref{DualWF*WF}) below). \ The form of these results can be seen most easily
through the use of formal density operator methods, as in Appendix E.

Perhaps it is useful to present the specific examples of $f_{n}$ and
$\widetilde{f_{n}}$, for $n=0,1,2,3$ and for $p=0,1,2,3$, in Table form.
\ This facilitates checking (\ref{WFDualWFOrthonormality}) for these few
cases, and illuminates the orthogonality mechanism.\bigskip

\begin{center}%
\begin{tabular}
[c]{|c|cccccccccc|}\hline
\multicolumn{11}{|c|}{$%
%TCIMACRO{\TeXButton{blue}{\color{blue}}}%
%BeginExpansion
\color{blue}%
%EndExpansion
\text{WFs}%
%TCIMACRO{\TeXButton{black}{\color{black}}}%
%BeginExpansion
\color{black}%
%EndExpansion
\text{ \& }%
%TCIMACRO{\TeXButton{red}{\color{red}}}%
%BeginExpansion
\color{red}%
%EndExpansion
\text{Dual WFs}%
%TCIMACRO{\TeXButton{black}{\color{black}}}%
%BeginExpansion
\color{black}%
%EndExpansion
\ \ \text{(non-zero values)}$}\\\hline
&  & $p=0$ &  & $p=1$ &  & $p=2$ &  & $p=3$ &  & $\cdots$\\\hline
&  &  &  &  &  &  &  &  &  & \\
$%
%TCIMACRO{\TeXButton{blue}{\color{blue}}}%
%BeginExpansion
\color{blue}%
%EndExpansion
f_{0}%
%TCIMACRO{\TeXButton{black}{\color{black}}}%
%BeginExpansion
\color{black}%
%EndExpansion
,%
%TCIMACRO{\TeXButton{red}{\color{red}}}%
%BeginExpansion
\color{red}%
%EndExpansion
\widetilde{f_{0}}%
%TCIMACRO{\TeXButton{black}{\color{black}}}%
%BeginExpansion
\color{black}%
%EndExpansion
$ &  & $%
%TCIMACRO{\TeXButton{blue}{\color{blue}}}%
%BeginExpansion
\color{blue}%
%EndExpansion
1%
%TCIMACRO{\TeXButton{black}{\color{black}}}%
%BeginExpansion
\color{black}%
%EndExpansion
\ ,\
%TCIMACRO{\TeXButton{red}{\color{red}}}%
%BeginExpansion
\color{red}%
%EndExpansion
1%
%TCIMACRO{\TeXButton{black}{\color{black}}}%
%BeginExpansion
\color{black}%
%EndExpansion
$ &  & $%
%TCIMACRO{\TeXButton{blue}{\color{blue}}}%
%BeginExpansion
\color{blue}%
%EndExpansion
-\dfrac{1}{2}\cos2x%
%TCIMACRO{\TeXButton{black}{\color{black}}}%
%BeginExpansion
\color{black}%
%EndExpansion
\ ,\
%TCIMACRO{\TeXButton{red}{\color{red}}}%
%BeginExpansion
\color{red}%
%EndExpansion
0%
%TCIMACRO{\TeXButton{black}{\color{black}}}%
%BeginExpansion
\color{black}%
%EndExpansion
$ &  & $%
%TCIMACRO{\TeXButton{blue}{\color{blue}}}%
%BeginExpansion
\color{blue}%
%EndExpansion
\dfrac{1}{32}\cos4x+\dfrac{1}{16}%
%TCIMACRO{\TeXButton{black}{\color{black}}}%
%BeginExpansion
\color{black}%
%EndExpansion
\ ,\
%TCIMACRO{\TeXButton{red}{\color{red}}}%
%BeginExpansion
\color{red}%
%EndExpansion
0%
%TCIMACRO{\TeXButton{black}{\color{black}}}%
%BeginExpansion
\color{black}%
%EndExpansion
$ &  & $%
%TCIMACRO{\TeXButton{blue}{\color{blue}}}%
%BeginExpansion
\color{blue}%
%EndExpansion
-\dfrac{1}{1152}\cos6x-\dfrac{1}{128}\cos2x%
%TCIMACRO{\TeXButton{black}{\color{black}}}%
%BeginExpansion
\color{black}%
%EndExpansion
\ ,\
%TCIMACRO{\TeXButton{red}{\color{red}}}%
%BeginExpansion
\color{red}%
%EndExpansion
0%
%TCIMACRO{\TeXButton{black}{\color{black}}}%
%BeginExpansion
\color{black}%
%EndExpansion
$ &  & $\cdots$\\
&  &  &  &  &  &  &  &  &  & \\
$%
%TCIMACRO{\TeXButton{blue}{\color{blue}}}%
%BeginExpansion
\color{blue}%
%EndExpansion
f_{1}%
%TCIMACRO{\TeXButton{black}{\color{black}}}%
%BeginExpansion
\color{black}%
%EndExpansion
,%
%TCIMACRO{\TeXButton{red}{\color{red}}}%
%BeginExpansion
\color{red}%
%EndExpansion
\widetilde{f_{1}}%
%TCIMACRO{\TeXButton{black}{\color{black}}}%
%BeginExpansion
\color{black}%
%EndExpansion
$ &  &  &  & $%
%TCIMACRO{\TeXButton{blue}{\color{blue}}}%
%BeginExpansion
\color{blue}%
%EndExpansion
\dfrac{1}{4}%
%TCIMACRO{\TeXButton{black}{\color{black}}}%
%BeginExpansion
\color{black}%
%EndExpansion
\ ,\
%TCIMACRO{\TeXButton{red}{\color{red}}}%
%BeginExpansion
\color{red}%
%EndExpansion
4%
%TCIMACRO{\TeXButton{black}{\color{black}}}%
%BeginExpansion
\color{black}%
%EndExpansion
$ &  & $%
%TCIMACRO{\TeXButton{blue}{\color{blue}}}%
%BeginExpansion
\color{blue}%
%EndExpansion
-\dfrac{1}{16}\cos2x%
%TCIMACRO{\TeXButton{black}{\color{black}}}%
%BeginExpansion
\color{black}%
%EndExpansion
\ ,\
%TCIMACRO{\TeXButton{red}{\color{red}}}%
%BeginExpansion
\color{red}%
%EndExpansion
0%
%TCIMACRO{\TeXButton{black}{\color{black}}}%
%BeginExpansion
\color{black}%
%EndExpansion
$ &  & $%
%TCIMACRO{\TeXButton{blue}{\color{blue}}}%
%BeginExpansion
\color{blue}%
%EndExpansion
\dfrac{1}{384}\cos4x+\dfrac{1}{256}%
%TCIMACRO{\TeXButton{black}{\color{black}}}%
%BeginExpansion
\color{black}%
%EndExpansion
\ ,\
%TCIMACRO{\TeXButton{red}{\color{red}}}%
%BeginExpansion
\color{red}%
%EndExpansion
0%
%TCIMACRO{\TeXButton{black}{\color{black}}}%
%BeginExpansion
\color{black}%
%EndExpansion
$ &  & $\cdots$\\
&  &  &  &  &  &  &  &  &  & \\
$%
%TCIMACRO{\TeXButton{blue}{\color{blue}}}%
%BeginExpansion
\color{blue}%
%EndExpansion
f_{2}%
%TCIMACRO{\TeXButton{black}{\color{black}}}%
%BeginExpansion
\color{black}%
%EndExpansion
,%
%TCIMACRO{\TeXButton{red}{\color{red}}}%
%BeginExpansion
\color{red}%
%EndExpansion
\widetilde{f_{2}}%
%TCIMACRO{\TeXButton{black}{\color{black}}}%
%BeginExpansion
\color{black}%
%EndExpansion
$ &  & $%
%TCIMACRO{\TeXButton{blue}{\color{blue}}}%
%BeginExpansion
\color{blue}%
%EndExpansion
0%
%TCIMACRO{\TeXButton{black}{\color{black}}}%
%BeginExpansion
\color{black}%
%EndExpansion
\ ,\
%TCIMACRO{\TeXButton{red}{\color{red}}}%
%BeginExpansion
\color{red}%
%EndExpansion
4%
%TCIMACRO{\TeXButton{black}{\color{black}}}%
%BeginExpansion
\color{black}%
%EndExpansion
$ &  & $%
%TCIMACRO{\TeXButton{blue}{\color{blue}}}%
%BeginExpansion
\color{blue}%
%EndExpansion
0%
%TCIMACRO{\TeXButton{black}{\color{black}}}%
%BeginExpansion
\color{black}%
%EndExpansion
\ ,\
%TCIMACRO{\TeXButton{red}{\color{red}}}%
%BeginExpansion
\color{red}%
%EndExpansion
32\cos2x%
%TCIMACRO{\TeXButton{black}{\color{black}}}%
%BeginExpansion
\color{black}%
%EndExpansion
$ &  & $%
%TCIMACRO{\TeXButton{blue}{\color{blue}}}%
%BeginExpansion
\color{blue}%
%EndExpansion
\dfrac{1}{64}%
%TCIMACRO{\TeXButton{black}{\color{black}}}%
%BeginExpansion
\color{black}%
%EndExpansion
\ ,\
%TCIMACRO{\TeXButton{red}{\color{red}}}%
%BeginExpansion
\color{red}%
%EndExpansion
64%
%TCIMACRO{\TeXButton{black}{\color{black}}}%
%BeginExpansion
\color{black}%
%EndExpansion
$ &  & $%
%TCIMACRO{\TeXButton{blue}{\color{blue}}}%
%BeginExpansion
\color{blue}%
%EndExpansion
-\dfrac{1}{384}\cos2x%
%TCIMACRO{\TeXButton{black}{\color{black}}}%
%BeginExpansion
\color{black}%
%EndExpansion
\ ,\
%TCIMACRO{\TeXButton{red}{\color{red}}}%
%BeginExpansion
\color{red}%
%EndExpansion
0%
%TCIMACRO{\TeXButton{black}{\color{black}}}%
%BeginExpansion
\color{black}%
%EndExpansion
$ &  & $\cdots$\\
&  &  &  &  &  &  &  &  &  & \\
$%
%TCIMACRO{\TeXButton{blue}{\color{blue}}}%
%BeginExpansion
\color{blue}%
%EndExpansion
f_{3}%
%TCIMACRO{\TeXButton{black}{\color{black}}}%
%BeginExpansion
\color{black}%
%EndExpansion
,%
%TCIMACRO{\TeXButton{red}{\color{red}}}%
%BeginExpansion
\color{red}%
%EndExpansion
\widetilde{f_{3}}%
%TCIMACRO{\TeXButton{black}{\color{black}}}%
%BeginExpansion
\color{black}%
%EndExpansion
$ &  &  &  & $%
%TCIMACRO{\TeXButton{blue}{\color{blue}}}%
%BeginExpansion
\color{blue}%
%EndExpansion
0%
%TCIMACRO{\TeXButton{black}{\color{black}}}%
%BeginExpansion
\color{black}%
%EndExpansion
\ ,\
%TCIMACRO{\TeXButton{red}{\color{red}}}%
%BeginExpansion
\color{red}%
%EndExpansion
36%
%TCIMACRO{\TeXButton{black}{\color{black}}}%
%BeginExpansion
\color{black}%
%EndExpansion
$ &  & $%
%TCIMACRO{\TeXButton{blue}{\color{blue}}}%
%BeginExpansion
\color{blue}%
%EndExpansion
0%
%TCIMACRO{\TeXButton{black}{\color{black}}}%
%BeginExpansion
\color{black}%
%EndExpansion
\ ,\
%TCIMACRO{\TeXButton{red}{\color{red}}}%
%BeginExpansion
\color{red}%
%EndExpansion
576\cos2x%
%TCIMACRO{\TeXButton{black}{\color{black}}}%
%BeginExpansion
\color{black}%
%EndExpansion
$ &  & $%
%TCIMACRO{\TeXButton{blue}{\color{blue}}}%
%BeginExpansion
\color{blue}%
%EndExpansion
\dfrac{1}{2304}%
%TCIMACRO{\TeXButton{black}{\color{black}}}%
%BeginExpansion
\color{black}%
%EndExpansion
\ ,\
%TCIMACRO{\TeXButton{red}{\color{red}}}%
%BeginExpansion
\color{red}%
%EndExpansion
2304%
%TCIMACRO{\TeXButton{black}{\color{black}}}%
%BeginExpansion
\color{black}%
%EndExpansion
$ &  & $\cdots$\\
&  &  &  &  &  &  &  &  &  & \\
$\vdots$ &  & $\vdots$ &  & $\vdots$ &  & $\vdots$ &  & $\vdots$ &  & $\ddots
$\\
&  &  &  &  &  &  &  &  &  & \\\cline{1-3}\cline{2-11}%
\end{tabular}

\end{center}

\section{Wigner transform of the bilocal metric}

It is explained in \cite{CM} -- as well as in the classic literature on the
subject -- how a scalar product for a biorthogonal system such as $\left\{
A_{k},J_{n}\right\}  $ can always be written as an integral over a doubled
configuration space involving a \textquotedblleft bilocal
metric\textquotedblright\ $K\left(  x,y\right)  $.%
\begin{equation}
\left(  \phi,\psi\right)  =\iint\phi\left(  x\right)  K\left(  x,y\right)
\psi\left(  y\right)  dxdy \label{ScalarProduct}%
\end{equation}

\subsection{Bilocal $\leftrightarrow$\ phase space}

When a scalar product is so expressed as a bilocal bilinear form then it is
naturally and very easily re-expressed in phase space (which we suppose to be
$\mathbb{R}^{2}$ in this paragraph) through the use of a Wigner transform
\cite{Folland}.
\begin{equation}
f_{\psi\phi}\left(  x,p\right)  \equiv\frac{1}{2\pi}\int\psi\left(  x-\frac
{1}{2}y\right)  \phi\left(  x+\frac{1}{2}y\right)  e^{iyp}dy
\label{WignerTransform}%
\end{equation}
We have chosen the normalization here so that for $p$\ on the real line
\begin{equation}
\psi\left(  x\right)  \phi\left(  x\right)  =\int_{-\infty}^{\infty}%
f_{\psi\phi}\left(  x,p\right)  dp
\end{equation}
More generally, Fourier inverting (\ref{WignerTransform}) gives the
point-split product%
\begin{equation}
\phi\left(  x\right)  \psi\left(  y\right)  =\int_{-\infty}^{\infty
}e^{i\left(  y-x\right)  p}f_{\psi\phi}\left(  \frac{x+y}{2},p\right)  dp
\end{equation}
Thus the scalar product (\ref{ScalarProduct}) can be re-written as
\begin{equation}
\left(  \phi,\psi\right)  =\iint R\left(  x,p\right)  f_{\psi\phi}\left(
x,p\right)  dxdp \label{PhaseSpaceScalarProduct}%
\end{equation}
where the phase-space metric is the Wigner transform of the bilocal metric.%
\begin{equation}
R\left(  x,p\right)  =\int e^{iyp}K\left(  x-\frac{1}{2}y,x+\frac{1}%
{2}y\right)  dy \label{PhaseSpaceMetric}%
\end{equation}
and inversely%
\begin{equation}
K\left(  x,y\right)  =\frac{1}{2\pi}\int_{-\infty}^{\infty}e^{i\left(
x-y\right)  p}R\left(  \frac{x+y}{2},p\right)  dp
\end{equation}
In a more abstract notation (as in Appendix E) the form of
(\ref{PhaseSpaceScalarProduct}) is $\left(  \overline{\phi},\psi\right)
=Tr\left(  \mathsf{R}\left\vert \psi\right\rangle \left\langle \phi\right\vert
\right)  =Tr\left(  \left\vert \psi\right\rangle \widetilde{\left\langle
\phi\right\vert }\right)  $.

\subsection{Liouville dual metric}

The preceding results are quite general. \ To be more specific, for $2\pi
$-periodic \emph{dual} functions of imaginary Liouville quantum mechanics, the
scalar product was shown in \cite{CM} to be $\frac{1}{\left(  2\pi\right)
^{2}}\iint_{0}^{2\pi}dxdy\overline{\chi_{j}\left(  x\right)  }J\left(
x,y\right)  \chi_{k}\left(  y\right)  $ $=\delta_{j,k}$ where%
\begin{equation}
J\left(  x,y\right)  =J_{0}\left(  e^{-ix}-e^{iy}\right)  =J_{0}\left(
e^{-ix}\right)  J_{0}\left(  e^{iy}\right)  +2\sum_{n=1}^{\infty}J_{n}\left(
e^{-ix}\right)  J_{n}\left(  e^{iy}\right)  \label{DualBilocalMetric}%
\end{equation}
Again, just a Bessel function. \ Or, re-expressed in a form which is
immediately useful in the following,%
\begin{equation}
J\left(  x,y\right)  =J_{0}\left(  -2ie^{i\left(  y-x\right)  /2}\sin\left(
\frac{x+y}{2}\right)  \right)
\end{equation}
Up to a normalization the corresponding metric in phase space is given by the
Wigner transform of this bilocal.\footnote{To take the free particle limit,
the parameter $m$ in (\ref{ImaginaryLiouvilleHamiltonian}) must first be
restored. \ See Appendix A.} \ 

A bit of care is needed since the Wigner transforms $\widetilde{f_{n}}\left(
x,p\right)  $\ on which this metric will act are actually defined so that a
dual function $\chi$ plays the role of $\phi$ and a conjugate dual function
$\overline{\chi}$ plays the role of $\psi$ in the above (compare
(\ref{WignerTransform})\ to (\ref{DualWF})). \ Thus, acting on $\widetilde
{f_{n}}\left(  x,p\right)  $ the metric would be the Wigner transform of $J$
as above, only with first and second arguments interchanged. \ We also adjust
the normalization here (and again later, in (\ref{DualWFNorms})) to take into
account our conventions and the fact that we are dealing with $2\pi$-periodic
functions (see Appendix C). \ In view of all this, we finally obtain a dual
phase-space metric given by%
\begin{align}
\widetilde{R}\left(  x,p\right)   &  =\frac{1}{2\pi}\int_{0}^{2\pi}J\left(
x+w,x-w\right)  e^{2iwp}dw=\frac{1}{2\pi}\int_{0}^{2\pi}J_{0}\left(
-2ie^{-iw}\sin x\right)  e^{2iwp}dw\nonumber\\
&  =\frac{1}{2\pi}\sum_{k=0}^{\infty}\frac{\left(  \sin x\right)  ^{2k}%
}{\left(  k!\right)  ^{2}}\int_{0}^{2\pi}e^{2iw\left(  p-k\right)  }%
dw=\sum_{k=0}^{\infty}\frac{\left(  \sin x\right)  ^{2k}}{\left(  k!\right)
^{2}}\delta_{p,k}%
\end{align}
Hence the simple final answer.%
\begin{equation}
\widetilde{R}\left(  x,p\right)  =\frac{\left(  \sin^{2}x\right)  ^{p}%
}{\left(  p!\right)  ^{2}}\text{ \ \ for integer\ }p\geq0\text{,\ but vanishes
for integer\ }p<0 \label{DualPhaseSpaceMetric}%
\end{equation}
An equivalent operator expression can be obtained by the method of Weyl
transforms. \ (See Appendix F.)

Another way to characterize $\widetilde{R}\left(  x,p\right)  $ is to note
that it satisfies the differential-difference equation%
\begin{equation}
p\partial_{x}\widetilde{R}\left(  x,p\right)  =\sin\left(  2x\right)
\widetilde{R}\left(  x,p-1\right)  \label{DualDifferentialDifference}%
\end{equation}
even when $p=0$, since $\partial_{x}\widetilde{R}\left(  x,p=0\right)  =0$ as
well as $\widetilde{R}\left(  x,p=-1\right)  =0$. \ (This should be compared
to (\ref{DifferentialDifference}) given below. \ Note that $\widetilde{R}$
actually corresponds to $R^{-1}$ in that later discussion.)

In fact, there is \emph{an} obvious continuation of
(\ref{DualPhaseSpaceMetric}) to all real $p$, or even to complex $x$ and $p$.
\ Namely
\begin{equation}
\widetilde{R}\left(  x,p\right)  =\frac{\left(  \sin^{2}x\right)  ^{p}%
}{\left(  \Gamma\left(  p+1\right)  \right)  ^{2}} \label{ContinuedMetric}%
\end{equation}
with its manifest zeroes and singularities (poles \emph{and} cuts). \ This
continuation also transparently satisfies (\ref{DualDifferentialDifference}).

Taking into account all our conventions, we may now express the normalizations
of pure states in terms of the dual WFs and the dual phase-space metric
as$\ $
\begin{equation}
\varepsilon_{n}=\frac{1}{2\pi}\int_{x,p}\hspace{-0.25in}%
%TCIMACRO{\tsum }%
%BeginExpansion
{\textstyle\sum}
%EndExpansion
~\widetilde{R}\left(  x,p\right)  \widetilde{f_{n}}\left(  x,p\right)
=\frac{1}{2\pi}\sum_{p=0}^{n}\int_{0}^{2\pi}\frac{\left(  \sin^{2}x\right)
^{p}}{\left(  p!\right)  ^{2}}\widetilde{f_{n}}\left(  x,p\right)  dx
\label{DualWFNorms}%
\end{equation}
where as usual $\varepsilon_{0}=1$ and $\varepsilon_{n}=2$ for $n>0$. \ It is
tedious but straightforward to use (\ref{LiouvilleDualWF})\ to check this
normalization and confirm that the dual metric does its job. \ More
importantly, (\ref{DualWFNorms}) is consistent with
(\ref{WFDualWFOrthonormality}) for the simple reason that%
\begin{equation}
\widetilde{R}\left(  x,p\right)  =\sum_{k=0}^{\infty}\varepsilon_{k}%
f_{k}\left(  x,p\right)  \label{DualMetricAsWFs}%
\end{equation}
This in turn follows from the expansion of the bilocal metric in terms of
Bessel bilinears, in (\ref{DualBilocalMetric}).

\section{Homogeneous versus inhomogeneous $\bigstar$genvalue equations}

Were the dual functions just the complex conjugates of the wave functions, the
two types of WFs that we have defined would be identical, $\widetilde{f_{n}%
}\left(  x,p\right)  =\left.  f_{n}\left(  x,p\right)  \right\vert
_{\overline{\psi}=\chi}$, but this is obviously \emph{not} true for the case
at hand. \ 

\subsection{WFs as eigenfunctions}

As an important distinguishing feature, for the first of these WFs we have%
\begin{equation}
H\star f_{n}=n^{2}f_{n}=f_{n}\star\overline{H} \label{WF*genvalue}%
\end{equation}
where the associative Groenewold star product operation \cite{Groenewold} is
($\hbar\equiv1$)
\begin{equation}
\star\ \equiv\exp\left(  \frac{i}{2}~\overleftarrow{\partial_{x}%
}~\overrightarrow{\partial_{p}}-\frac{i}{2}~\overleftarrow{\partial_{p}%
}~\overrightarrow{\partial_{x}}\right)  \label{GroenewoldStar}%
\end{equation}
Thus the wave function WFs are energy $\bigstar$genfunctions. \ 

The $\bigstar$genvalue equation (\ref{WF*genvalue}) for the WFs is not only a
distinguishing feature but can actually be used to define the $f_{n}$ as real
functions, without any prior knowledge of the underlying wave functions.
\ Taking imaginary and real parts of (\ref{WF*genvalue})\ we obtain an
equation like (\ref{DualDifferentialDifference}), namely%
\begin{equation}
p\partial_{x}f_{n}\left(  x,p\right)  =\sin\left(  2x\right)  f_{n}\left(
x,p-1\right)  \label{WFDifferentialDifferenceImag}%
\end{equation}
as well as
\begin{equation}
\left(  p^{2}-\frac{1}{4}\partial_{x}^{2}\right)  f_{n}\left(  x,p\right)
+\cos\left(  2x\right)  f_{n}\left(  x,p-1\right)  =n^{2}f_{n}\left(
x,p\right)  \label{WFDifferentialDifferenceReal}%
\end{equation}
Using (\ref{WFDifferentialDifferenceImag}) twice, the second derivative in
(\ref{WFDifferentialDifferenceReal}) becomes%
\begin{equation}
\partial_{x}^{2}f_{n}\left(  x,p\right)  =\frac{2\cos\left(  2x\right)  }%
{p}f_{n}\left(  x,p-1\right)  +\frac{\sin^{2}\left(  2x\right)  }{p\left(
p-1\right)  }f_{n}\left(  x,p-2\right)
\end{equation}
Hence (\ref{WFDifferentialDifferenceReal}) reduces to a second-order
difference equation in the momentum.%
\begin{equation}
\left(  p^{2}-n^{2}\right)  f_{n}\left(  x,p\right)  +\frac{2p-1}{2p}%
\cos\left(  2x\right)  f_{n}\left(  x,p-1\right)  -\frac{\sin^{2}\left(
2x\right)  }{4p\left(  p-1\right)  }f_{n}\left(  x,p-2\right)  =0
\label{WF2ndOrderDifference}%
\end{equation}
We may solve this second-order equation by forward recursion under the
condition\footnote{Allowing support for negative integer $p$ leads to
singularities at $p=0$ and $p=1$. \ Therefore we rule out this possibility.}
that $f_{n}\left(  x,p<0\right)  =0$. \ The resulting WFs have support only
for non-negative integer $p$. We find that $f_{n}\left(  x,p<n\right)  =0$,
$f_{n}\left(  x,p=n\right)  $ is arbitrary, and all $f_{n}\left(
x,p>n\right)  $ are uniquely determined by (\ref{WF2ndOrderDifference}) in
terms of our choice for $f_{n}\left(  x,p=n\right)  $. \ For example, for the
ground state $n=0$, the choice $f_{0}\left(  x,0\right)  =1$ immediately
reproduces the terms in (\ref{GroundStateWF}). \ Similarly the choice (with
\emph{no} $x$ dependence)%
\begin{equation}
f_{n}\left(  x,p=n\right)  =\frac{1}{4^{n}\left(  n!\right)  ^{2}}%
\end{equation}
reproduces the series (\ref{LiouvilleWF}). \ The choice for $f_{n}\left(
x,p=n\right)  $ must be independent of $x$ so that $f_{n}\left(  x,p<n\right)
=0$ is consistent with (\ref{WFDifferentialDifferenceImag}). \ Thus the WFs
are determined by the $\bigstar$genvalue equation.

\subsection{Entwining the dual metric}

It follows immediately from (\ref{DualMetricAsWFs}) and (\ref{WF*genvalue}%
)\ that%
\begin{equation}
H\star\widetilde{R}\left(  x,p\right)  =\widetilde{R}\left(  x,p\right)
\star\overline{H} \label{HEntwinedWithDualMetric}%
\end{equation}
which amounts to just (\ref{DualDifferentialDifference}). \ (This should be
compared to (\ref{HEntwinedWithMetric}) given below.) \ The equivalent
operator statement is obvious, and follows directly from the Weyl
correspondence. \ (See Appendix F.) \ As direct verification of
(\ref{HEntwinedWithDualMetric}), we compute%
\begin{equation}
H\star\widetilde{R}\left(  x,p\right)  =\frac{p}{2\sin^{2}x}~\widetilde
{R}\left(  x,p\right)  =\widetilde{R}\left(  x,p\right)  \star\overline{H}%
\end{equation}
The relation (\ref{HEntwinedWithDualMetric}) is actually a special case of the
\textquotedblleft two star equation\textquotedblright\ given in \cite{CFZ},
Eqn(73). \ In the language of that paper, (\ref{HEntwinedWithDualMetric}) is
the ultra-local version for which $\mathcal{T}\left(  x,p;X,P\right)
=\delta\left(  x-X\right)  \delta\left(  p-P\right)  \widetilde{R}\left(
x,p\right)  $ and one Hamiltonian function ($H$) is the complex conjugate of
the other ($\mathcal{H}$). \ 

\subsection{Inhomogeneities for dual WFs}

For the dual WFs (\ref{DualWF}), by direct calculation using standard methods
(as in an Appendix B) we find
\begin{gather}
\widetilde{f_{n}}\star H=\left(  \left(  p+\frac{1}{2}i\partial_{x}\right)
^{2}+e^{2ix}e^{\partial_{p}}\right)  \frac{1}{2\pi}\int_{0}^{2\pi}%
\overline{\chi_{n}}\left(  x-y\right)  \chi_{n}\left(  x+y\right)
e^{2iyp}dy\nonumber\\
=\frac{1}{2\pi}\int_{0}^{2\pi}e^{2iyp}~\overline{\chi_{n}}\left(  x-y\right)
\left(  -\frac{1}{4}\left(  \partial_{y}+\partial_{x}\right)  ^{2}%
+e^{2i\left(  x+y\right)  }\right)  \chi_{n}\left(  x+y\right)  dy
\end{gather}
Then from (\ref{InhomoBesselEqn}) we have%
\begin{equation}
\widetilde{f_{n}}\star H=n^{2}\widetilde{f_{n}}+\frac{1}{2\pi}\int_{0}^{2\pi
}dy~e^{2iyp}~\overline{\chi_{n}}\left(  x-y\right)  \left\{
\begin{array}
[c]{cc}%
2ne^{i\left(  x+y\right)  } & \text{for \ \ }n\in\mathbb{N}_{\text{odd}}\\
& \\
\varepsilon_{n}e^{2i\left(  x+y\right)  } & \text{for \ \ }n\in\mathbb{N}%
_{\text{even}}%
\end{array}
\right.
\end{equation}
Taking $\chi_{n}\left(  x\right)  =A_{n}\left(  e^{ix}\right)  $\ and using
(\ref{ASeries}), this gives%
\begin{gather}
\widetilde{f_{n}}\star H=n^{2}\widetilde{f_{n}}+2^{n}n\sum_{k=0}^{\left\lfloor
n/2\right\rfloor }\frac{\left(  n-k-1\right)  !}{4^{k}k!}e^{i\left(
n-2k\right)  x}\frac{1}{2\pi}\int_{0}^{2\pi}dy\ e^{iy\left(  2p+2k-n\right)
}\left\{
\begin{array}
[c]{c}%
2ne^{i\left(  x+y\right)  }\\
\varepsilon_{n}e^{2i\left(  x+y\right)  }%
\end{array}
\right. \nonumber\\
=n^{2}\widetilde{f_{n}}+2^{n}n\sum_{k=0}^{\left\lfloor n/2\right\rfloor }%
\frac{\left(  n-k-1\right)  !}{4^{k}k!}e^{i\left(  n-2k\right)  x}\left\{
\begin{array}
[c]{c}%
2ne^{ix}\delta_{2p+2k,n-1}\\
\varepsilon_{n}e^{2ix}\delta_{2p+2k,n-2}%
\end{array}
\right.
\end{gather}
That is to say, in complete analogy with the dual wave functions, the dual WFs
obey \emph{inhomogeneous} $\bigstar$genvalue equations. \ 

In particular, for the dual ground state WF
\begin{equation}
\widetilde{f_{0}}\star H=e^{2ix}\delta_{p,-1}\ ,\ \ \ \overline{H}%
\star\widetilde{f_{0}}=e^{-2ix}\delta_{p,-1} \label{DualGroundStateWF*H}%
\end{equation}
and more generally%
\begin{equation}
\widetilde{f_{n}}\star H=n^{2}\widetilde{f_{n}}+\left\{
\begin{array}
[c]{cc}%
n^{2}\frac{\left(  \left\lfloor n/2\right\rfloor +p\right)  !}{\left(
\left\lfloor n/2\right\rfloor -p\right)  !}4^{1+p}e^{2i\left(  1+p\right)  x}
& \text{for\ odd\ }n\text{, if }\left\lfloor n/2\right\rfloor \geq
p\geq0\text{, otherwise }0\\
& \\
2n\frac{\left(  \left\lfloor n/2\right\rfloor +p\right)  !}{\left(
\left\lfloor n/2\right\rfloor -p-1\right)  !}4^{1+p}e^{2i\left(  2+p\right)
x} & \text{for even }n>0\text{, if }\left\lfloor n/2\right\rfloor \geq
1+p\geq0\text{, otherwise }0
\end{array}
\right.  \label{DualWF*H}%
\end{equation}

\begin{equation}
\overline{H}\star\widetilde{f_{n}}=n^{2}\widetilde{f_{n}}+\left\{
\begin{array}
[c]{cc}%
n^{2}\frac{\left(  \left\lfloor n/2\right\rfloor +p\right)  !}{\left(
\left\lfloor n/2\right\rfloor -p\right)  !}4^{1+p}e^{-2i\left(  1+p\right)  x}
& \text{for\ odd\ }n\text{, if }\left\lfloor n/2\right\rfloor \geq
p\geq0\text{, otherwise }0\\
& \\
2n\frac{\left(  \left\lfloor n/2\right\rfloor +p\right)  !}{\left(
\left\lfloor n/2\right\rfloor -p-1\right)  !}4^{1+p}e^{-2i\left(  2+p\right)
x} & \text{for even }n>0\text{, if }\left\lfloor n/2\right\rfloor \geq
1+p\geq0\text{, otherwise }0
\end{array}
\right.  \label{Hbar*DualWF}%
\end{equation}
As an alternative derivation, this can be checked by acting directly on the
dual WFs as explicit sums, (\ref{LiouvilleDualWF}).

These equations for the dual WFs do \emph{not} determine them uniquely, when
they are only required to be finite real solutions to the implied
differential-difference equations (again with $\widetilde{f_{n}}\left(
x,p=n\right)  $ independent of $x$) since we could always add to any
$\widetilde{f_{n}}$\ a solution to the homogeneous equation, namely an $f_{n}%
$. \ However, if we require the stronger conditions that we seek solutions
with \emph{finite} momentum support, such that $\widetilde{f_{n}}\left(
x,p<0\right)  =0$ and $\widetilde{f_{n}}\left(  x,p>n\right)  =0$,\ then the
solutions are uniquely determined, up to normalization. \ In fact, the form of
the inhomogeneities is also fixed by these requirements. \ This is similar to
what happens in the analysis of the dual wave functions, which may be
constructed without knowledge of the wave functions if we require that they be
finite polynomials in $\exp\left(  -ix\right)  $. \ In that analysis the form
of the inhomogeneities is also fixed (see \cite{CM} and \cite{CMS}).

In (\ref{DualGroundStateWF*H}), and in (\ref{DualWF*H}) and (\ref{Hbar*DualWF}%
) for other even $n$, we see that the inhomogeneity has support for $p=-1$,
even though $\widetilde{f_{n}}$ has support only for non-negative $p$. \ The
star product with the Hamiltonian has spread out the distribution, in a
typical quantum fashion, just slightly into the realm of negative $p$. \ Thus
our original expectation that we should be able to ignore all integer $p<0$
was not quite correct.

\subsection{$\left\langle H\right\rangle $ from dual WFs}

For imaginary Liouville QM the difference between (\ref{DualWF*H}) and
(\ref{Hbar*DualWF}) is%
\begin{equation}
\widetilde{f_{n}}\star H-\overline{H}\star\widetilde{f_{n}}=\left\{
\begin{array}
[c]{cc}%
2in^{2}\frac{\left(  \left\lfloor n/2\right\rfloor +p\right)  !}{\left(
\left\lfloor n/2\right\rfloor -p\right)  !}4^{1+p}\sin\left(  2x\left(
1+p\right)  \right)  & \text{for\ odd\ }n\text{, if }\left\lfloor
n/2\right\rfloor \geq p\geq0\\
& \\
4in\frac{\left(  \left\lfloor n/2\right\rfloor +p\right)  !}{\left(
\left\lfloor n/2\right\rfloor -p-1\right)  !}4^{1+p}\sin\left(  2x\left(
2+p\right)  \right)  & \text{for even }n>0\text{, if }\left\lfloor
n/2\right\rfloor \geq1+p\geq0
\end{array}
\right.
\end{equation}
The peculiar structure of the RHS combines with that of the metric
(\ref{DualPhaseSpaceMetric})\ to establish \emph{specifically} for the
Liouville model that the expectation of the Hamiltonian is real within the
phase-space framework. \
\begin{equation}
0=\frac{1}{2\pi}\int_{x,p}\hspace{-0.25in}%
%TCIMACRO{\tsum }%
%BeginExpansion
{\textstyle\sum}
%EndExpansion
~\widetilde{R}\left(  x,p\right)  \left(  \widetilde{f_{n}}\star
H-\overline{H}\star\widetilde{f_{n}}\right)  \label{Real<H>}%
\end{equation}
This holds just because $0=\int_{0}^{2\pi}\left(  \sin^{2}x\right)  ^{p}%
\sin\left(  2x\left(  1+p\right)  \right)  dx$ as well as $0=\int_{0}^{2\pi
}\left(  \sin^{2}x\right)  ^{p}\sin\left(  2x\left(  2+p\right)  \right)  dx$,
for integer $p\geq0$, and because the vanishing of the metric for negative $p$
conveniently eliminates the $p=-1$\ possibility for even $n$. \ 

Indeed, for all $n$%
\begin{equation}
n^{2}=\frac{%
%TCIMACRO{\dint _{x,p}}%
%BeginExpansion
{\displaystyle\int_{x,p}}
%EndExpansion
\hspace{-0.31in}\sum~\widetilde{R}\left(  x,p\right)  \left(  \widetilde
{f_{n}}\left(  x,p\right)  \star H\right)  }{%
%TCIMACRO{\dint _{x,p}}%
%BeginExpansion
{\displaystyle\int_{x,p}}
%EndExpansion
\hspace{-0.31in}\sum~\widetilde{R}\left(  x,p\right)  \widetilde{f_{n}}\left(
x,p\right)  }=\frac{%
%TCIMACRO{\dint _{x,p}}%
%BeginExpansion
{\displaystyle\int_{x,p}}
%EndExpansion
\hspace{-0.31in}\sum~\widetilde{R}\left(  x,p\right)  \left(  \overline
{H}\star\widetilde{f_{n}}\left(  x,p\right)  \right)  }{%
%TCIMACRO{\dint _{x,p}}%
%BeginExpansion
{\displaystyle\int_{x,p}}
%EndExpansion
\hspace{-0.31in}\sum~\widetilde{R}\left(  x,p\right)  \widetilde{f_{n}}\left(
x,p\right)  } \label{<H>}%
\end{equation}
despite the inhomogeneities in (\ref{DualWF*H}) and (\ref{Hbar*DualWF}),
because\ $0=\int_{0}^{2\pi}\left(  \sin^{2}x\right)  ^{p}\exp\left(
2ix\left(  1+p\right)  \right)  dx$ as well as $0=\int_{0}^{2\pi}\left(
\sin^{2}x\right)  ^{p}\exp\left(  2ix\left(  2+p\right)  \right)  dx$ for
integer $p\geq0$. \ Again the vanishing of the metric for negative $p$
eliminates the possibility of anguish at $p=-1$.\footnote{Otherwise the RHS
of, say, (\ref{DualGroundStateWF*H}), would be troublesome when multiplied by
$\left.  \left(  \sin x\right)  ^{2p}\right\vert _{p=-1}$ and integrated over
$x$, although not impossible to handle (cf. Cauchy's principal value
prescription).}

However it must be said that (\ref{Real<H>}) can also be established more
\emph{generally} by the method of combining (\ref{HEntwinedWithDualMetric}%
)\ with the \textquotedblleft\emph{Lone} Star Lemma\textquotedblright%
\ \cite{ZFC}\ which allows us the option of inserting or removing a
\emph{single} $\star$ from the phase space summand/integrand, hence to write
\begin{align}
\frac{1}{2\pi}\int_{x,p}\hspace{-0.25in}%
%TCIMACRO{\tsum }%
%BeginExpansion
{\textstyle\sum}
%EndExpansion
~\widetilde{R}\left(  x,p\right)  \left(  \widetilde{f_{n}}\star H\right)   &
=\frac{1}{2\pi}\int_{x,p}\hspace{-0.25in}%
%TCIMACRO{\tsum }%
%BeginExpansion
{\textstyle\sum}
%EndExpansion
~\widetilde{R}\left(  x,p\right)  \star\widetilde{f_{n}}\star H=\frac{1}{2\pi
}\int_{x,p}\hspace{-0.25in}%
%TCIMACRO{\tsum }%
%BeginExpansion
{\textstyle\sum}
%EndExpansion
~H\star\widetilde{R}\left(  x,p\right)  \star\widetilde{f_{n}}\nonumber\\
&  =\frac{1}{2\pi}\int_{x,p}\hspace{-0.25in}%
%TCIMACRO{\tsum }%
%BeginExpansion
{\textstyle\sum}
%EndExpansion
~\widetilde{R}\left(  x,p\right)  \star\overline{H}\star\widetilde{f_{n}%
}=\frac{1}{2\pi}\int_{x,p}\hspace{-0.25in}%
%TCIMACRO{\tsum }%
%BeginExpansion
{\textstyle\sum}
%EndExpansion
~\widetilde{R}\left(  x,p\right)  \left(  \overline{H}\star\widetilde{f_{n}%
}\right)
\end{align}

\section{A sesquilinear star product and bracket}

There is a compelling formal way to express the entwining relation
(\ref{HEntwinedWithDualMetric}) between the dual metric and the Hamiltonian
which suggests it would be appropriate to abbreviate quasi-hermitian to
q-hermitian in the title of this paper. \ Certainly, the structure of
(\ref{HEntwinedWithDualMetric}) brings to mind the deformations of commutators
and braiding relations such as occur in q-algebras and quantum groups
\cite{Chari}. \ To pursue that statement, let us rewrite the entwining
relation as \
\begin{equation}
H\star\widetilde{R}\left(  x,p\right)  =\widetilde{R}\left(  x,p\right)
\star\mathbb{K}H\mathbb{K}%
\end{equation}
where $\mathbb{K}$\ is the anti-linear operation of complex conjugation.
\ Since $\widetilde{R}$\ is in fact a \emph{real} function, this may also be
written as%
\begin{equation}
H\star\mathbb{K}\widetilde{R}\left(  x,p\right)  =\widetilde{R}\left(
x,p\right)  \star\mathbb{K}H
\end{equation}
By defining a \textquotedblleft sesquilinear star product\textquotedblright%
\begin{equation}
\circledS\equiv\star~\mathbb{K}%
\end{equation}
and a corresponding modification of the Moyal bracket \cite{Moyal}, the
entwining relation becomes simply\footnote{For the less symmetric Voros
product $\triangleleft\ \equiv\exp\left(  i\overleftarrow{\partial}%
_{x}\overrightarrow{\partial}_{p}\right)  $ (see the Appendix in
\cite{Voros1978}) an analogous construction is not quite so elegant, since in
that case the metric would not be a real function.}%
\begin{equation}
0=\left[  H,\widetilde{R}\right]  _{\circledS}\equiv H\circledS\widetilde
{R}-\widetilde{R}\circledS H
\end{equation}
A form of sesquilinearity is evident, although left$\leftrightarrow$right
ordering dependent, in the properties%
\begin{align}
C\circledS\left(  \alpha A+\beta B\right)   &  =\left(  C\circledS A\right)
\alpha+\left(  C\circledS B\right)  \beta=\overline{\alpha}\left(  C\circledS
A\right)  +\overline{\beta}\left(  C\circledS B\right) \\
\left(  \alpha A+\beta B\right)  \circledS C  &  =\alpha\left(  A\circledS
C\right)  +\beta\left(  B\circledS C\right)  =\left(  A\circledS C\right)
\overline{\alpha}+\left(  B\circledS C\right)  \overline{\beta}%
\end{align}
for constants $\alpha$\ and\ $\beta$. \ Also note that $\star\mathbb{K\neq
K}\star$, rather%
\begin{equation}
\overline{\star}=\mathbb{K}\star\mathbb{K}=\exp\left(  -\frac{i}%
{2}~\overleftarrow{\partial_{x}}~\overrightarrow{\partial_{p}}+\frac{i}%
{2}~\overleftarrow{\partial_{p}}~\overrightarrow{\partial_{x}}\right)
\end{equation}
This \textquotedblleft star-bar\textquotedblright\ is again an associative but
non-commutative product, like (\ref{GroenewoldStar}), and just amounts to
flipping the sign of the deformation parameter, $\hbar$ ($\equiv1$ throughout
this paper).

\section{Other phase space distributions}

We may also take star products of the various WFs. \ This leads in a routine
way to a larger class of distributions on the phase space, although the
details for the present circumstances are rather novel not only because the
functions are periodic in $x$ but also because $\chi_{n}\neq\overline{\psi
_{n}}$. \ This means that in general there is a third class of
\textquotedblleft hybrid\textquotedblright\ WFs involving mixed bilinears in
the wave functions and their duals. \ 

\subsection{Hybrid WFs}

By direct calculation, with $p\in\mathbb{Z}$, we find%
\begin{align}
f_{k}\left(  x,p\right)  \star\widetilde{f_{n}}\left(  x,p\right)   &
=g_{n}\left(  x,p\right)  ~\delta_{k,n}\label{WF*DualWF}\\
&  \ \ \ \nonumber\\
g_{n}\left(  x,p\right)   &  \equiv\frac{1}{2\pi}\int_{0}^{2\pi}\psi
_{n}\left(  x-y\right)  ~\chi_{n}\left(  x+y\right)  ~e^{2iyp}%
dy\label{MixedUpWF}\\
&  \ \ \ \nonumber\\
g_{n}\left(  x,p\right)   &  =\frac{\left(  -1\right)  ^{p-n}}{4^{p-n}\left(
n-1\right)  !}~e^{2i\left(  p-n\right)  x}\sum_{k=0}^{\left\lfloor
n/2\right\rfloor }\frac{\left(  n-k-1\right)  !\left(  -1\right)  ^{k}}%
{4^{2k}k!\left(  p+k\right)  !\left(  p+k-n\right)  !}~e^{4ikx}%
\end{align}
the latter from the series (\ref{JSeries}) and (\ref{ASeries}). \ So
$g_{n}\left(  x,p\right)  \neq0$ if and only if $p\geq n-\left\lfloor
n/2\right\rfloor $, and thus has infinite support in $p$ similar to $f_{n}$.
\ In particular, again for any phase-space point $\left(  x,p\right)
\in\mathbb{S}^{1}\times\mathbb{Z}$,
\begin{subequations}
\begin{align}
g_{0}\left(  x,p\right)   &  =\frac{\left(  -1\right)  ^{p}}{4^{p}\left(
p!\right)  ^{2}}~e^{2ipx}\text{ \ \ for \ \ }p\geq0\\
g_{1}\left(  x,p\right)   &  =\frac{\left(  -1\right)  ^{p-1}}{4^{p-1}%
p!\left(  p-1\right)  !}~e^{2i\left(  p-1\right)  x}\text{ \ \ for \ \ }%
p\geq1\\
g_{2}\left(  x,p\right)   &  =\frac{\left(  -1\right)  ^{p}}{4^{p}\left(
p+1\right)  !\left(  p-1\right)  !}~\left(  16\left(  p^{2}-1\right)
e^{2i\left(  p-2\right)  x}-e^{2ixp}\right)  \text{ \ \ for \ \ }p\geq1
\end{align}
etc. \ Without further calculation we also find
\end{subequations}
\begin{align}
\widetilde{f_{k}}\left(  x,p\right)  \star f_{n}\left(  x,p\right)   &
=\overline{g_{n}\left(  x,p\right)  }~\delta_{k,n}\label{DualWF*WF}\\
&  \ \ \nonumber\\
\overline{g_{n}\left(  x,p\right)  }  &  \equiv\frac{1}{2\pi}\int_{0}^{2\pi
}\overline{\chi_{n}\left(  x-y\right)  }~\overline{\psi_{n}\left(  x+y\right)
}~e^{2iyp}dy \label{MixedUpWFConjugate}%
\end{align}
as consequences of the previous relations, upon making use of $\widetilde
{f_{n}}\star f_{n}=\overline{f_{n}\star\widetilde{f_{n}}}$. \ Summing and
integrating over $p$ and $x$ the RHS of either (\ref{WF*DualWF}) or
(\ref{DualWF*WF}) gives back (\ref{WFDualWFOrthonormality}). \ A sum over
integer $p$ suffices in this case.\footnote{However, in other situations it
may be necessary to sum over \emph{semi-integer} $p$, i.e. all $p$ such that
$2p\in\mathbb{Z}$. \ While $e_{n}$, $f_{n}$, $\widetilde{f_{n}}$, and $g_{n}$
only have support for integer $p$, there are \textquotedblleft
non-diagonal\textquotedblright\ variants of these distributions, namely
$e_{k,n}$, $f_{k,n}$, $\widetilde{f_{k,n}}$, and $g_{k,n}$ for $k\neq n$\ (see
Appendix D), which have support for semi-integer $p$ when $k-n$ is an odd
integer. \ In dealing with such nondiagonal cases, some care is required to
respect the positivity of the metric.} \ These hybrids also obey a form of the
completeness relation that follows from the corresponding statement for wave
functions and their duals.%
\begin{equation}
\frac{w}{w-z}=\sum_{n=0}^{\infty}A_{n}\left(  w\right)  J_{n}\left(  z\right)
\label{Cauchy}%
\end{equation}
However, the Wigner transform requires some regulation to give a precise
meaning to $\sum_{k=0}^{\infty}g_{k}\left(  x,p\right)  $.

Finally, we find the remaining simple star products%
\begin{align}
g_{k}\left(  x,p\right)  \star g_{n}\left(  x,p\right)   &  =g_{n}\left(
x,p\right)  ~\delta_{k,n}\\
g_{k}\left(  x,p\right)  \star f_{n}\left(  x,p\right)   &  =f_{n}\left(
x,p\right)  ~\delta_{k,n}\\
\widetilde{f_{k}}\left(  x,p\right)  \star g_{n}\left(  x,p\right)   &
=\widetilde{f_{n}}\left(  x,p\right)  ~\delta_{k,n}%
\end{align}
The first of these is of the usual esthetic form that holds for non-hybrid WFs
in familiar hermitian quantum mechanical systems (e.g. the free particle WFs
satisfy $e_{k}\left(  x,p\right)  \star e_{n}\left(  x,p\right)  =e_{n}\left(
x,p\right)  ~\delta_{k,n}$). \ However, unlike in that non-hybrid hermitian
situation, the hybrid $g_{n}$ is \emph{not} real. \ A direct consequence of
the second of these star products is that all the $g_{n}$ are right-$\star
$-mapped by the dual metric into $f_{n}$.
\begin{equation}
g_{n}\left(  x,p\right)  \star\widetilde{R}\left(  x,p\right)  =\varepsilon
_{n}f_{n}\left(  x,p\right)
\end{equation}
This follows immediately from (\ref{DualMetricAsWFs}).

Once again, the form of the various star products of WFs can most easily be
seen through the use of density operator methods, as in Appendix E. \ However,
for devotees of phase-space methods, we make the following technical points.
\ The results are obtained directly from the structure of the Wigner
transforms (\ref{WF}), (\ref{DualWF}), and (\ref{MixedUpWF}), and the
orthogonality of the wave functions and their duals. \ For example, consider
(\ref{WF*DualWF}). \ Applying the star product to each of the integral
representations (\ref{WF}) and (\ref{DualWF}), and carrying out the requisite
variable shifts on the integrands, the resulting coupled integrals $\int
_{0}^{2\pi}dy_{1}\int_{0}^{2\pi}dy_{2}\cdots$ can be split into uncoupled
integrals $\int_{2\pi}^{4\pi}d\left(  y_{1}+y_{2}\right)  \cdots\times\frac
{1}{2}\int_{-2\pi}^{2\pi}d\left(  y_{1}-y_{2}\right)  \cdots$ upon making use
of the periodicity of the integrands, hence the final result can be obtained.
\ (Further details are in Appendix C.)

Moving on, we compute $H\star g_{k}$ and $g_{k}\star H$. \ The first of these
is easily seen to be%
\begin{equation}
H\star g_{n}=n^{2}g_{n}%
\end{equation}
but the other has an inhomogeneity.%
\begin{equation}
g_{n}\star H=n^{2}g_{n}+\left\{
\begin{array}
[c]{cc}%
2nh_{n,-1} & \text{for \ \ }n\in\mathbb{N}_{\text{odd}}\\
& \\
\varepsilon_{n}h_{n,-2} & \text{for \ \ }n\in\mathbb{N}_{\text{even}}%
\end{array}
\right.  \label{gInhomo}%
\end{equation}
where the RHS involves non-diagonal versions of hybrid WFs, of the type
indicated (see Appendix D and set $\phi_{l}\left(  x\right)  =\exp ilx$).
\ Explicitly%
\begin{align}
h_{n,l}\left(  x,p\right)   &  =\frac{1}{2\pi}\int_{0}^{2\pi}\psi_{n}\left(
x-y\right)  ~e^{-il\left(  x+y\right)  }~e^{2iyp}dy=\frac{1}{2^{n}%
}e^{2i\left(  p-l\right)  x}\sum_{k=0}^{\infty}\frac{\left(  -1\right)  ^{k}%
}{4^{k}k!\left(  k+n\right)  !}\delta_{2p,n+l+2k}\nonumber\\
&  =e^{2i\left(  p-l\right)  x}\frac{\left(  -1\right)  ^{p-\frac{l+n}{2}}%
}{2^{2p-l}\left(  p-\frac{l+n}{2}\right)  !\left(  p-\frac{l-n}{2}\right)  !}
\label{hNondiagonal}%
\end{align}
for \emph{integer} $\left(  p-\tfrac{1}{2}\left(  l+n\right)  \right)  \geq0$,
but zero otherwise, upon using the series (\ref{JSeries}). \ We note that the
right action of $H$ has spread the $g_{n}$ distributions in momentum, through
the effects of the inhomogeneities, and in particular has produced a
contribution at $p=-1$ from $g_{0}$.

In similar language, we could have abbreviated the explicit results in
(\ref{DualWF*H}) and (\ref{Hbar*DualWF}) as
\begin{equation}
\overline{H}\star\widetilde{f_{n}}=n^{2}\widetilde{f_{n}}+\left\{
\begin{array}
[c]{cc}%
2n~\widetilde{h_{-1,n}} & \text{for \ \ }n\in\mathbb{N}_{\text{odd}}\\
& \\
\varepsilon_{n}~\widetilde{h_{-2,n}} & \text{for \ \ }n\in\mathbb{N}%
_{\text{even}}%
\end{array}
\right.  \label{DualWFInhomo1}%
\end{equation}

\begin{equation}
\widetilde{f_{n}}\star H=n^{2}\widetilde{f_{n}}+\left\{
\begin{array}
[c]{cc}%
2n~\overline{\widetilde{h_{-1,n}}} & \text{for \ \ }n\in\mathbb{N}%
_{\text{odd}}\\
& \\
\varepsilon_{n}~\overline{\widetilde{h_{-2,n}}} & \text{for \ \ }%
n\in\mathbb{N}_{\text{even}}%
\end{array}
\right.  \label{DualWFInhomo2}%
\end{equation}
where again the RHSs involve non-diagonal versions of hybrid WFs, of the type
indicated (again see Appendix D and set $\phi_{l}\left(  x\right)  =\exp
ilx$). \ Explicitly, upon using the series (\ref{ASeries}), \
\begin{align}
\widetilde{h_{l,n}}\left(  x,p\right)   &  =\frac{1}{2\pi}\int_{0}^{2\pi
}e^{il\left(  x-y\right)  }~\chi_{n}\left(  x+y\right)  ~e^{2iyp}%
dy=2^{n}ne^{2i\left(  l-p\right)  x}\sum_{k=0}^{\left\lfloor n/2\right\rfloor
}\frac{\left(  n-k-1\right)  !}{4^{k}k!}\delta_{2p,l+n-2k}\nonumber\\
&  =2^{2p-l}ne^{2i\left(  l-p\right)  x}\frac{\left(  p-\frac{l-n}%
{2}-1\right)  !}{\left(  \frac{l+n}{2}-p\right)  !} \label{iNondiagonal}%
\end{align}
for \emph{integer} $\left(  \tfrac{1}{2}\left(  l+n\right)  -p\right)  \geq0$
and $\left(  \tfrac{1}{2}\left(  l+n\right)  -p\right)  \leq\left\lfloor
n/2\right\rfloor $, but zero otherwise.

It is significant that the RHSs of (\ref{gInhomo}), (\ref{DualWFInhomo1}), and
(\ref{DualWFInhomo2}) only have support for $p\in\mathbb{Z}$, and so conform
to our original expectation about momentum quantization. \ But in general,
(\ref{hNondiagonal}) and (\ref{iNondiagonal}) do not vanish when
$2p\in\mathbb{Z}$, where contributions at semi-integer $p$ can occur when
$l+n$ is an \emph{odd} integer. \ So, upon considering these more general,
nondiagonal WFs, \emph{the phase space must be expanded} to include all points
$\left(  x,p\right)  \in\mathbb{S}^{1}\times\mathbb{Z}/2$.

\subsection{More hybrid WFs}

As anticipated in our abbreviated expressions for the various inhomogeneities,
(\ref{hNondiagonal}) and (\ref{iNondiagonal}), we may continue the process of
hybridizing WFs by constructing Wigner transforms of other pairs of functions.
\ If one function is an imaginary Liouville eigenfunction while the other is a
free particle solution, we are led to define the hybrid%
\begin{equation}
h_{n}\left(  x,p\right)  \equiv\frac{1}{2\pi}\int_{0}^{2\pi}\psi_{n}\left(
x-y\right)  ~\overline{\phi_{n}}\left(  x+y\right)  ~e^{2iyp}dy
\label{LiouvilleFreeHybridWFs}%
\end{equation}
where $\psi_{n}$ are the Bessels, and $\phi_{n}$ are free solutions, $2\pi
$-periodic, and orthonormal in the usual sense.%
\begin{equation}
-\frac{d^{2}}{dx^{2}}\phi_{n}\left(  x\right)  =n^{2}\phi_{n}\left(  x\right)
\ ,\ \ \ \frac{1}{2\pi}\int_{0}^{2\pi}\overline{\phi_{k}}\left(  x\right)
~\phi_{n}\left(  x\right)  dx=\delta_{k,n} \label{FreeParticleWaveFunctions}%
\end{equation}
Then by standard techniques we have%
\begin{align}
H\star h_{n}\left(  x,p\right)   &  =n^{2}~h_{n}\left(  x,p\right)
=h_{n}\left(  x,p\right)  \star p^{2}\label{H*h}\\
& \nonumber\\
\overline{h_{n}\left(  x,p\right)  }\star\overline{H}  &  =n^{2}%
~\overline{h_{n}\left(  x,p\right)  }=p^{2}\star\overline{h_{n}\left(
x,p\right)  } \label{h*Hbar}%
\end{align}
as well as%
\begin{equation}
h_{k}\left(  x,p\right)  \star\overline{h_{n}\left(  x,p\right)  }%
=f_{n}\left(  x,p\right)  ~\delta_{k,n} \label{h*hbar=f}%
\end{equation}
and so on. \ However, since the Bessels on the circle are not orthonormal, we
do not have similar simple results for $\overline{h_{n}\left(  x,p\right)
}\star h_{k}\left(  x,p\right)  $, although this does not really seem to
matter in practice. \ 

These $\star$ product results can be established without explicit forms for
either the $h_{n}$ or the $f_{n}$, but it is perhaps useful to have such
expressions in hand. \ Unlike the situation for the $f_{n}$, as given in
(\ref{LiouvilleWF}), there is more freedom in the construction of the $h_{n}$
since we have a choice between taking $\phi_{n}\left(  x\right)  $ to be
$e^{inx}$ or $e^{-inx}$, for $n\geq0$, or some linear combination of the two.
\ Explicit results depend on how we make this choice. \ Perhaps the simplest
choice, conceptually, is to take only right- or left-moving plane waves for
the $\phi_{n}$. \ For example, when $\phi_{n}$ is right-moving, then
$\overline{\phi_{n}}\left(  x\right)  =e^{-inx}$ for $n\geq0$ is left-moving,
and
\begin{equation}
h_{n}^{\text{R}}\left(  x,p\right)  =\frac{1}{2\pi}\int_{0}^{2\pi}J_{n}\left(
e^{i\left(  x-y\right)  }\right)  ~e^{-in\left(  x+y\right)  }~e^{2iyp}%
dy=\frac{2^{n}\left(  -1\right)  ^{p-n}}{4^{p}p!\left(  p-n\right)
!}e^{2i\left(  p-n\right)  x}\text{ \ \ for \ \ }p\geq n
\label{RChiralHybridWFs}%
\end{equation}
but vanishes for $p<n$. \ On the other hand, if we take only left-moving plane
waves for the $\phi_{n}$, then $\overline{\phi_{n}}\left(  x\right)  =e^{inx}$
for $n\geq0$ is right-moving, and%
\begin{equation}
h_{n}^{\text{L}}\left(  x,p\right)  =\frac{1}{2\pi}\int_{0}^{2\pi}J_{n}\left(
e^{i\left(  x-y\right)  }\right)  ~e^{in\left(  x+y\right)  }~e^{2iyp}%
dy=\frac{\left(  -1\right)  ^{p}}{4^{p}2^{n}p!\left(  p+n\right)
!}e^{2i\left(  p+n\right)  x}\text{ \ \ for \ \ }p\geq0
\label{LChiralHybridWFs}%
\end{equation}
but vanishes for $p<0$.

In conjunction with the $h_{n}\left(  x,p\right)  $, we also define the duals%
\begin{equation}
\widetilde{h_{n}}\left(  x,p\right)  \equiv\frac{1}{2\pi}\int_{0}^{2\pi}%
\phi_{n}\left(  x-y\right)  ~\chi_{n}\left(  x+y\right)  ~e^{2iyp}dy
\end{equation}
where $\chi_{n}$ are the duals to the $\psi_{n}$. \ Various star products
follow immediately.%
\begin{equation}
\widetilde{h_{n}}\left(  x,p\right)  \star H=n^{2}~\widetilde{h_{n}}\left(
x,p\right)  +\frac{1}{2\pi}\int_{0}^{2\pi}\phi_{n}\left(  x-y\right)
~e^{2iyp}\left\{
\begin{array}
[c]{cc}%
\varepsilon_{n}e^{2i\left(  x+y\right)  } & \text{for even }n\geq0\\
2ne^{i\left(  x+y\right)  } & \text{for odd }n>0
\end{array}
\right.
\end{equation}%
\begin{equation}
p^{2}\star\widetilde{h_{n}}\left(  x,p\right)  =n^{2}~\widetilde{h_{n}}\left(
x,p\right)
\end{equation}%
\begin{align}
\widetilde{h_{k}}\left(  x,p\right)  \star f_{n}\left(  x,p\right)   &
=\overline{h_{n}\left(  x,p\right)  }~\delta_{k,n}\label{hDual*f}\\
& \nonumber\\
f_{k}\left(  x,p\right)  \star\overline{\widetilde{h_{n}}\left(  x,p\right)
}  &  =h_{n}\left(  x,p\right)  ~\delta_{k,n}\label{f*hDual}\\
& \nonumber\\
h_{k}\left(  x,p\right)  \star\widetilde{h_{n}}\left(  x,p\right)   &
=g_{n}\left(  x,p\right)  ~\delta_{k,n}\\
& \nonumber\\
\overline{\widetilde{h_{k}}\left(  x,p\right)  }\star\widetilde{h_{n}}\left(
x,p\right)   &  =\widetilde{f_{n}}\left(  x,p\right)  ~\delta_{k,n}\\
& \nonumber\\
\widetilde{h_{k}}\left(  x,p\right)  \star h_{n}\left(  x,p\right)   &
=e_{n}\left(  x,p\right)  ~\delta_{k,n}%
\end{align}
where the last of these brings us back to the plane wave WFs we introduced
initially, in (\ref{FreeWF}). \ Also, all the $\overline{\widetilde{h_{n}}}$
and $\widetilde{h_{n}}$ are respectively left- and right-$\star$-mapped by the
dual metric into the $h_{n}$ and their conjugates.
\begin{align}
\widetilde{R}\left(  x,p\right)  \star\overline{\widetilde{h_{n}}\left(
x,p\right)  }  &  =\varepsilon_{n}h_{n}\left(  x,p\right) \\
\widetilde{h_{n}}\left(  x,p\right)  \star\widetilde{R}\left(  x,p\right)   &
=\varepsilon_{n}\overline{h_{n}\left(  x,p\right)  }%
\end{align}
This follows immediately from (\ref{f*hDual}), (\ref{hDual*f}), and
(\ref{DualMetricAsWFs}).

More explicit results again depend on how we choose the free particle
solutions. \ Choosing right-moving plane waves, with $n\geq0$ as above, we
have%
\begin{equation}
\widetilde{h_{n}^{\text{R}}}\left(  x,p\right)  \equiv\frac{1}{2\pi}\int
_{0}^{2\pi}e^{in\left(  x-y\right)  }~\chi_{n}\left(  x+y\right)
~e^{2iyp}dy=2^{2p-n}ne^{2i\left(  n-p\right)  x}\frac{\left(  p-1\right)
!}{\left(  n-p\right)  !}%
\end{equation}
for $0\leq n-p\leq\left\lfloor n/2\right\rfloor $, but zero otherwise.
\ Choosing left-moving plane waves, with $n\geq0$ , we have%
\begin{equation}
\widetilde{h_{n}^{\text{L}}}\left(  x,p\right)  \equiv\frac{1}{2\pi}\int
_{0}^{2\pi}e^{-in\left(  x-y\right)  }~\chi_{n}\left(  x+y\right)
~e^{2iyp}dy=2^{2p+n}ne^{-2ix\left(  n+p\right)  }\frac{\left(  n+p-1\right)
!}{\left(  -p\right)  !}%
\end{equation}
for $0\leq-p\leq\left\lfloor n/2\right\rfloor $, but zero otherwise. \ For
these specific choices we compute the right action of $H$ to be%
\begin{align}
\widetilde{h_{n}^{\text{R}}}\left(  x,p\right)  \star H  &  =n^{2}%
~\widetilde{h_{n}^{\text{R}}}\left(  x,p\right)  +\left\{
\begin{array}
[c]{cc}%
\varepsilon_{n}e^{i\left(  n+2\right)  x}\delta_{2p,n-2} & \text{for even
}n\geq0\\
2ne^{i\left(  n+1\right)  x}\delta_{2p,n-1} & \text{for odd }n>0
\end{array}
\right. \\
& \nonumber\\
\widetilde{h_{n}^{\text{L}}}\left(  x,p\right)  \star H  &  =n^{2}%
~\widetilde{h_{n}^{\text{L}}}\left(  x,p\right)  +\left\{
\begin{array}
[c]{cc}%
\varepsilon_{n}e^{i\left(  -n+2\right)  x}\delta_{2p,-n-2} & \text{for even
}n\geq0\\
2ne^{i\left(  -n+1\right)  x}\delta_{2p,-n-1} & \text{for odd }n>0
\end{array}
\right.
\end{align}
The inhomogeneities here can be identified with nondiagonal free particle WFs,
$e_{n,-2}^{\text{R}}$ \& $e_{n,-1}^{\text{R}}$, and $e_{n,2}^{\text{L}}$ \&
$e_{n,1}^{\text{L}}$. \ It is significant that the inhomogeneities have
support only for integer $p$.\footnote{Although in general the nondiagonal
free WFs have support at semi-integer $p$ when $l+n$ is odd: \ $e_{n,l}%
^{\text{R}}\left(  x,p\right)  =e^{i\left(  n-l\right)  x}\delta_{2p,n+l}$ and
$e_{n,l}^{\text{L}}\left(  x,p\right)  =e^{i\left(  -n+l\right)  x}%
\delta_{2p,-n-l}$. \ So once again, considering these more general,
nondiagonal WFs, the phase space must be expanded to include all points
$\left(  x,p\right)  \in\mathbb{S}^{1}\times\mathbb{Z}/2$.} \ But note that
this includes, e.g., $p=-1$ when $n=0$ even though only $\widetilde
{h_{0}^{\text{R,L}}}\left(  x,p=0\right)  \neq0$. \ Once again the star
product of $H$ with the $\widetilde{h_{n}}\left(  x,p\right)  $ has spread the
distributions on the phase space to give contributions outside their initial
momentum support through the effects of the inhomogeneities. \ (For
$\widetilde{h_{n>0}^{\text{L}}}\left(  x,p\right)  $ cases, the initial
support was only for negative momentum, and the star action of $H$ also gives
only negative momentum contributions, but spread out nonetheless.)

\section{Direct solutions of the dual metric equation}

Basic solutions to (\ref{HEntwinedWithDualMetric}), or equivalently
(\ref{DualDifferentialDifference}), are obtained by separation of variables.
\ We find two classes of solutions. \ The first of these is non-singular for
all real $p$, although there are zeroes for negative integer $p$.
\begin{equation}
\widetilde{R}\left(  x,p;s\right)  =\frac{1}{s^{p}\Gamma\left(  1+p\right)
}\exp\left(  -\frac{1}{2}s\cos2x\right)  \label{BasicDualWFMetric}%
\end{equation}
For real $s$ this is real and positive definite on the positive momentum
half-line. \ The other class of solutions has poles for all positive integer
$p$.%
\begin{equation}
\widetilde{R}_{\text{other}}\left(  x,p;s\right)  =\frac{\Gamma\left(
-p\right)  }{s^{p}}\exp\left(  \frac{1}{2}s\cos2x\right)
\label{OtherBasicDualWFMetric}%
\end{equation}
For later use we also compute the left- and right-actions of $H$ and
$\overline{H}$ on $\widetilde{R}\left(  x,p;s\right)  $.%
\begin{equation}
H\star\widetilde{R}\left(  x,p;s\right)  =\left(  p^{2}-\frac{1}{8}%
s^{2}+\left(  p-\frac{1}{2}\right)  s\cos2x+\frac{1}{8}s^{2}\cos4x\right)
\widetilde{R}\left(  x,p;s\right)  =\widetilde{R}\left(  x,p;s\right)
\star\overline{H} \label{H*DualR}%
\end{equation}

Linear combinations of (\ref{BasicDualWFMetric})\ and/or
(\ref{OtherBasicDualWFMetric})\ are also solutions. \ This permits us to build
a \textquotedblleft composite\textquotedblright\ metric from members of the
first class by using a contour integral representation. \ For $t>0$%
\begin{equation}
\widetilde{R}\left(  x,p;s,t\right)  \equiv\frac{1}{2\pi i}\int_{-\infty
}^{\left(  0+\right)  }\widetilde{R}\left(  x,p;s\tau\right)  \frac{e^{t\tau}%
}{\tau}d\tau\label{DualRContour}%
\end{equation}
The contour begins at $-\infty$, with $\arg\tau=-\pi$, proceeds below the real
$\tau$\ axis towards the origin, loops in the positive, counterclockwise sense
around the origin (hence the $\left(  0+\right)  $ notation), and then
continues above the real $\tau$ axis back to $-\infty$, with $\arg\tau=+\pi$.
\ By construction, $\widetilde{R}\left(  x,p;s,t\right)  $ actually depends
only on the ratio $t/s$. \ Evaluation of the contour integral gives%
\begin{equation}
\widetilde{R}\left(  x,p;s,t\right)  =\left(  \frac{t}{s}-\tfrac{1}{2}%
\cos2x\right)  ^{p}\frac{1}{\left(  \Gamma\left(  1+p\right)  \right)  ^{2}}
\label{DualRComposite}%
\end{equation}
where we have made use of
\begin{equation}
\frac{1}{\Gamma\left(  1+p\right)  }=\frac{1}{2\pi i}\int_{-\infty}^{\left(
0+\right)  }\tau^{-p-1}e^{\tau}d\tau\label{1/Gamma}%
\end{equation}
From $\widetilde{R}\left(  x,p;s,t\right)  $ we therefore recover our original
dual metric by setting $s=2t$. \
\begin{equation}
\widetilde{R}\left(  x,p;2t,t\right)  =\frac{\left(  \sin^{2}x\right)  ^{p}%
}{\left(  \Gamma\left(  p+1\right)  \right)  ^{2}}=\widetilde{R}\left(
x,p\right)  \label{DualRAsDualRComposite}%
\end{equation}

\section{The $\bigstar$ root of the dual metric}

\subsection{$\widetilde{S}$ as a direct solution of an entwining equation}

We look for an equivalence between the Liouville $H=p^{2}+e^{2ix}$ and the
free particle $\mathbb{H}=p^{2}$ as given by solutions of%
\begin{equation}
\widetilde{S}\left(  x,p\right)  ^{-1}\star H\star\widetilde{S}\left(
x,p\right)  =p^{2}%
\end{equation}
or, barring complete invertibility, as solutions of the entwining equation
(cf. (\ref{SEntwinedWithH}) below)%
\begin{equation}
H\ \star\ \widetilde{S}\left(  x,p\right)  =\widetilde{S}\left(  x,p\right)
\ \star\ p^{2} \label{DualSEntwinedWithH}%
\end{equation}
This is again a special case of the ultra-local two star equation\ as given in
\cite{CFZ}. \ For the Liouville--free-particle case, this amounts to an
equation similar to that for $\widetilde{R}$, but inherently complex.
\begin{equation}
2ip\partial_{x}\widetilde{S}\left(  x,p\right)  =e^{2ix}e^{-\partial_{p}%
}\widetilde{S}\left(  x,p\right)  =e^{2ix}\widetilde{S}\left(  x,p-1\right)
\label{DualSDifferentialDifference}%
\end{equation}
Once again solutions are easily found through the use of a product ansatz.
\ For any value of a parameter $s$, we find two immediate solutions:%
\begin{equation}
\widetilde{S}\left(  x,p;s\right)  =\frac{1}{s^{p}\Gamma\left(  1+p\right)
}\exp\left(  -\frac{1}{4}s\exp\left(  2ix\right)  \right)
\label{BasicDualSSolution}%
\end{equation}%
\begin{equation}
\widetilde{S}_{\text{other}}\left(  x,p;s\right)  =\frac{1}{s^{p}}%
\Gamma\left(  -p\right)  \exp\left(  \frac{1}{4}s\exp\left(  2ix\right)
\right)  \label{OtherBasicDualSSolution}%
\end{equation}
The first of these is a \textquotedblleft good\textquotedblright\ solution for
$p\in\left(  -1,\infty\right)  $, say, while the second is good for
$p\in\left(  -\infty,0\right)  $, thereby providing an overlapping pair of
solutions that cover the entire real $p$ axis, but \emph{not} smoothly or
continuously. \ These solutions could always be multiplied by periodic
functions of $p$, $\exp\left(  2i\pi np\right)  $ for $n\in\mathbb{Z}$, but
for integer-valued $p$ (which is our primary interest) this has no effect. \ 

The two solutions for $\widetilde{S}$\ are brought closer in appearance by
shifting $s\rightarrow s\exp\left(  \pm i\pi/2\right)  $ and using the
reflection relation for the $\Gamma$: $\ \Gamma\left(  -p\right)  =\dfrac
{-\pi}{\Gamma\left(  1+p\right)  \sin\pi p}$ \ \ Thus we may take as our two
basic solutions the alternate forms%
\begin{align}
\widetilde{S}\left(  x,p;s\exp\left(  +i\pi/2\right)  \right)   &  =e^{-i\pi
p/2}\frac{1}{s^{p}\Gamma\left(  1+p\right)  }\exp\left(  -\frac{i}{4}%
s\exp\left(  2ix\right)  \right) \\
&  \ \ \ \nonumber\\
\frac{i}{\pi}\widetilde{S}_{\text{other}}\left(  x,p;s\exp\left(
-i\pi/2\right)  \right)   &  =e^{i\pi\left(  1+p\right)  /2}\left(  \dfrac
{-1}{\sin\pi p}\right)  \frac{1}{s^{p}\Gamma\left(  1+p\right)  }\exp\left(
-\frac{i}{4}s\exp\left(  2ix\right)  \right) \\
&  =e^{i\pi\left(  1+2p\right)  /2}\left(  \dfrac{-1}{\sin\pi p}\right)
\widetilde{S}\left(  x,p;s\exp\left(  +i\pi/2\right)  \right)
\end{align}
which now coincide at the point $p=-1/2$, since we have also adjusted the
normalization and phase of the 2nd solution. \ However, their derivatives with
respect to $p$ still do \emph{not} match at $p=-1/2$, due to $\left.
\frac{\partial}{\partial p}\left(  -e^{i\pi\left(  1+2p\right)  /2}%
/\sin\left(  \pi p\right)  \right)  \right\vert _{p=-1/2}=i\pi$. \ 

Taking all this into account, and exploiting the linearity of the equation,
(\ref{DualSDifferentialDifference}), a more general solution would be%
\begin{equation}
\widetilde{S}_{\text{general}}\left(  x,p\right)  =\sum_{n}\int ds\ \left(
c_{n}\left(  s\right)  e^{-i\pi p/4}+d_{n}\left(  s\right)  e^{i\pi\left(
1+p\right)  /4}\dfrac{-1}{\sin\pi p}\right)  \times e^{2\pi inp}\frac{1}%
{s^{p}\Gamma\left(  1+p\right)  }\exp\left(  -\frac{i}{4}s\exp\left(
2ix\right)  \right)
\end{equation}
For integer-valued $p$ the sum over $n$ may be omitted.

\subsection{The dual metric as an absolute $\bigstar$\ square}

Each such solution for $\widetilde{S}$\ leads to a candidate real metric,
given by%
\begin{equation}
\widetilde{R}=\widetilde{S}\star\overline{\widetilde{S}}
\label{DualR=DualS*DualSBar}%
\end{equation}
To verify this, we note that the entwining equation for $\widetilde{S}$, and
its conjugate $\overline{\widetilde{S}}$,%
\begin{equation}
H\star\widetilde{S}=\widetilde{S}\left(  x,p\right)  \star p^{2}%
\ ,\ \ \ p^{2}\star\overline{\widetilde{S}}=\overline{\widetilde{S}}%
\star\overline{H}%
\end{equation}
may be combined with the associativity of the star product to obtain
\begin{equation}
H\star\widetilde{S}\star\overline{\widetilde{S}}=\widetilde{S}\left(
x,p\right)  \star p^{2}\star\overline{\widetilde{S}}=\widetilde{S}%
\star\overline{\widetilde{S}}\star\overline{H}%
\end{equation}
Thus the form in (\ref{DualR=DualS*DualSBar})\ yields a solution to
(\ref{HEntwinedWithDualMetric}). \ We need to work out a representative
$\widetilde{S}\star\overline{\widetilde{S}}$ star product to understand the
relation to the previous solutions for $\widetilde{R}$. \ 

We do this for the first form of the basic $\widetilde{S}$ solutions,
(\ref{BasicDualSSolution}), using the standard integral representation
(\ref{1/Gamma}) for $1/\Gamma$. \ We find a result that coincides with one of
the composite dual metrics (\ref{DualRComposite}).
\begin{align}
\widetilde{S}\left(  x,p;s\right)  \star\overline{\widetilde{S}}\left(
x,p;s\right)   &  =\left(  1+\frac{1}{16}s^{4}-\frac{1}{2}s^{2}\cos2x\right)
^{p}\frac{1}{s^{2p}\left(  \Gamma\left(  1+p\right)  \right)  ^{2}}\nonumber\\
&  =\widetilde{R}\left(  x,p;s^{2},1+\frac{1}{16}s^{4}\right)
\end{align}
By choosing $s=\pm2$, we again obtain the original dual metric.%
\begin{equation}
\widetilde{S}\left(  x,p;\pm2\right)  \star\overline{\widetilde{S}}\left(
x,p;\pm2\right)  =\widetilde{R}\left(  x,p;4,2\right)  =\frac{\left(  \sin
^{2}x\right)  ^{p}}{\left(  \Gamma\left(  p+1\right)  \right)  ^{2}%
}=\widetilde{R}\left(  x,p\right)
\end{equation}
This provides a greater appreciation of the information contained in
$\widetilde{S}$, and motivates us to consider an alternative construction of
such $\bigstar$\ roots of $\widetilde{R}$.

\subsection{$\widetilde{S}$ as a sum of hybrid WFs}

The basic ideas here are essentially the same as used in the construction of
$\widetilde{R}$ as a sum of $f_{n}$, only the bilinears appearing in the
Wigner transforms involve two different types of functions: \ One wave
function is an imaginary Liouville eigenfunction while the other is a free
particle solution, precisely the $h_{n}\left(  x,p\right)  $ defined earlier
in (\ref{LiouvilleFreeHybridWFs}). \ 

We form the sums%
\begin{equation}
\widetilde{S}\left(  x,p\right)  =\sum_{n=0}^{\infty}\sqrt{\varepsilon_{n}%
}~h_{n}\left(  x,p\right)  \ ,\ \ \ \overline{\widetilde{S}\left(  x,p\right)
}=\sum_{n=0}^{\infty}\sqrt{\varepsilon_{n}}~\overline{h_{n}\left(  x,p\right)
} \label{DualSAsSumOfHybrids}%
\end{equation}
Then from (\ref{H*h}) and (\ref{h*Hbar}) we immediately obtain%
\begin{equation}
H\star\widetilde{S}\left(  x,p\right)  =\widetilde{S}\left(  x,p\right)  \star
p^{2}\ ,\ \ \ p^{2}\star\overline{\widetilde{S}\left(  x,p\right)  }%
=\overline{\widetilde{S}\left(  x,p\right)  }\star\overline{H}%
\end{equation}
and, upon using (\ref{h*hbar=f}), we also obtain $\widetilde{R}$ as the
absolute star-square of $\widetilde{S}$.%
\begin{equation}
\widetilde{S}\left(  x,p\right)  \star\overline{\widetilde{S}\left(
x,p\right)  }=\sum_{n=0}^{\infty}\sqrt{\varepsilon_{n}}~h_{n}\left(
x,p\right)  \star\sum_{k=0}^{\infty}\sqrt{\varepsilon_{k}}~\overline
{h_{k}\left(  x,p\right)  }=\sum_{n=0}^{\infty}\varepsilon_{n}~f_{n}\left(
x,p\right)  =\widetilde{R}\left(  x,p\right)
\label{DualMetricAsDualS*DualSBar}%
\end{equation}
However, since the Bessels on the circle are not orthonormal, we do not have a
similar relation for $\overline{\widetilde{S}\left(  x,p\right)  }%
\star\widetilde{S}\left(  x,p\right)  $. \ While these results can be
established without explicit forms for the $h_{n}\left(  x,p\right)  $ and for
$\widetilde{S}\left(  x,p\right)  $, it is perhaps useful to have such
expressions in hand. \ 

Like the individual hybrid WFs, explicit results for the sum depend on how we
choose the free particle solutions. \ In particular, we may take the hybrid
WFs built from right-moving and left-moving plane waves in
(\ref{RChiralHybridWFs}) and (\ref{LChiralHybridWFs}). \ This gives rise to
the phase-space equivalent of the \textquotedblleft chiral
kernel\textquotedblright\ in \cite{CM}.%
\begin{align}
\widetilde{S}_{\text{R}}\left(  x,p\right)   &  =\sum_{n=0}^{\infty}%
\sqrt{\varepsilon_{n}}~h_{n}^{\text{R}}\left(  x,p\right)  =\frac{\left(
-1\right)  ^{p}}{4^{p}p!}e^{2ipx}\sum_{n=0}^{p}\sqrt{\varepsilon_{n}}%
\frac{\left(  -1\right)  ^{-n}2^{n}}{\left(  p-n\right)  !}e^{-2inx}%
\nonumber\\
&  =\left(  1-\sqrt{2}\right)  \frac{\left(  -1\right)  ^{p}}{4^{p}\left(
p!\right)  ^{2}}e^{2ipx}+\sqrt{2}\frac{1}{2^{p}\left(  p!\right)  ^{2}%
}e^{-\frac{1}{2}e^{2ix}}\Gamma\left(  p+1,-\frac{1}{2}e^{2ix}\right)
\label{RChiralDualS}%
\end{align}
where the incomplete $\Gamma$ function has made an appearance (e.g. see
\cite{Abram}).%
\begin{align}
\sum_{k=0}^{p}\frac{1}{k!}z^{k}  &  =e^{z}\frac{\Gamma\left(  p+1,z\right)
}{\Gamma\left(  p+1\right)  }\\
\Gamma\left(  p+1,z\right)   &  =\int_{z}^{\infty}s^{p}e^{-s}ds=\Gamma\left(
p+1\right)  -\frac{z^{p+1}}{p+1}\left.  _{1}F_{1}\right.  \left(
p+1;p+2;-z\right)
\end{align}
We leave it as an exercise for the reader to use the properties of the
incomplete $\Gamma$\ to check directly that $\widetilde{S_{\text{R}}}\left(
x,p\right)  \star\overline{\widetilde{S_{\text{R}}}\left(  x,p\right)
}=\widetilde{R}\left(  x,p\right)  $.

Similarly, we have%
\begin{align}
\widetilde{S}_{\text{L}}\left(  x,p\right)   &  =\sum_{n=0}^{\infty}%
\sqrt{\varepsilon_{n}}~h_{n}^{\text{L}}\left(  x,p\right)  =\frac{\left(
-1\right)  ^{p}}{4^{p}p!}e^{2ipx}\sum_{n=0}^{\infty}\sqrt{\varepsilon_{n}%
}\frac{1}{2^{n}\left(  p+n\right)  !}e^{2inx}\\
&  =\frac{\left(  -1\right)  ^{p}}{4^{p}\left(  p!\right)  ^{2}}e^{2ipx}%
+\sqrt{2}\frac{\left(  -1\right)  ^{p}}{2^{p}\left(  p!\right)  ^{2}}%
e^{\frac{1}{2}e^{2ix}}\left(  \Gamma\left(  p+1\right)  -\Gamma\left(
p+1,\frac{1}{2}e^{2ix}\right)  \right)
\end{align}
Comparison with the results in \cite{CM}, \S 4, Eqn's (67) and (69), shows
that $\widetilde{S}_{\text{R,L}}\left(  x,p\right)  $ are essentially Fourier
transforms with respect to \emph{one} variable of a particular combination of
Lommel's functions of \emph{two} variables. \ Who knew?

Alternatively, we could take the free particle solutions to be non-chiral: $1
$ for $n=0$ but for $n>0$, $\left(  e^{-inx}+\left(  -1\right)  ^{n}%
e^{inx}\right)  /\sqrt{2}$. \ \ That is%
\begin{equation}
\phi_{n}\left(  x\right)  =\frac{\sqrt{\varepsilon_{n}}}{2}\left(
e^{-inx}+\left(  -1\right)  ^{n}e^{inx}\right)
\end{equation}
which are properly normalized, as in (\ref{FreeParticleWaveFunctions}). \ In
this case we would be led to the Wigner transform of the well-known generating
function.
\begin{align}
e^{iz\sin x}  &  =\sum_{n=-\infty}^{\infty}J_{n}\left(  z\right)
e^{inx}=J_{0}\left(  z\right)  +\sum_{n=1}^{\infty}J_{n}\left(  z\right)
\left(  e^{inx}+\left(  -1\right)  ^{n}e^{-inx}\right) \label{Generating}\\
&  =\sum_{n=0}^{\infty}\sqrt{\varepsilon_{n}}J_{n}\left(  z\right)
\overline{\phi_{n}\left(  x\right)  }%
\end{align}
Thus we obtain a remarkably simple result for the non-chiral kernel.%
\begin{equation}
\widetilde{S}_{\text{NC}}\left(  x,p\right)  \equiv\frac{1}{2\pi}\int
_{0}^{2\pi}e^{ie^{i\left(  x-y\right)  }\sin\left(  x+y\right)  }%
e^{2ipy}dy=\frac{\left(  \frac{-1}{2}\right)  ^{p}}{p!}\exp\left(  \frac{1}%
{2}e^{2ix}\right)  \label{NonchiralDualS}%
\end{equation}
which we recognize as a particular case of our basic solutions,
(\ref{BasicDualSSolution}).%
\begin{equation}
\widetilde{S}_{\text{NC}}\left(  x,p\right)  =\widetilde{S}\left(
x,p;s=-2\right)
\end{equation}
The selected non-chiral free particle solutions have WFs with symmetric
momentum support.
\begin{equation}
e_{n}\left(  x,p\right)  =\frac{\varepsilon_{n}}{4}\left(  \delta_{p,n}%
+\delta_{p,-n}+2\left(  -1\right)  ^{n}\cos\left(  2nx\right)  \times
\delta_{p,0}\right)  \label{NonchiralFreeWFs}%
\end{equation}
The corresponding hybrids are linear combinations of those in
(\ref{RChiralHybridWFs}) and (\ref{LChiralHybridWFs}), as given by%
\begin{align}
h_{n}^{\text{NC}}\left(  x,p\right)   &  =\frac{\sqrt{\varepsilon_{n}}}%
{2}\left(  h_{n}^{\text{L}}\left(  x,p\right)  +\left(  -1\right)  ^{n}%
h_{n}^{\text{R}}\left(  x,p\right)  \right) \nonumber\\
&  =\frac{\sqrt{\varepsilon_{n}}}{2}\left\{
\begin{array}
[c]{c}%
\frac{\left(  -1\right)  ^{p}}{4^{p}2^{n}p!\left(  p+n\right)  !}e^{2i\left(
p+n\right)  x}\text{ \ \ for \ \ }n>p\geq0\\
\\
\frac{\left(  -1\right)  ^{p}}{4^{p}2^{n}p!\left(  p+n\right)  !}e^{2i\left(
p+n\right)  x}+\left(  -1\right)  ^{n}\frac{2^{n}\left(  -1\right)  ^{p-n}%
}{4^{p}p!\left(  p-n\right)  !}+e^{2i\left(  p-n\right)  x}\text{ \ \ for
\ \ }p\geq n
\end{array}
\right.
\end{align}
and vanish for $p<0$. \ All this leads us once again to the expected relation
(\ref{DualMetricAsDualS*DualSBar}).

\section{Meanwhile, back at the metric}

We might also wish to systematically determine all metrics, rather than dual
metrics, through which the Liouville Hamiltonian is rendered hermitian. \ This
is perhaps a more conventional problem to attack in the framework of
quasi-hermitian theories \cite{Scholtz1992,Scholtz2005,Scholtz2006}. \ That is
to say, we seek all solutions to%
\begin{equation}
\overline{H}\star R=R\star H \label{HEntwinedWithMetric}%
\end{equation}
for real functions $R\left(  x,p\right)  =\overline{R\left(  x,p\right)  }$.
\ (This should be compared to (\ref{HEntwinedWithDualMetric}) given above.)
\ In this context it is somewhat repetitious but perhaps instructive to go
through a few details omitted in the previous two sections, including some
forays into calculational cul-de-sacs.

\subsection{$R$ as a formal sum of the dual WFs (Not!)}

In parallel to the previous construction of the dual metric as a sum of WFs,
(\ref{DualMetricAsWFs}), we might try%
\begin{equation}
Q\left(  x,p\right)  =\sum_{k=0}^{\infty}\varepsilon_{k}\widetilde{f_{k}%
}\left(  x,p\right)  \label{Trouble}%
\end{equation}
which gives, at least formally, the expected phase-space orthogonality
relation%
\begin{equation}
\varepsilon_{n}=\frac{1}{2\pi}\int_{x,p}\hspace{-0.25in}%
%TCIMACRO{\tsum }%
%BeginExpansion
{\textstyle\sum}
%EndExpansion
~Q\left(  x,p\right)  f_{n}\left(  x,p\right)
\end{equation}
However, $Q$ does \emph{not} satisfy the homogeneous equation
(\ref{HEntwinedWithMetric}) due to the inhomogeneities resulting when $H$ and
$\overline{H}$\ act on the individual $\widetilde{f_{k}}$. \ Moreover, the
best we can do with the sum (\ref{Trouble}), so far, is to interpret it as an
asymptotic series related to the Wigner transform of an integral whose
asymptotic expansion is a formal generating function for the dual wave
functions. Rather than pursue this here, we turn to direct solutions of the
differential-difference equation corresponding to (\ref{HEntwinedWithMetric}).

\subsection{Solving directly for $R$}

We find the following left- and right-sided star products, and their
difference.%
\begin{align}
\overline{H}\star R\left(  x,p\right)   &  =\left(  p^{2}+e^{-2ix}\right)
\star R\left(  x,p\right)  =\left(  \left(  p-\frac{i}{2}\partial_{x}\right)
^{2}+e^{-2ix}e^{\partial_{p}}\right)  R\left(  x,p\right) \\
R\left(  x,p\right)  \star H  &  =R\left(  x,p\right)  \star\left(
p^{2}+e^{2ix}\right)  =\left(  \left(  p+\frac{i}{2}\partial_{x}\right)
^{2}+e^{2ix}e^{\partial_{p}}\right)  R\left(  x,p\right) \\
R\left(  x,p\right)  \star H-\overline{H}\star R\left(  x,p\right)   &
=2ip\partial_{x}R\left(  x,p\right)  +2i\sin\left(  2x\right)  R\left(
x,p+1\right)
\end{align}
So then, solving (\ref{HEntwinedWithMetric}) amounts to solving the linear
differential-difference equation\footnote{This first-order equation becomes a
second-order equation if the Voros product is used. \ That is, if we demand
$\overline{H}\triangleleft R=R\triangleleft H$ then we have to solve:
\ $\left(  \partial_{x}^{2}-2ip\partial_{x}-\exp\left(  2ix\right)  \right)
R\left(  x,p\right)  +\exp\left(  -2ix\right)  R\left(  x,p+2\right)  =0$.
\ Moreover, the solutions of this second-order equation are not real. \ So we
prefer to use $\star$\ and not $\triangleleft$.}%
\begin{equation}
-p\partial_{x}R\left(  x,p\right)  =\sin\left(  2x\right)  R\left(
x,p+1\right)  \label{DifferentialDifference}%
\end{equation}
This should be compared to (\ref{DualDifferentialDifference}) given above,
which it becomes upon letting $p\rightarrow-p$. (Note that $R$ actually
corresponds to $\widetilde{R}^{-1}$ in that earlier discussion.) \ This is
immediately solved upon assuming a product form, $R\left(  x,p\right)
=q\left(  x\right)  r\left(  p\right)  $. \ There are two distinct sets of
solutions corresponding to positive and negative constants of separation, $\pm
s$. \ 

The first one-parameter ($s$) set of solutions is%
\begin{equation}
R\left(  x,p;s\right)  =s^{p}\Gamma\left(  p\right)  \exp\left(  \frac{1}%
{2}s\cos\left(  2x\right)  \right)  \label{Solutions}%
\end{equation}
up to an overall multiplicative constant. \ Identification with the previous
dual metric basic solution is given by $\widetilde{R}_{\text{other}}\left(
x,-p;s\right)  =R\left(  x,p;s\right)  $. \ Also, the duplication formula
$\Gamma\left(  p\right)  =\frac{2^{p}}{\sqrt{4\pi}}\Gamma\left(  \tfrac{1}%
{2}p\right)  \Gamma\left(  \tfrac{1}{2}+\tfrac{1}{2}p\right)  $ gives%
\begin{equation}
R\left(  x,p;s\right)  =\frac{\left(  2s\right)  ^{p}}{\sqrt{4\pi}}%
\Gamma\left(  \tfrac{1}{2}p\right)  \Gamma\left(  \tfrac{1}{2}+\tfrac{1}%
{2}p\right)  \exp\left(  \frac{1}{2}s\cos\left(  2x\right)  \right)
\end{equation}
for whatever that's worth,\footnote{These solutions immediately call to mind
the Mellin transforms of Bessel functions, $\ \mathcal{J}_{\nu}\left(
p\right)  \equiv\frac{\Gamma\left(  \frac{\nu-p}{2}\right)  }{2^{p}%
\Gamma\left(  1+\frac{\nu+p}{2}\right)  }$, which are solutions to the
Liouville energy eigenvalue problem in momentum space, $\left(  p^{2}%
+e^{-2\partial_{p}}\right)  \mathcal{J}_{\nu}\left(  p\right)  =\nu
^{2}\mathcal{J}_{\nu}\left(  p\right)  $.} while the reflection relation
$\Gamma\left(  p\right)  =\frac{\pi}{\Gamma\left(  1-p\right)  \sin\pi p}$
gives%
\begin{equation}
R\left(  x,p;s\right)  =\frac{\pi s^{p}}{\Gamma\left(  1-p\right)  \sin\pi
p}\exp\left(  \frac{1}{2}s\cos\left(  2x\right)  \right)
\end{equation}
For real $s$, the solutions (\ref{Solutions}) are hermitian and
positive-definite functions of real variables $x,p$ so long as $p>0$.
\ However, $R\left(  x,p;s\right)  $ is not bounded either on the negative $p$
half-line, or on the positive $p$ half-line for any real $s$ (cf. Stirling's
approximation). \ Otherwise, $\left(  -1\right)  ^{k+1}R\left(  x,p\right)  $
is hermitian and positive-definite for $-k-1<p<-k$, where $k=0,1,2,\cdots$.

Now we know from previous considerations that for $n>0$ the periodic WFs
$f_{n}$ have support only for positive $p$, so (\ref{Solutions}) would seem to
be a preferred set of solutions, \emph{except }for the ground state: \ $f_{0}$
has support at $p=0$. \ This issue must still be addressed.

Another set of solutions is given by the form%
\begin{equation}
R_{\text{other}}\left(  x,p;s\right)  =\frac{s^{p}}{\Gamma\left(  1-p\right)
}\exp\left(  -\frac{1}{2}s\cos\left(  2x\right)  \right)
\label{OtherSolutions}%
\end{equation}
Identification with the previous dual metric basic solution is given by
$\widetilde{R}\left(  x,-p;s\right)  =R_{\text{other}}\left(  x,p;s\right)  $.
\ The duplication $\dfrac{1}{\Gamma\left(  1-p\right)  }=\frac{2^{p}\sqrt{\pi
}}{\Gamma\left(  \frac{1}{2}-\frac{1}{2}p\right)  \Gamma\left(  1-\frac{1}%
{2}p\right)  }$ and reflection relations $\dfrac{1}{\Gamma\left(  1-p\right)
}=\frac{\Gamma\left(  p\right)  \sin\pi p}{\pi}$ now give
\begin{align}
R_{\text{other}}\left(  x,p;s\right)   &  =\frac{\left(  2s\right)  ^{p}%
\sqrt{\pi}}{\Gamma\left(  \frac{1}{2}-\frac{1}{2}p\right)  \Gamma\left(
1-\frac{1}{2}p\right)  }\exp\left(  -\frac{1}{2}s\cos\left(  2x\right)
\right) \\
R_{\text{other}}\left(  x,p;s\right)   &  =\frac{s^{p}\Gamma\left(  p\right)
\sin\pi p}{\pi}\exp\left(  -\frac{1}{2}s\cos\left(  2x\right)  \right)
\end{align}
again for whatever that's worth. \ For real $s$, these other solutions are
hermitian and positive-definite functions of real variables $x,p$ so long as
$p<1$. \ Otherwise, $\left(  -1\right)  ^{k}R_{\text{other}}\left(
x,p;s\right)  $ is hermitian and positive-definite for $k<p<k+1$, where
$k=1,2,\cdots$. \ Also, $R_{\text{other}}$ is now bounded for all real $p$. \ 

An interesting problem now is to find the \emph{real} square-root of either
$R\left(  x,p;s\right)  $, or $R_{\text{other}}\left(  x,p;s\right)  $. \ This
is \emph{not} just $s^{p/2}\sqrt{\Gamma\left(  p\right)  }e^{\frac{1}{4}%
s\cos\left(  2x\right)  }$, say, since the square-root must be taken in a
$\star$\ sense. \ That is, we seek a real $S$ such that%
\begin{equation}
R\left(  x,p;s\right)  =S\left(  x,p;s\right)  \star S\left(  x,p;s\right)
\end{equation}
Were it not for the $\Gamma$ function, this would be easy since%
\begin{equation}
\exp\left(  a\cos\left(  2x\right)  +bp\right)  =\exp\left(  \frac{a}%
{2\cosh\left(  b/2\right)  }\cos\left(  2x\right)  +\frac{b}{2}p\right)
\star\exp\left(  \frac{a}{2\cosh\left(  b/2\right)  }\cos\left(  2x\right)
+\frac{b}{2}p\right)  \label{EasyProduct}%
\end{equation}
But the presence of the $\Gamma$ function makes the problem a little more
challenging. \ 

Recall for $p>0$, $\Gamma\left(  p\right)  =\int_{0}^{\infty}t^{p-1}e^{-t}dt$
while for all $p$ there is again the contour integral representation
(\ref{1/Gamma}), so%
\begin{equation}
\frac{1}{\Gamma\left(  1-p\right)  }=\frac{1}{2\pi i}\int_{-\infty}^{\left(
0+\right)  }t^{p-1}e^{t}dt
\end{equation}
The contour here is the same as in (\ref{DualRContour}). \ Thus%
\begin{align}
R_{\text{other}}\left(  x,p;s\right)   &  =\int_{-\infty}^{\left(  0+\right)
}t^{p-1}\exp\left(  t+ps-\frac{1}{2}e^{s}\cos\left(  2x\right)  \right)
dt\nonumber\\
&  =\int_{-\infty}^{\left(  0+\right)  }\exp\left(  e^{\ln t}+p\ln
t+ps-\frac{1}{2}e^{s}\cos\left(  2x\right)  \right)  d\ln t
\end{align}
Star composition of two such integrands is now possible, as in
(\ref{EasyProduct}), although we then have to deal with a double integral..
\ But taking the square-root in this approach is not transparent.

From another perspective, the problem involves computation of either%
\begin{align}
\Gamma\left(  p\right)  \star f\left(  x\right)   &  =\int_{0}^{\infty}%
t^{p-1}\star f\left(  x\right)  e^{t}dt=\int_{0}^{\infty}\exp\left(  p\ln
t\right)  \star f\left(  x\right)  e^{t}d\ln t\nonumber\\
&  =\int_{0}^{\infty}\exp\left(  p\ln t\right)  f\left(  x-\frac{1}{2}i\ln
t\right)  e^{t}d\ln t=\int_{0}^{\infty}t^{p-1}e^{t}f\left(  x-\frac{1}{2}i\ln
t\right)  dt
\end{align}
or%
\begin{align}
\frac{1}{\Gamma\left(  1-p\right)  }\star f\left(  x\right)   &
=\int_{-\infty}^{\left(  0+\right)  }t^{p-1}\star f\left(  x\right)
e^{t}dt=\int_{-\infty}^{\left(  0+\right)  }\exp\left(  p\ln t\right)  \star
f\left(  x\right)  e^{t}d\ln t\nonumber\\
&  =\int_{-\infty}^{\left(  0+\right)  }\exp\left(  p\ln t\right)  f\left(
x-\frac{1}{2}i\ln t\right)  e^{t}d\ln t=\int_{-\infty}^{\left(  0+\right)
}t^{p-1}e^{t}f\left(  x-\frac{1}{2}i\ln t\right)  dt
\end{align}
Alternatively, since $\Gamma\left(  p\right)  \star$ $e^{ikx}=e^{ikx}%
\Gamma\left(  p+\frac{1}{2}k\right)  $, then if $f\left(  x\right)  =\int
e^{ikx}F\left(  k\right)  dk$ the two computations become%
\begin{equation}
\Gamma\left(  p\right)  \star f\left(  x\right)  =\int\Gamma\left(  p\right)
\star e^{ikx}F\left(  k\right)  dk=\int e^{ikx}\Gamma\left(  p+\frac{1}%
{2}k\right)  F\left(  k\right)  dk
\end{equation}%
\begin{equation}
\frac{1}{\Gamma\left(  1-p\right)  }\star f\left(  x\right)  =\int\frac
{1}{\Gamma\left(  1-p\right)  }\star e^{ikx}F\left(  k\right)  dk=\int
e^{ikx}\frac{F\left(  k\right)  }{\Gamma\left(  1-p-\frac{1}{2}k\right)  }dk
\end{equation}
We also note that%
\begin{equation}
\exp\left(  a\cos2x\right)  =\sum_{n=-\infty}^{\infty}I_{n}\left(  a\right)
e^{2inx}\ ,\ \ \ I_{n}\left(  a\right)  =\frac{1}{\pi}\int_{0}^{\pi}%
e^{a\cos\theta}\cos\left(  n\theta\right)  d\theta
\end{equation}
with $I_{n}\left(  a\right)  =I_{-n}\left(  a\right)  $ and $I_{n}\left(
-a\right)  =\left(  -1\right)  ^{n}I_{n}\left(  a\right)  $. \ Therefore we
may write, as least formally, the ordinary product in $R$ as a star product%
\begin{equation}
\Gamma\left(  p\right)  \exp\left(  a\cos2x\right)  =\sum_{n=-\infty}^{\infty
}I_{n}\left(  a\right)  \Gamma\left(  p\right)  e^{2inx}=\sum_{n=-\infty
}^{\infty}I_{n}\left(  a\right)  \frac{\Gamma\left(  p\right)  }{\Gamma\left(
p+n\right)  }\Gamma\left(  p\right)  \star e^{2inx}=\Gamma\left(  p\right)
\star\sum_{n=-\infty}^{\infty}\frac{I_{n}\left(  a\right)  }{\left(  p\right)
_{n}}~e^{2inx}%
\end{equation}
where the Pochhammer symbol is $\left(  p\right)  _{n}=\left(  p\right)
\left(  p+1\right)  \cdots\left(  p+n-1\right)  =\Gamma\left(  p+n\right)
/\Gamma\left(  p\right)  $. \ Note the zeroes in the summand for integer $p>0$
when $n\leq-p$. \ Unfortunately, this is not of much use without a closed form
for the sum. \ This would appear to be one of the aforementioned cul-de-sacs,
so let us try a different route, parallel to that used previously to construct
$\widetilde{S}$.

\subsection{Solving for$\sqrt[\bigstar]{R}$}

We look for an equivalence between the Liouville $H=p^{2}+e^{2ix}$ and the
free particle $\mathbb{H}=p^{2}$ as given by%
\begin{equation}
\mathbb{H}=S\mathbb{\star}H\mathbb{\star}S^{-1}%
\end{equation}
That is to say, in phase-space we wish to solve the linear equation (cf.
(\ref{DualSEntwinedWithH}) above)%
\begin{equation}
p^{2}\mathbb{\ \mathbb{\star}\ }S=S\ \star\ H \label{SEntwinedWithH}%
\end{equation}
This is again a special case of the ultra-local two star equation\ as given in
\cite{CFZ}. \ For the Liouville--free-particle case, this again amounts to an
equation similar to that for $R$, but inherently complex.
\begin{equation}
-2ip\partial_{x}S\left(  x,p\right)  =e^{2ix}e^{\partial_{p}}S\left(
x,p\right)  =e^{2ix}S\left(  x,p+1\right)  \label{SEquation}%
\end{equation}
which becomes the previous (\ref{DualSDifferentialDifference}) upon
$p\rightarrow-p$. \ Once again solutions are easily found through the use of a
product ansatz. \ For any value of the parameter $s$ we have two immediate
solutions:%
\begin{align}
S\left(  x,p;s\right)   &  =s^{p}\Gamma\left(  p\right)  \exp\left(  \frac
{1}{4}s\exp\left(  2ix\right)  \right) \label{SSolutions}\\
& \nonumber\\
S_{\text{other}}\left(  x,p;s\right)   &  =\frac{s^{p}}{\Gamma\left(
1-p\right)  }\exp\left(  -\frac{1}{4}s\exp\left(  2ix\right)  \right)
\label{OtherSSolutions}%
\end{align}
The first of these is a good\ solution for $p\in\left(  0,\infty\right)  $,
say, while the second is good for $p\in\left(  -\infty,1\right)  $, thereby
providing a solution for the entire real $p$ axis, albeit a discontinuous one.
\ The relation to the previous dual solutions is obviously%
\begin{equation}
S\left(  x,p;s\right)  =\widetilde{S}_{\text{other}}\left(  x,-p;s\right)
\ ,\ \ \ S_{\text{other}}\left(  x,p;s\right)  =\widetilde{S}\left(
x,-p;s\right)
\end{equation}
As in the dual situation, these solutions could always be multiplied by
periodic functions of $p$, $\exp\left(  2i\pi np\right)  $ for $n\in
\mathbb{Z}$, but again, for integer-valued $p$ this has no effect. \ 

The two solutions for $S$\ are brought closer in appearance by shifting
$s\rightarrow se^{\pm i\pi/2}$ and using the reflection relation for the
$\Gamma$. \ Thus we may take as our two basic solutions the alternate forms%
\begin{align}
S\left(  x,p;s\exp\left(  +i\pi/2\right)  \right)   &  =e^{i\pi p/2}%
s^{p}\Gamma\left(  p\right)  \exp\left(  \frac{i}{4}s\exp\left(  2ix\right)
\right) \\
& \nonumber\\
\frac{i}{\pi}S_{\text{other}}\left(  x,p;s\exp\left(  -i\pi/2\right)  \right)
&  =e^{i\pi\left(  1-p\right)  /2}s^{p}\sin\left(  \pi p\right)  \Gamma\left(
p\right)  \exp\left(  \frac{i}{4}s\exp\left(  2ix\right)  \right) \\
&  =e^{i\pi\left(  1-2p\right)  /2}\sin\left(  \pi p\right)  ~S\left(
x,p;s\exp\left(  +i\pi/2\right)  \right)
\end{align}
which now coincide at the point $p=1/2$, since we have also rescaled the 2nd
solution. \ However, their derivatives with respect to $p$ still do \emph{not}
match at $p=1/2$, due to $\left.  \frac{\partial}{\partial p}\left(
e^{i\pi\left(  1-2p\right)  /2}\sin\left(  \pi p\right)  \right)  \right\vert
_{p=1/2}=-i\pi$. \ Taking all this into account, and exploiting the linearity
of the equation, a more general solution would be%
\begin{equation}
S_{\text{general}}\left(  x,p\right)  =\sum_{n}\int ds\ \left(  c_{n}\left(
s\right)  e^{i\pi p/2}+d_{n}\left(  s\right)  e^{i\pi\left(  1-p\right)
/2}\sin\left(  \pi p\right)  \right)  \times e^{2\pi inp}s^{p}\Gamma\left(
p\right)  \exp\left(  \frac{i}{4}s\exp\left(  2ix\right)  \right)
\end{equation}
Once again, for integer-valued $p$ the sum over $n$ may be omitted.

Since these solutions for $S$\ are complex, a suitable real metric would now
be given by%
\begin{equation}
R=\overline{S}\star S \label{R=Sbar*S}%
\end{equation}
and as before we have $\overline{H}\star R=R\star H$ consistent with $S\star
H\star S^{-1}=\mathbb{H}=\overline{\mathbb{H}}=\overline{S^{-1}}\star
\overline{H}\star\overline{S}$. \ For completeness, we extend this to a simple
theorem that can be used to construct other composite solutions for the metric
through multiple star products:

\noindent\textbf{[Lemma]} \ \emph{Any} solution of (\ref{SEntwinedWithH}),
i.e. (\ref{SEquation}), gives a real solution of (\ref{HEntwinedWithMetric}),
i.e. (\ref{DifferentialDifference}), namely $R\equiv\overline{S}\star S$.
\ Moreover, appropriately ordered odd star products of any solutions to
(\ref{HEntwinedWithMetric}), (\ref{SEntwinedWithH}), and the complementary
equation (\ref{HEntwinedWithDualMetric}) are also solutions to
(\ref{DifferentialDifference}) and (\ref{SEquation}). \ For example,
$\overline{H}\star R_{1}\star\widetilde{R}_{2}\star R_{3}=R_{1}\star
\widetilde{R}_{2}\star R_{3}\star H$. \ Similarly, $\mathbb{H}\star S_{1}%
\star\widetilde{R}_{2}\star R_{3}=S_{1}\star\widetilde{R}_{2}\star R_{3}\star
H$.

We need to work out a star product at least for a particular case just to
understand the relation to the previous solutions for $R$. \ So we do this for
the first form of the solutions, (\ref{SSolutions}), using the standard
integral representation for $\Gamma\left(  p\right)  $. \ We find a
composite\ metric%
\begin{align}
R  &  =\overline{S}\left(  x,p;s\right)  \star S\left(  x,p;s\right)
\nonumber\\
&  =\left(  1-\frac{1}{4}s^{2}e^{2ix}\right)  ^{-p}\left(  1-\frac{1}{4}%
s^{2}e^{-2ix}\right)  ^{-p}s^{2p}\Gamma^{2}\left(  p\right) \nonumber\\
&  =\left(  1+\frac{1}{16}s^{4}-\frac{1}{2}s^{2}\cos2x\right)  ^{-p}%
s^{2p}\Gamma^{2}\left(  p\right)  \label{CompositeR}%
\end{align}
assuming $s$ is real and provided $\operatorname{Re}\left(  s^{2}e^{\pm
2ix}\right)  <4 $ to avoid the singularities in the $\left(  1-\frac{1}%
{4}s^{2}e^{\pm2ix}\right)  ^{-p}$ factors. \ For example, if $e^{2s}<4$, then
all real $x$ satisfy this condition, and the composite metric is real and
positive definite for real $p>0$. \ The double pole at $p=0$ is somewhat
troubling, and again presents an issue for the ground state WF.
\ Nevertheless, we easily check that this composite $R$ satisfies the metric
equation (\ref{DifferentialDifference}).

Similarly, for the other solution (\ref{OtherSSolutions}), we find%
\begin{align}
R  &  =\overline{S}_{\text{other}}\left(  x,p\right)  \star S_{\text{other}%
}\left(  x,p\right) \nonumber\\
&  =\left(  1-\frac{1}{4}s^{2}e^{2ix}\right)  ^{-p}\left(  1-\frac{1}{4}%
s^{2}e^{-2ix}\right)  ^{-p}\frac{s^{2p}}{\Gamma^{2}\left(  1-p\right)
}\nonumber\\
&  =\left(  1+\frac{1}{16}s^{4}-\frac{1}{2}s^{2}\cos2x\right)  ^{-p}%
\frac{s^{2p}}{\Gamma^{2}\left(  1-p\right)  } \label{OtherCompositeR}%
\end{align}
again for real $s$ and provided $\operatorname{Re}\left(  e^{2s}e^{\pm
2ix}\right)  <4$. \ \ Again we easily verify that this composite $R$ satisfies
the metric equation (\ref{DifferentialDifference}). \ We also note that these
two composite metrics are more simply related than the previous forms of the
metric or even the pairs of solutions for $S$. \ Namely, we go from one
composite $R$ to the other just by the interchange $\Gamma^{2}\left(
p\right)  \leftrightarrow1/\Gamma^{2}\left(  1-p\right)  $.

We should also express the composite solution in terms of (sums) of the
previous solutions for $R$, if that is possible. \ Indeed, just as in the dual
metric situation, it is possible. \ We find%
\begin{equation}
\left(  1+\frac{1}{16}s^{4}-\frac{1}{2}s^{2}\cos2x\right)  ^{-p}s^{2p}%
\Gamma^{2}\left(  p\right)  =\int_{0}^{\infty}\frac{d\tau}{\tau}e^{-\left(
1+\frac{1}{16}s^{4}\right)  \tau}R\left(  x,p;s^{2}\tau\right)
\end{equation}
where $R\left(  x,p;s^{2}\tau\right)  $ is as defined in (\ref{Solutions}).
\ Also%
\begin{equation}
\left(  1+\frac{1}{16}s^{4}-\frac{1}{2}s^{2}\cos2x\right)  ^{-p}\frac{s^{2p}%
}{\Gamma^{2}\left(  1-p\right)  }=\int_{-\infty}^{\left(  0+\right)  }%
\frac{d\tau}{\tau}e^{\left(  1+\frac{1}{16}s^{4}\right)  \tau}R_{\text{other}%
}\left(  x,p;s^{2}\tau\right)
\end{equation}
where $R_{\text{other}}\left(  x,p;s^{2}\tau\right)  $ is as defined in
(\ref{OtherSolutions}).

\subsection{$\sqrt[\bigstar]{R}$ as a sum of hybrid WFs\ (Not!)}

In parallel to the previous construction of $\widetilde{S}$ as a sum of the
hybrid WFs $h_{n}$, as in (\ref{DualSAsSumOfHybrids}), we might try to
construct roots of $R$ as formal sums of the dual hybrids $\widetilde{h_{n}}$.
\
\begin{equation}
T\left(  x,p\right)  =\sum_{n=0}^{\infty}\sqrt{\varepsilon_{n}}~\widetilde
{h_{n}}\left(  x,p\right)  \ ,\ \ \ \overline{T\left(  x,p\right)  }%
=\sum_{n=0}^{\infty}\sqrt{\varepsilon_{n}}~\overline{\widetilde{h_{n}}\left(
x,p\right)  }%
\end{equation}
However, these are \emph{not} solutions to (\ref{SEntwinedWithH}) or its
conjugate due to the inhomogeneities resulting when $H$ and $\overline{H}%
$\ act on the individual $\widetilde{h_{n}}$ and $\overline{\widetilde{h_{n}}%
}$. \ Even so, these and similar sums do yield some interesting star products,
at least formally. \ For example,
\begin{equation}
\overline{T\left(  x,p\right)  }\star T\left(  x,p\right)  =\sum_{k=0}%
^{\infty}\varepsilon_{k}\widetilde{f_{k}}\left(  x,p\right)  =Q\left(
x,p\right)
\end{equation}
Rather than pursue this here, we consider other direct solutions of the
differential-difference equation corresponding to (\ref{SEntwinedWithH}).

\subsection{Additional solutions for $S$}

Re-instating the coupling constant $m$ via $\exp\left(  ix\right)  \rightarrow
m\exp\left(  ix\right)  $, as in (\ref{ImaginaryLiouvilleHamiltonian}), a
straightforward series solution in powers of $m$ gives another form for
solutions to (\ref{SEntwinedWithH}).%
\begin{align}
S\left(  m;x,p\right)   &  =\Gamma\left(  p\right)  \left(  \frac{2}%
{m}\right)  ^{p}e^{-ipx}I_{p}\left(  me^{ix}\right) \label{MoreSSolutionsI}\\
S_{\text{other}}\left(  m;x,p\right)   &  =\frac{1}{\Gamma\left(  1-p\right)
}\left(  \frac{2}{m}\right)  ^{p}\left(  -i\right)  ^{p}e^{-ipx}I_{p}\left(
ime^{ix}\right)  \label{MoreSSolutionsIi}%
\end{align}
the latter obtained from the first form by $x\rightarrow x+\pi/2$ and
$\Gamma\left(  p\right)  \rightarrow1/\Gamma\left(  1-p\right)  $. \ We also
note that (\ref{MoreSSolutionsIi})\ may be written as
\begin{equation}
S_{\text{other}}\left(  m;x,p\right)  =\frac{1}{\Gamma\left(  1-p\right)
}\left(  \frac{2}{m}\right)  ^{p}e^{-ipx}J_{p}\left(  me^{ix}\right)
\label{MoreSSolutionsJ}%
\end{equation}
Here we have made use of the modified Bessel function $I_{\nu}\left(
z\right)  $ with properties
\begin{subequations}
\begin{align}
I_{\nu}\left(  z\right)   &  =\left(  \frac{z}{2}\right)  ^{\nu}\sum
_{k=0}^{\infty}\frac{\left(  z^{2}/4\right)  ^{k}}{k!\Gamma\left(
\nu+k+1\right)  }\ ,\ \ \ \\
J_{\nu}\left(  z\right)   &  =\left(  -i\right)  ^{\nu}I_{\nu}\left(
iz\right)  =\left(  \frac{z}{2}\right)  ^{\nu}\sum_{k=0}^{\infty}\frac{\left(
-z^{2}/4\right)  ^{k}}{k!\Gamma\left(  \nu+k+1\right)  }\\
z\frac{d}{dz}I_{\nu}\left(  z\right)   &  =zI_{\nu\pm1}\left(  z\right)
\pm\nu I_{\nu}\left(  z\right) \label{RecursionI}\\
I_{n}\left(  z\right)   &  =\frac{1}{2\pi}\int_{0}^{2\pi}e^{z\cos\theta}e^{\pm
in\theta}d\theta\text{ \ \ for \ \ }n\in\mathbb{N}\label{IntegralRepI}\\
\exp\left(  z\cos\theta\right)   &  =I_{0}\left(  z\right)  +2\sum
_{k=1}^{\infty}I_{k}\left(  z\right)  \cos\left(  k\theta\right)
\end{align}
Either the first or the third of these properties leads to the most direct
verification that (\ref{MoreSSolutionsI}) and (\ref{MoreSSolutionsIi}) satisfy
the equation (\ref{SEntwinedWithH}). \ 

We also recall the elegant contour integral representations of Schl\"{a}fli
and Sonine which are valid for \emph{all} $\nu$ and $z$.
\end{subequations}
\begin{align}
I_{\nu}\left(  z\right)   &  =\left(  \frac{z}{2}\right)  ^{\nu}\frac{1}{2\pi
i}\int_{-\infty}^{\left(  0+\right)  }w^{-\nu-1}\exp\left(  w+\frac{z^{2}}%
{4w}\right)  dw\label{Sonine}\\
J_{\nu}\left(  z\right)   &  =\left(  \frac{z}{2}\right)  ^{\nu}\frac{1}{2\pi
i}\int_{-\infty}^{\left(  0+\right)  }w^{-\nu-1}\exp\left(  w-\frac{z^{2}}%
{4w}\right)  dw
\end{align}
The contour is the same as that used in (\ref{DualRContour}). \ Thus we
identify (\ref{MoreSSolutionsI}) and (\ref{MoreSSolutionsJ}) as linear
combinations of the basic solutions (\ref{SSolutions}) and
(\ref{OtherSSolutions}) (after re-instating the coupling constant $m$ via
$x\rightarrow x-i\ln m$), expressed as integrals over $w=1/s$.%
\begin{align}
S\left(  m;x,p\right)   &  =\frac{1}{2\pi i}\int_{-\infty}^{\left(  0+\right)
}S\left(  x-i\ln m,p;s=\frac{1}{w}\right)  e^{w}\frac{dw}{w}\\
S_{\text{other}}\left(  m;x,p\right)   &  =\frac{1}{2\pi i}\int_{-\infty
}^{\left(  0+\right)  }S_{\text{other}}\left(  x-i\ln m,p;s=\frac{1}%
{w}\right)  e^{w}\frac{dw}{w}%
\end{align}
In the same manner, we can construct additional solutions to
(\ref{DualSEntwinedWithH}), with $m$ re-instated, to obtain $\widetilde
{S}\left(  m;x,p\right)  =S_{\text{other}}\left(  m;x,-p\right)  $ and
$\widetilde{S}_{\text{other}}\left(  m;x,p\right)  =S\left(  m;x,-p\right)  $.
\ We leave as an exercise the calculation of the corresponding composite
metric, $R\left(  m;x,p\right)  =\overline{S\left(  m;x,p\right)  }\star
S\left(  m;x,p\right)  $.

\section{Metric dependence of expectation values}

It is important to realize that expectation values are metric-dependent, in
general. \ As a first illustration of this, we consider expectations of WFs
for energy eigenstates using the basic solutions (\ref{Solutions})\ for the
phase-space metric.%
\begin{equation}
\mathcal{N}_{n}\left(  s\right)  \equiv\frac{1}{2\pi}\int_{x,p}\hspace
{-0.25in}%
%TCIMACRO{\tsum }%
%BeginExpansion
{\textstyle\sum}
%EndExpansion
~R\left(  x,p;s\right)  f_{n}\left(  x,p\right)  =\frac{1}{2\pi}\sum_{p}%
s^{p}\Gamma\left(  p\right)  \int_{0}^{2\pi}f_{n}\left(  x,p\right)
\exp\left(  \frac{1}{2}s\cos2x\right)  dx
\end{equation}
For the ground state case, $n=0$, there is a problem here. The sum over $p$
must include $p=0$ and this diverges due to the $\Gamma\left(  p\right)  $\ in
$R\left(  x,p;s\right)  $. \ The individual basic metric solutions are not
useful in this one case. \ It would appear to be necessary to take a linear
combination of the basic metrics -- an integral over $s$, perhaps -- to tame
this singularity. \ We leave this as an unsolved problem.

For all $n>0$, the individual $R\left(  x,p;s\right)  $ give finite,
reasonably well-behaved $\mathcal{N}_{n}\left(  s\right)  $. \ We have%
\begin{equation}
\mathcal{N}_{n}\left(  s\right)  \equiv\frac{1}{2\pi}\int_{x,p}\hspace
{-0.25in}%
%TCIMACRO{\tsum }%
%BeginExpansion
{\textstyle\sum}
%EndExpansion
~R\left(  x,p;s\right)  f_{n}\left(  x,p\right)  =\frac{1}{2\pi}\sum
_{p=n}^{\infty}s^{p}\Gamma\left(  p\right)  \int_{0}^{2\pi}f_{n}\left(
x,p\right)  \exp\left(  \frac{1}{2}s\cos2x\right)  dx
\end{equation}
From the series (\ref{LiouvilleWF})\ and the integral representation of the
modified Bessels, (\ref{IntegralRepI}),\ these basic norms are%
\begin{equation}
\mathcal{N}_{n}\left(  s\right)  =\sum_{p=n}^{\infty}\frac{s^{p}\Gamma\left(
p\right)  }{4^{p}}\sum_{k=0}^{p-n}\frac{1}{k!\left(  n+k\right)  !\left(
p-k\right)  !\left(  p-k-n\right)  !}\times I_{n-p+2k}\left(  \frac{s}%
{2}\right)
\end{equation}
Alternatively,%
\begin{align}
\mathcal{N}_{n}\left(  s\right)   &  =\sum_{p=n}^{\infty}\frac{s^{p}}{p\left(
p-n\right)  !}\frac{1}{2\pi}\int_{0}^{2\pi}\left(  \frac{i\sin2x}{2}\right)
^{p}\operatorname{LegendreP}\left(  p,-n,i\cot2x\right)  \exp\left(  \frac
{1}{2}s\cos2x\right)  dx\\
&  =\frac{1}{n!}\sum_{k=0}^{\infty}\frac{1}{k!\left(  n+k\right)  }\left(
\frac{1}{4}s\right)  ^{n+k}\frac{1}{2\pi}\int_{0}^{2\pi}e^{2ikx}%
\operatorname{KummerM}\left(  n+k,1+n,\frac{s}{4}e^{-2ix}\right)  \exp\left(
\frac{1}{2}s\cos2x\right)  dx\nonumber
\end{align}
The infinite sums here are convergent, but we have not found a closed-form.
\ Nevertheless, the norms are clearly $s$- and hence metric-dependent. \ If we
switch our attention to the dual WFs, we encounter only finite sums of Bessel functions.

\subsection{Diagonal dual WFs}

As further illustration of the metric dependence of phase-space expectations,
we compute expectations of dual WFs for energy eigenstates using the basic
solutions (\ref{BasicDualWFMetric})\ for the phase-space dual metric. \
\begin{equation}
\widetilde{\mathcal{N}}_{n}\left(  s\right)  \equiv\frac{1}{2\pi}\int
_{x,p}\hspace{-0.25in}%
%TCIMACRO{\tsum }%
%BeginExpansion
{\textstyle\sum}
%EndExpansion
~\widetilde{R}\left(  x,p;s\right)  \widetilde{f_{n}}\left(  x,p\right)
=\frac{1}{2\pi}\sum_{p=0}^{n}\frac{1}{s^{p}\Gamma\left(  1+p\right)  }\int
_{0}^{2\pi}\widetilde{f_{n}}\left(  x,p\right)  \exp\left(  -\frac{1}{2}%
s\cos2x\right)  dx
\end{equation}
From the series (\ref{LiouvilleDualWF})\ and the integral representation
(\ref{IntegralRepI})\ these dual norms are%
\begin{equation}
\widetilde{\mathcal{N}}_{n}\left(  s\right)  =\sum_{p=0}^{n}\frac{1}%
{s^{p}\Gamma\left(  1+p\right)  }4^{p}n^{2}\sum_{k=\max\left(
0,n-p-\left\lfloor n/2\right\rfloor \right)  }^{\min\left(  \left\lfloor
n/2\right\rfloor ,n-p\right)  }\frac{\left(  n-k-1\right)  !\left(
k+p-1\right)  !}{k!\left(  n-p-k\right)  !}\times\left(  -1\right)
^{2k+p-n}I_{2k+p-n}\left(  \frac{s}{2}\right)
\end{equation}
Note that $\widetilde{\mathcal{N}}_{n}\left(  s\right)  =\widetilde
{\mathcal{N}}_{n}\left(  -s\right)  $. \ For example:%
\begin{align}
\widetilde{\mathcal{N}}_{0}\left(  s\right)   &  =I_{0}\left(  \frac{s}%
{2}\right) \\
\widetilde{\mathcal{N}}_{1}\left(  s\right)   &  =\frac{4}{s}I_{0}\left(
\frac{s}{2}\right) \\
\widetilde{\mathcal{N}}_{2}\left(  s\right)   &  =\frac{4}{s^{2}}\left(
s^{2}+8\right)  I_{0}\left(  \frac{s}{2}\right)  -\frac{32}{s}I_{1}\left(
\frac{s}{2}\right) \nonumber\\
&  =8I_{2}\left(  \frac{s}{2}\right)  +\frac{4}{s^{2}}\left(  8-s^{2}\right)
I_{0}\left(  \frac{s}{2}\right) \\
\widetilde{\mathcal{N}}_{3}\left(  s\right)   &  =\frac{12}{s^{3}}\left(
3s^{2}+32\right)  I_{0}\left(  \frac{s}{2}\right)  -\frac{288}{s^{2}}%
I_{1}\left(  \frac{s}{2}\right) \nonumber\\
&  =\frac{72}{s}I_{2}\left(  \frac{s}{2}\right)  +\frac{12}{s^{3}}\left(
32-3s^{2}\right)  I_{0}\left(  \frac{s}{2}\right)
\end{align}
These results should be compared to the much simpler norms (\ref{DualWFNorms})
obtained upon computing phase-space averages using (\ref{DualPhaseSpaceMetric}%
). \ The basic norms are indeed positive for all $s$, but $s$ dependent, as is
evident upon graphing the chosen examples.

\subsection{Non-diagonal dual WFs}

We also compute, for $k\neq n$,%
\begin{equation}
\widetilde{\mathcal{N}}_{k,n}\left(  s\right)  \equiv\frac{1}{2\pi}\int
_{x,p}\hspace{-0.25in}%
%TCIMACRO{\tsum }%
%BeginExpansion
{\textstyle\sum}
%EndExpansion
~\widetilde{R}\left(  x,p;s\right)  \widetilde{f_{k,n}}\left(  x,p\right)
=\overline{\widetilde{\mathcal{N}}_{n,k}\left(  s\right)  }%
\end{equation}
to see if they vanish. \ As a consequence of the inhomogeneities in the dual
energy eigenvalue equations, it is \emph{not} obvious that they should vanish
for $k\neq n$. \ In fact, they do \emph{not}.

To verify this statement and to compute $\widetilde{\mathcal{N}}_{k,n}\left(
s\right)  $, for general $k\neq n$, we first determine the non-diagonal dual
WFs.%
\begin{align}
\widetilde{f_{k,n}}\left(  x,p\right)   &  \equiv\frac{1}{2\pi}\int_{0}^{2\pi
}\overline{\chi_{k}}\left(  x-y\right)  ~\chi_{n}\left(  x+y\right)
~e^{2iyp}dy\nonumber\\
&  =4^{p}kne^{ix\left(  k-n\right)  }\sum_{j=0}^{\left\lfloor k/2\right\rfloor
}\sum_{l=0}^{\left\lfloor n/2\right\rfloor }\frac{\left(  k-j-1\right)
!\left(  n-l-1\right)  !}{j!l!}\ e^{2ix\left(  l-j\right)  }\delta
_{k+n-2p,2j+2l} \label{NondiagonalDualWFDoubleSum}%
\end{align}
The sums result from using the series in (\ref{ASeries}). \ Given the range of
the sums, we must have $0\leq2p\leq k+n$ for a non-zero result. \ So, as for
the $k=n$ case, the support in $p$ is \emph{finite}. \ Also, for a
contribution to the sums, we must have $k+n=2p\operatorname{mod}2$. \ Thus
$\widetilde{f_{k,n}}\left(  x,p\right)  $ can be non-zero for semi-integer
$p$, when $k+n$ is an odd integer. \ Let us simplify the double sum in
(\ref{NondiagonalDualWFDoubleSum}). \ First, suppose $k+n$ is even. \ Then
$\delta_{k+n-2p,2j+2l}$ under the double sum implies $p$ must be integer and
$\leq\frac{k+n}{2}$. \ The upper limits of the sums further imply
$k/2-\left\lfloor k/2\right\rfloor +n/2-\left\lfloor n/2\right\rfloor \leq p$.
\ That is to say, for $k+n$ even, we must have $p$ integer and $0\leq\left(
k+n\right)  /2-\left\lfloor k/2\right\rfloor -\left\lfloor n/2\right\rfloor
\leq p\leq\frac{k+n}{2}$. \ On the other hand, for $k+n$ odd, we must have $p$
semi-integer and $\frac{1}{2}\leq p\leq\frac{k+n}{2}$. \ Incorporating
these\ conditions and using the Kronecker delta to eliminate one sum, for both
even and odd $k+n$ cases, we find%
\begin{equation}
\widetilde{f_{k,n}}\left(  x,p\right)  =4^{p}kne^{i\left(  k-n\right)  x}%
\sum_{j=\max\left(  0,n+\left\lfloor \frac{k-n}{2}\right\rfloor -\left\lfloor
p\right\rfloor -\left\lfloor \frac{n}{2}\right\rfloor \right)  }^{\min\left(
\left\lfloor \frac{k}{2}\right\rfloor ,n+\left\lfloor \frac{k-n}%
{2}\right\rfloor -\left\lfloor p\right\rfloor \right)  }\frac{\left(
k-j-1\right)  !\left(  \left\lfloor p\right\rfloor +j-\left\lfloor \frac
{k-n}{2}\right\rfloor -1\right)  !}{j!\left(  n+\left\lfloor \frac{k-n}%
{2}\right\rfloor -\left\lfloor p\right\rfloor -j\right)  !}\ e^{2i\left(
n+\left\lfloor \frac{k-n}{2}\right\rfloor -\left\lfloor p\right\rfloor
-2j\right)  x} \label{NondiagonalDualWFs}%
\end{equation}
Note that this coincides with (\ref{LiouvilleDualWF})\ when $k=n$. \ Also note
we do not need the floor function in the $\left\lfloor \frac{k-n}%
{2}\right\rfloor $ factors when $k+n$ is even, but we do need it when $k+n$ is
odd. \ Similarly, we do not need the floor function in $\left\lfloor
p\right\rfloor $ when $k+n$ is even, but we do need it when $k+n$ is odd. \ In
particular
\begin{subequations}
\begin{align}
\widetilde{f_{0,1}}\left(  x,p\right)   &  =2e^{-ix}\delta_{p,1/2}\\
\widetilde{f_{0,2}}\left(  x,p\right)   &  =2\delta_{p,0}+8e^{-2ix}%
\delta_{p,1}\label{DualWF02}\\
\widetilde{f_{1,2}}\left(  x,p\right)   &  =4e^{ix}\delta_{p,1/2}%
+16e^{-ix}\delta_{p,3/2}\\
\widetilde{f_{1,3}}\left(  x,p\right)   &  =12\delta_{p,1}+96e^{-2ix}%
\delta_{p,2}\\
\widetilde{f_{2,3}}\left(  x,p\right)   &  =12e^{-ix}\delta_{p,1/2}+\left(
48e^{ix}+96e^{-3ix}\right)  \delta_{p,3/2}+384e^{-ix}\delta_{p,5/2}\\
\widetilde{f_{2,4}}\left(  x,p\right)   &  =4\delta_{p,0}+\left(
16e^{2ix}+64e^{-2ix}\right)  \delta_{p,1}+\left(  256+768e^{-4ix}\right)
\delta_{p,2}+3072e^{-2ix}\delta_{p,3}%
\end{align}
We now finish checking nondiagonal orthogonality for the basic metrics, i.e.
whether $\widetilde{\mathcal{N}}_{k,n}\left(  s\right)  =0$ for $k\neq n$.
\ This is not true, in general. \ 

First, suppose $k+n$ is even, and consider the simplest case. The average over
position is
\end{subequations}
\begin{align}
\frac{1}{2\pi}\int_{0}^{2\pi}\widetilde{f_{0,2}}\left(  x,p\right)
\widetilde{R}\left(  x,p;s\right)  dx  &  =\frac{1}{s^{p}\Gamma\left(
1+p\right)  }\frac{1}{2\pi}\int_{0}^{2\pi}\left(  2\delta_{p,0}+8e^{-2ix}%
\delta_{p,1}\right)  \exp\left(  -\frac{s}{2}\cos2x\right)  dx\nonumber\\
&  =\frac{1}{s^{p}\Gamma\left(  1+p\right)  }\left(  2I_{0}\left(  \frac{s}%
{2}\right)  \delta_{p,0}-8I_{1}\left(  \frac{s}{2}\right)  \delta
_{p,1}\right)
\end{align}
This is real for real $s$, so%
\begin{equation}
\frac{1}{2\pi}\int_{0}^{2\pi}\widetilde{f_{0,2}}\left(  x,p\right)
\widetilde{R}\left(  x,p;s\right)  dx=\frac{1}{2\pi}\int_{0}^{2\pi}%
\widetilde{f_{2,0}}\left(  x,p\right)  \widetilde{R}\left(  x,p;s\right)  dx
\end{equation}
Summing over $p$ now gives%
\begin{equation}
\widetilde{\mathcal{N}}_{0,2}\left(  s\right)  =\widetilde{\mathcal{N}}%
_{2,0}\left(  s\right)  =2I_{0}\left(  \frac{s}{2}\right)  -\frac{8}{s}%
I_{1}\left(  \frac{s}{2}\right)  =2I_{2}\left(  \frac{s}{2}\right)
\end{equation}
This does not vanish for $s\neq0$. \ Similarly for other even $k+n$ cases:
\begin{equation}
\widetilde{\mathcal{N}}_{k,n}\left(  s\right)  =\widetilde{\mathcal{N}}%
_{n,k}\left(  s\right)  =\frac{4^{p}kn}{s^{p}}\sum_{j=\max\left(  0,\frac
{k+n}{2}-p-\left\lfloor \frac{n}{2}\right\rfloor \right)  }^{\min\left(
\left\lfloor \frac{k}{2}\right\rfloor ,\frac{k+n}{2}-p\right)  }\frac{\left(
k-j-1\right)  !\left(  p+j-\frac{k-n}{2}-1\right)  !}{j!\left(  \frac{k+n}%
{2}-p-j\right)  !p!}I_{k-p-2j}\left(  -\frac{s}{2}\right)
\label{BasicNondiagonalExpectations}%
\end{equation}
We leave odd $k+n$ as an exercise for the interested reader.

In contrast to (\ref{BasicNondiagonalExpectations}), all the non-diagonal dual
WFs vanish when summed/integrated over phase space using the dual metric of
(\ref{DualPhaseSpaceMetric}). \
\begin{equation}
\frac{1}{2\pi}\int_{x,p}\hspace{-0.25in}%
%TCIMACRO{\tsum }%
%BeginExpansion
{\textstyle\sum}
%EndExpansion
~\widetilde{R}\left(  x,p\right)  \widetilde{f_{k,n}}\left(  x,p\right)
=0\text{ \ \ for \ \ }k\neq n \label{NondiagonalOrthogonalityUnderDualMetric}%
\end{equation}
as follows most easily from (\ref{DualMetricAsWFs}). \ For the case when $k+n
$ is even, this involves cancellations among the various terms for different
$p$ after integrating over $x$. \ For example,%
\begin{gather}
\frac{1}{2\pi}\int_{0}^{2\pi}\sum_{p}\widetilde{f_{0,2}}\left(  x,p\right)
\widetilde{R}\left(  x,p\right)  dx=\frac{1}{2\pi}\int_{0}^{2\pi}%
\widetilde{f_{0,2}}\left(  x,p\right)  \frac{\left(  \sin^{2}x\right)  ^{p}%
}{\left(  p!\right)  ^{2}}dx\nonumber\\
=\frac{1}{2\pi}\int_{0}^{2\pi}2dx+\frac{1}{2\pi}\int_{0}^{2\pi}8e^{-2ix}%
\sin^{2}xdx=2-2=0
\end{gather}
However, when $k+n$ is odd, and therefore $p$ is semi-integer, we interpret
the metric as integer powers of $\left\vert \sin x\right\vert $. \ Then the
individual terms contributed by $\widetilde{f_{k,n}}\left(  x,p\right)  $ to
the sum over $p$ each vanish separately upon integration over $x$. \ In this
regard we note that $\frac{\left\vert \sin x\right\vert ^{2p}}{\left(
\Gamma\left(  p+1\right)  \right)  ^{2}}\ $is still a solution of
(\ref{DualDifferentialDifference}), for $p>0$, given that $\left(  \sin
x\right)  ^{2p}\delta\left(  x\right)  =0$. \ (That is, all test functions are
required to be non-singular at $x=0$.)

Thus it is evident that requiring the phase-space averages of $\widetilde
{f_{k,n}}$ to vanish for $k\neq n$ imposes additional conditions on the dual
metric, beyond those specified in (\ref{DualDifferentialDifference}).
\ However, these additional conditions are \emph{not} sufficient to force the
dual metric to be proportional to $\widetilde{R}\left(  x,p\right)  $ in
(\ref{DualPhaseSpaceMetric}). \ Any dual metric of the form given in
(\ref{DualMetricAsWFs}) but with arbitrary coefficients, i.e. $\widetilde
{R}\left(  x,p\right)  =\sum_{k=0}^{\infty}c_{k}f_{k}\left(  x,p\right)  $,
will give a vanishing phase-space average for nondiagonal dual WFs. \ This
follows immediately from%
\begin{equation}
\frac{1}{2\pi}\int_{x,p}\hspace{-0.25in}%
%TCIMACRO{\tsum }%
%BeginExpansion
{\textstyle\sum}
%EndExpansion
~f_{j}\left(  x,p\right)  \widetilde{f_{k,l}}\left(  x,p\right)  =\delta
_{j,k}\delta_{j,l}%
\end{equation}
From (\ref{BasicNondiagonalExpectations}) we can only conclude that
$\widetilde{R}\left(  x,p;s\right)  $ cannot be so expressed as a linear
combination of the diagonal $f_{k}$. \ Of course, this can also be established
by other means.

As a consequence of $\widetilde{\mathcal{N}}_{k,n}\left(  s\right)  \neq0$ the
norms of general pure state dual WFs are even more complicated. \ For example,
consider a linear combination of dual functions, $\chi=\alpha A_{0}+\beta
A_{2}$, and form the corresponding dual WF. $\ $%
\begin{equation}
\widetilde{f}=\frac{1}{2\pi}\int_{0}^{2\pi}\overline{\chi}\left(  x-y\right)
~\chi\left(  x+y\right)  ~e^{2iyp}dy=\left\vert \alpha\right\vert
^{2}\widetilde{f_{0}}+\overline{\alpha}\beta\widetilde{f_{0,2}}+\alpha
\overline{\beta}\widetilde{f_{2,0}}+\left\vert \beta\right\vert ^{2}%
\widetilde{f_{2}} \label{DualWFLinearCombo}%
\end{equation}
Under the action of $\widetilde{R}\left(  x,p;s\right)  $\ the normalization
of this pure state dual WF is%
\begin{gather}
\widetilde{\mathcal{N}}\left(  s\right)  =\frac{1}{2\pi}\int_{0}^{2\pi
}\widetilde{R}\left(  x,p;s\right)  \widetilde{f}\left(  x,p\right)
dx=\left\vert \alpha\right\vert ^{2}\widetilde{\mathcal{N}}_{0}\left(
s\right)  +\overline{\alpha}\beta\widetilde{\mathcal{N}}_{0,2}\left(
s\right)  +\alpha\overline{\beta}\widetilde{\mathcal{N}}_{2,0}\left(
s\right)  +\left\vert \beta\right\vert ^{2}\widetilde{\mathcal{N}}_{2}\left(
s\right) \nonumber\\
=\left(  \left\vert \alpha\right\vert ^{2}+\frac{4}{s^{2}}\left(
8-s^{2}\right)  \left\vert \beta\right\vert ^{2}\right)  I_{0}\left(  \frac
{s}{2}\right)  +\left(  2\overline{\alpha}\beta+2\alpha\overline{\beta
}+8\left\vert \beta\right\vert ^{2}\right)  I_{2}\left(  \frac{s}{2}\right)
\end{gather}
\ On the other hand, under $\widetilde{R}\left(  x,p\right)  $ in
(\ref{DualPhaseSpaceMetric}), the norm of this dual WF would be just
$\left\vert \alpha\right\vert ^{2}+2\left\vert \beta\right\vert ^{2}$.

Ultimately, $\widetilde{\mathcal{N}}_{k,n}\neq0$ for $k\neq n$ originates in
the inhomogeneities in the dual eigenvalue equations. \ By contrast, a similar
calculation for the metric, in those cases where is it is well-defined
($k\neq0\neq n$), would give%
\begin{equation}
\mathcal{N}_{k,n}\left(  s\right)  \equiv\frac{1}{2\pi}\int_{x,p}%
\hspace{-0.25in}%
%TCIMACRO{\tsum }%
%BeginExpansion
{\textstyle\sum}
%EndExpansion
~R\left(  x,p;s\right)  f_{k,n}\left(  x,p\right)  =0
\end{equation}
for $k\neq n$. \ This is a simple consequence of the Lone Star Lemma, the
homogeneous equations $H\star f_{k,n}\left(  x,p\right)  =k^{2}f_{k,n}\left(
x,p\right)  $ and $f_{k,n}\left(  x,p\right)  \star\overline{H}=n^{2}%
f_{k,n}\left(  x,p\right)  $, and (\ref{HEntwinedWithMetric}). \ Hence, for a
linear combination of WFs corresponding to the wave function $\psi=\alpha
J_{k}+\beta J_{n}$, i.e. $\ $%
\begin{equation}
f=\frac{1}{2\pi}\int_{0}^{2\pi}\psi\left(  x-y\right)  ~\overline{\psi}\left(
x+y\right)  ~e^{2iyp}dy=\left\vert \alpha\right\vert ^{2}f_{k}+\overline
{\alpha}\beta f_{n,k}+\alpha\overline{\beta}f_{k,n}+\left\vert \beta
\right\vert ^{2}f_{n} \label{WFLinearCombo}%
\end{equation}
the action of $R\left(  x,p;s\right)  $\ would produce the normalization%
\begin{equation}
\mathcal{N}\left(  s\right)  =\frac{1}{2\pi}\int_{0}^{2\pi}\widetilde
{R}\left(  x,p;s\right)  \widetilde{f}\left(  x,p\right)  dx=\left\vert
\alpha\right\vert ^{2}\mathcal{N}_{k}\left(  s\right)  +\left\vert
\beta\right\vert ^{2}\mathcal{N}_{n}\left(  s\right)
\end{equation}
We have assumed that neither $k$ nor $n$ are zero.

\subsection{Expectations of $H$}

The corresponding expectations of $H$ under the action of $\widetilde
{R}\left(  x,p;s\right)  $ are given by%
\begin{align}
\widetilde{\mathcal{H}}_{k,n}\left(  s\right)   &  \equiv\frac{1}{2\pi}%
\int_{x,p}\hspace{-0.25in}%
%TCIMACRO{\tsum }%
%BeginExpansion
{\textstyle\sum}
%EndExpansion
~\widetilde{R}\left(  x,p;s\right)  \left(  \widetilde{f_{k,n}}\left(
x,p\right)  \star H\right) \nonumber\\
&  =\frac{1}{2\pi}\int_{x,p}\hspace{-0.25in}%
%TCIMACRO{\tsum }%
%BeginExpansion
{\textstyle\sum}
%EndExpansion
~\widetilde{R}\left(  x,p;s\right)  \left(  n^{2}\widetilde{f_{k,n}}+\left\{
\begin{array}
[c]{cc}%
2n~\overline{\widetilde{h_{-1,k}}} & \text{for \ \ }n\in\mathbb{N}%
_{\text{odd}}\\
\varepsilon_{n}~\overline{\widetilde{h_{-2,k}}} & \text{for \ \ }%
n\in\mathbb{N}_{\text{even}}%
\end{array}
\right.  \right)
\end{align}
where diagonal cases will be denoted by $\mathcal{H}_{n,n}\equiv
\mathcal{H}_{n}$. \ Alternatively, we may use the Lone Star Lemma and
(\ref{H*DualR}) to write%
\begin{align}
\widetilde{\mathcal{H}}_{k,n}\left(  s\right)   &  =\frac{1}{2\pi}\int
_{x,p}\hspace{-0.25in}%
%TCIMACRO{\tsum }%
%BeginExpansion
{\textstyle\sum}
%EndExpansion
~\left(  H\star\widetilde{R}\left(  x,p;s\right)  \right)  \widetilde{f_{k,n}%
}\left(  x,p\right) \nonumber\\
&  =\frac{1}{2\pi}\int_{x,p}\hspace{-0.25in}%
%TCIMACRO{\tsum }%
%BeginExpansion
{\textstyle\sum}
%EndExpansion
~\left(  p^{2}-\frac{1}{8}s^{2}+\left(  p-\frac{1}{2}\right)  s\cos2x+\frac
{1}{8}s^{2}\cos4x\right)  \widetilde{R}\left(  x,p;s\right)  \widetilde
{f_{k,n}}\left(  x,p\right)
\end{align}
Since $\widetilde{R}\left(  x,p;s\right)  $ is real for real $s$, and
$\widetilde{f_{n,k}}=\overline{\widetilde{f_{k,n}}}$,
(\ref{HEntwinedWithDualMetric}) then gives%
\begin{equation}
\overline{\widetilde{\mathcal{H}}_{k,n}}\left(  s\right)  =\widetilde
{\mathcal{H}}_{n,k}\left(  s\right)
\end{equation}

For example, using (\ref{DualWF02}) we have%
\begin{align}
\widetilde{\mathcal{H}}_{0,2}\left(  s\right)   &  =\frac{1}{2\pi}\int
_{0}^{2\pi}\left(  -\frac{1}{8}s^{2}-\frac{1}{2}s\cos2x+\frac{1}{8}s^{2}%
\cos4x\right)  \widetilde{R}\left(  x,0;s\right)  \times2dx\nonumber\\
&  +\frac{1}{2\pi}\int_{0}^{2\pi}\left(  1-\frac{1}{8}s^{2}+\frac{1}{2}%
s\cos2x+\frac{1}{8}s^{2}\cos4x\right)  \widetilde{R}\left(  x,1;s\right)
\times8e^{-2ix}dx
\end{align}
From (\ref{IntegralRepI}) and (\ref{RecursionI})\ we then obtain
\begin{align}
\widetilde{\mathcal{H}}_{0,2}\left(  s\right)   &  =\left(  -\frac{1}{8}%
s^{2}I_{0}\left(  \frac{s}{2}\right)  +\frac{1}{2}sI_{1}\left(  \frac{s}%
{2}\right)  +\frac{1}{8}s^{2}I_{2}\left(  \frac{s}{2}\right)  \right)
\times2\nonumber\\
&  +\left(  -\left(  1-\frac{1}{8}s^{2}\right)  I_{1}\left(  \frac{s}%
{2}\right)  +\frac{1}{4}sI_{0}\left(  \frac{s}{2}\right)  +\frac{1}{4}%
sI_{2}\left(  \frac{s}{2}\right)  -\frac{1}{16}s^{2}I_{1}\left(  \frac{s}%
{2}\right)  -\frac{1}{16}s^{2}I_{3}\left(  \frac{s}{2}\right)  \right)
\times\frac{8}{s}\nonumber\\
&  =8I_{2}\left(  \frac{s}{2}\right)
\end{align}
So, for real $s$, $\widetilde{\mathcal{H}}_{0,2}\left(  s\right)
=\overline{\widetilde{\mathcal{H}}_{0,2}}\left(  s\right)  =\widetilde
{\mathcal{H}}_{2,0}\left(  s\right)  $. \ Similarly, the $n=0$ and $2$
diagonal cases are given by%
\begin{align}
\widetilde{\mathcal{H}}_{0}\left(  s\right)   &  =\frac{1}{8}s^{2}I_{2}\left(
\frac{s}{2}\right) \\
\widetilde{\mathcal{H}}_{2}\left(  s\right)   &  =48I_{2}\left(  \frac{s}%
{2}\right)  +\frac{16}{s^{2}}\left(  8-s^{2}\right)  I_{0}\left(  \frac{s}%
{2}\right)
\end{align}
to be compared to $\widetilde{\mathcal{N}}_{n}\left(  s\right)  $ given above.
\ The important point here is that the average of $H$ for definite energy dual
WFs, when computed as the ratio $\widetilde{\mathcal{H}}_{n}\left(  s\right)
/\widetilde{\mathcal{N}}_{n}\left(  s\right)  $ using the basic metric
$\widetilde{R}\left(  x,p;s\right)  $, is far from a simple expression. \ It
is certainly not $n^{2}$, as it was in (\ref{<H>}), when computed using the
dual metric of (\ref{DualPhaseSpaceMetric}).

With these cases in hand, we can also consider the linear combination of dual
WFs in (\ref{DualWFLinearCombo}).%
\begin{equation}
\widetilde{\mathcal{H}}\left(  s\right)  =\frac{1}{2\pi}\int_{x,p}%
\hspace{-0.25in}%
%TCIMACRO{\tsum }%
%BeginExpansion
{\textstyle\sum}
%EndExpansion
~\widetilde{R}\left(  x,p;s\right)  \left(  \widetilde{f}\left(  x,p\right)
\star H\right)  =\left\vert \alpha\right\vert ^{2}\mathcal{H}_{0}\left(
s\right)  +\overline{\alpha}\beta\mathcal{H}_{0,2}\left(  s\right)
+\alpha\overline{\beta}\mathcal{H}_{2,0}\left(  s\right)  +\left\vert
\beta\right\vert ^{2}\mathcal{H}_{2}\left(  s\right)
\end{equation}
We find%
\begin{equation}
\widetilde{\mathcal{H}}\left(  s\right)  =\frac{16}{s^{2}}\left(
8-s^{2}\right)  \left\vert \beta\right\vert ^{2}I_{0}\left(  \frac{s}%
{2}\right)  +\left(  \frac{1}{8}s^{2}\left\vert \alpha\right\vert
^{2}+8\left(  \overline{\alpha}\beta+\alpha\overline{\beta}\right)
+48\left\vert \beta\right\vert ^{2}\right)  I_{2}\left(  \frac{s}{2}\right)
\end{equation}
to be compared to $\widetilde{\mathcal{N}}\left(  s\right)  $ given above.
\ Again, the important point is that the ratio $\widetilde{\mathcal{H}}\left(
s\right)  /\widetilde{\mathcal{N}}\left(  s\right)  $ is far from a simple
expression, and is not $4\times2\left\vert \beta\right\vert ^{2}/\left(
\left\vert \alpha\right\vert ^{2}+2\left\vert \beta\right\vert ^{2}\right)  $,
as it would have been had we used the dual metric of
(\ref{DualPhaseSpaceMetric}).

Other cases of even $k+n$ are handled similarly, using
(\ref{NondiagonalDualWFs}). \ We have%
\begin{align}
&  \frac{1}{2\pi}\int_{0}^{2\pi}\left(  p^{2}-\frac{1}{8}s^{2}+\left(
p-\frac{1}{2}\right)  s\cos2x+\frac{1}{8}s^{2}\cos4x\right)  \widetilde
{R}\left(  x,p;s\right)  \widetilde{f_{k,n}}\left(  x,p\right)  dx\nonumber\\
&  =\frac{4^{p}kn}{s^{p}\Gamma\left(  1+p\right)  }\sum_{j=\max\left(
0,n+\left\lfloor \frac{k-n}{2}\right\rfloor -\left\lfloor p\right\rfloor
-\left\lfloor \frac{n}{2}\right\rfloor \right)  }^{\min\left(  \left\lfloor
\frac{k}{2}\right\rfloor ,n+\left\lfloor \frac{k-n}{2}\right\rfloor
-\left\lfloor p\right\rfloor \right)  }\frac{\left(  k-j-1\right)  !\left(
p+j-\frac{k-n}{2}-1\right)  !}{j!\left(  \frac{k+n}{2}-p-j\right)  !}%
\times\nonumber\\
&  \times\frac{1}{2\pi}\int_{0}^{2\pi}\left(  p^{2}-\frac{1}{8}s^{2}+\left(
p-\frac{1}{2}\right)  s\cos2x+\frac{1}{8}s^{2}\cos4x\right)  e^{2i\left(
k-p-2j\right)  x}\exp\left(  -\frac{1}{2}s\cos2x\right)
\end{align}
Then (\ref{IntegralRepI}) and (\ref{RecursionI}) permit evaluation of the
spatial integral to obtain
\begin{align}
&  \frac{1}{2\pi}\int_{0}^{2\pi}\left(  p^{2}-\frac{1}{8}s^{2}+\left(
p-\frac{1}{2}\right)  s\cos2x+\frac{1}{8}s^{2}\cos4x\right)  e^{2i\left(
k-p-2j\right)  x}\exp\left(  -\frac{1}{2}s\cos2x\right) \nonumber\\
&  =p^{2}I_{k-p-2j}\left(  -\frac{s}{2}\right)  +\frac{s}{2}p\left(
I_{k-p-2j+1}\left(  -\frac{s}{2}\right)  +I_{k-p-2j-1}\left(  -\frac{s}%
{2}\right)  \right) \nonumber\\
&  +\frac{s}{4}\left(  k-p-2j\right)  \left(  I_{k-p-2j+1}\left(  -\frac{s}%
{2}\right)  -I_{k-p-2j-1}\left(  -\frac{s}{2}\right)  \right)
\end{align}
Hence, for even $k+n$,%
\begin{align}
\widetilde{\mathcal{H}}_{k,n}\left(  s\right)   &  =%
%TCIMACRO{\tsum _{p}}%
%BeginExpansion
{\textstyle\sum_{p}}
%EndExpansion
\frac{4^{p}kn}{s^{p}p!}\sum_{j=\max\left(  0,\frac{k+n}{2}-p-\left\lfloor
\frac{n}{2}\right\rfloor \right)  }^{\min\left(  \left\lfloor \frac{k}%
{2}\right\rfloor ,\frac{k+n}{2}-p\right)  }\frac{\left(  k-j-1\right)
!\left(  p+j-\frac{k-n}{2}-1\right)  !}{j!\left(  \frac{k+n}{2}-p-j\right)
!}\times\nonumber\\
& \nonumber\\
&  \times\left(
\begin{array}
[c]{c}%
p^{2}I_{k-p-2j}\left(  -\frac{s}{2}\right)  +\frac{s}{2}p\left(
I_{k-p-2j+1}\left(  -\frac{s}{2}\right)  +I_{k-p-2j-1}\left(  -\frac{s}%
{2}\right)  \right) \\
\\
+\frac{s}{4}\left(  k-p-2j\right)  \left(  I_{k-p-2j+1}\left(  -\frac{s}%
{2}\right)  -I_{k-p-2j-1}\left(  -\frac{s}{2}\right)  \right)
\end{array}
\right)
\end{align}
Again, this is real for real $s$, so $\widetilde{\mathcal{H}}_{k,n}%
=\widetilde{\mathcal{H}}_{n,k}$. \ In particular, for $k=n$ we obtain%
\begin{align}
\widetilde{\mathcal{H}}_{n}\left(  s\right)   &  =%
%TCIMACRO{\tsum _{p}}%
%BeginExpansion
{\textstyle\sum_{p}}
%EndExpansion
\frac{4^{p}n^{2}}{s^{p}p!}\sum_{j=\max\left(  0,n-p-\left\lfloor \frac{n}%
{2}\right\rfloor \right)  }^{\min\left(  \left\lfloor \frac{n}{2}\right\rfloor
,n-p\right)  }\frac{\left(  n-j-1\right)  !\left(  p+j-1\right)  !}{j!\left(
n-p-j\right)  !}\times\nonumber\\
& \nonumber\\
&  \times\left(
\begin{array}
[c]{c}%
p^{2}I_{n-p-2j}\left(  -\frac{s}{2}\right)  +\frac{s}{2}p\left(
I_{n-p-2j+1}\left(  -\frac{s}{2}\right)  +I_{n-p-2j-1}\left(  -\frac{s}%
{2}\right)  \right) \\
\\
+\frac{s}{4}\left(  n-p-2j\right)  \left(  I_{n-p-2j+1}\left(  -\frac{s}%
{2}\right)  -I_{n-p-2j-1}\left(  -\frac{s}{2}\right)  \right)
\end{array}
\right)
\end{align}
We leave odd $k+n$ as an exercise for the reader.

Ultimately, $\widetilde{\mathcal{H}}_{k,n}\neq0$ for $k\neq n$ can be traced
back to its origins in the inhomogeneities in the dual eigenvalue equations,
as was also true for $\widetilde{\mathcal{N}}_{k,n}$. \ A similar calculation
for the metric, instead of the dual metric, would give%
\begin{equation}
\mathcal{H}_{k,n}\left(  s\right)  \equiv\frac{1}{2\pi}\int_{x,p}%
\hspace{-0.25in}%
%TCIMACRO{\tsum }%
%BeginExpansion
{\textstyle\sum}
%EndExpansion
~R\left(  x,p;s\right)  H\star f_{k,n}\left(  x,p\right)  =k^{2}%
\mathcal{N}_{k,n}\left(  s\right)  =n^{2}\mathcal{N}_{n}\left(  s\right)
~\delta_{k,n}%
\end{equation}
at least in those cases where is it is well-defined. \ Hence, for definite
energy WFs, we would obtain the expected averages: \ $\left\langle
H\right\rangle =\mathcal{H}_{n}\left(  s\right)  /\mathcal{N}_{n}\left(
s\right)  =n^{2}$. \ Also, for a linear combination of WFs corresponding to
the wave function $\psi=\alpha J_{k}+\beta J_{n}$, as in (\ref{WFLinearCombo})
above, we would have the more intuitive result:%
\begin{equation}
\mathcal{H}\left(  s\right)  =\frac{1}{2\pi}\int_{x,p}\hspace{-0.25in}%
%TCIMACRO{\tsum }%
%BeginExpansion
{\textstyle\sum}
%EndExpansion
~R\left(  x,p;s\right)  H\star f\left(  x,p\right)  =\left\vert \alpha
\right\vert ^{2}k^{2}\mathcal{N}_{k}\left(  s\right)  +\left\vert
\beta\right\vert ^{2}n^{2}\mathcal{N}_{n}\left(  s\right)
\end{equation}
We have again assumed that neither $k$ nor $n$ vanishes.

All this raises the question of how physics can be extracted free from any
metric dependence, in general, especially in situations where calculations are
performed using dual functions that obey inhomogeneous equations.
\ Unfortunately, at this time we have no model-independent answer to this
question. \ (However, for a brief discussion of possible (semi)classical
behavior in the specific $\exp\left(  2ix\right)  $ model, please see the last
subsection in Section IV of \cite{CM}.)

\section{Conclusions}

There are several types of functions involved in the analysis of imaginary
Liouville quantum mechanics: \ Energy eigenfunctions, their duals, and
free-particle plane waves related to the eigenfunctions by various equivalence
maps. \ By cross-breeding pairs of these functions we have produced a wide
variety of hybrid Wigner transforms that exhibit a rich diversity of star
product relations. \ Through the use of sums of these WFs, and star products
of those sums, we have also constructed various phase space equivalence
transformations and metrics. \ We have further shown how these phase space
kernels can be built from direct, elementary solutions of their controlling
star product equations. \ Finally, we have discussed, albeit briefly, the
import of the metrics for the calculation of physical expectation values.
\ Later, we hope to extend the analysis of this paper to include the general
class of Hamiltonians of the form $H=\left(  p+\nu\right)  ^{2}+\sum_{k>0}%
\mu_{k}\exp\left(  ikx\right)  $. \ 

\paragraph{Acknowledgements}

We thank the Institute for Advanced Study for its hospitality and support, and
for providing a stimulating environment in which most of this work was
completed. \ We also thank Luca Mezincescu for useful discussions. \ One of us
(TC) acknowledges the beneficial natural setting of the Aspen Center for
Physics in which this work was initiated during the summer of 2005. \ This
material is based upon work supported by the National Science Foundation under
Grant No's. 0303550 and 0555603.

\newpage%
%TCIMACRO{\TeXButton{TeX}{\addcontentsline{toc}{section}{References}}}%
%BeginExpansion
\addcontentsline{toc}{section}{References}%
%EndExpansion

\newpage%

%TCIMACRO{\TeXButton{TeX}{\section*{Appendix A.  The free particle limit}
%\addcontentsline{toc}{section}{Appendix A.  The free particle limit}}}%
%BeginExpansion
\section*{Appendix A.  The free particle limit}
\addcontentsline{toc}{section}{Appendix A.  The free particle limit}%
%EndExpansion

This is achieved by taking $m\rightarrow0$, but first we have to rescale the
Bessels and the Neumann polynomials to have sensible limits as $m$ vanishes.
\ So we take the biorthonormal set to be $\left\{  m^{n}A_{n}\left(
me^{ix}\right)  ,m^{-n}J_{n}\left(  me^{ix}\right)  \right\}  $ which becomes
$\left\{  2^{n}n!e^{-inx},\frac{1}{2^{n}n!}e^{inx}\right\}  $ as
$m\rightarrow0$. \ Indeed, the $2^{n}n!$ factors look weird, but we must
attribute that convention to Bessel and Neumann. \ 

Now, what happens as $m\rightarrow0$ to the metric on the space of dual
functions as in \cite{CM}? \ After rescaling the Bessels, it is just%
\begin{equation}
J\left(  x,y\right)  =\sum_{n=0}^{\infty}m^{-n}J_{n}\left(  me^{-ix}\right)
m^{-n}J_{n}\left(  me^{iy}\right)  \ _{\overrightarrow{m\rightarrow0}}%
\ \sum_{n=0}^{\infty}\frac{1}{2^{n}n!}e^{-inx}\frac{1}{2^{n}n!}e^{iny}%
=I_{0}\left(  e^{in\left(  y-x\right)  /2}\right)
\end{equation}
Another Bessel function. \ This integral kernel is not unity (i.e. a Dirac
delta) but a projector. \ 

When integrated with (dual) functions that are constants or negative powers of
$e^{iy}$, and their conjugates, it gives the expected orthonormality.%
\begin{equation}
\frac{1}{\left(  2\pi\right)  ^{2}}\int dxdy\ m^{n}A_{n}^{\ast}\left(
me^{ix}\right)  J\left(  x,y\right)  m^{k}A_{k}\left(  me^{iy}\right)
=\delta_{n,k}%
\end{equation}
even as $m\rightarrow0$. \ But when integrated with positive powers of
$e^{iy}$ (not constants) it gives zero.%

%TCIMACRO{\TeXButton{TeX}{\section
%*{Appendix B.  Arbitrary functions acting on WFs through star products}
%\addcontentsline{toc}{section}%
%{Appendix B.  Arbitrary functions acting on WFs through star products}}}%
%BeginExpansion
\section
*{Appendix B.  Arbitrary functions acting on WFs through star products}
\addcontentsline{toc}{section}%
{Appendix B.  Arbitrary functions acting on WFs through star products}%
%EndExpansion

For any two functions $\psi$ and $\phi$, we define the Wigner transform
$f_{\psi\phi}$ somewhat unconventionally (we do \emph{not} conjugate $\phi$)
as%
\begin{equation}
f_{\psi\phi}\left(  x,p\right)  \equiv\frac{1}{2^{N}\pi}\int\psi\left(
x-y\right)  e^{2iyp}\phi\left(  x+y\right)  dy
\end{equation}
where $N$ will be chosen for convenience, later. \ Then for any other function
$G$\ on the phase-space, we have the star product%
\begin{align}
G\left(  x,p\right)  \star f_{\psi\phi}\left(  x,p\right)   &  =G\left(
x,p-\frac{1}{2}i\overrightarrow{\partial_{x}}\right)  \frac{1}{2^{N}\pi}\int
e^{-\overleftarrow{\partial_{x}}y}e^{2iyp}\psi\left(  x-y\right)  \phi\left(
x+y\right)  dy\nonumber\\
&  =\frac{1}{2^{N}\pi}\int e^{2iyp}G\left(  x-y,p-\frac{1}{2}i\overrightarrow
{\partial_{x}}\right)  \psi\left(  x-y\right)  \phi\left(  x+y\right)  dy
\label{JustOneDerivative}%
\end{align}
Similarly%
\begin{align}
f_{\psi\phi}\left(  x,p\right)  \star G\left(  x,p\right)   &  =G\left(
x,p+\frac{1}{2}i\overrightarrow{\partial_{x}}\right)  \frac{1}{2^{N}\pi}\int
e^{\overleftarrow{\partial_{x}}y}e^{2iyp}\psi\left(  x-y\right)  \phi\left(
x+y\right)  e^{y}dy\nonumber\\
&  =\frac{1}{2^{N}\pi}\int e^{2iyp}G\left(  x+y,p+\frac{1}{2}i\overrightarrow
{\partial_{x}}\right)  \psi\left(  x-y\right)  \phi\left(  x+y\right)  dy
\label{JustOneDerivativeAgain}%
\end{align}
For two classes of special cases things simplify considerably here. \ 

In the first class, if $G\left(  x,p\right)  =V\left(  x\right)  $ then%
\begin{align}
V\left(  x\right)  \star f_{\psi\phi}\left(  x,p\right)   &  =\frac{1}%
{2^{N}\pi}\int V\left(  x-y\right)  \psi\left(  x-y\right)  e^{2iyp}%
\phi\left(  x+y\right)  dy\label{V*f}\\
f_{\psi\phi}\left(  x,p\right)  \star V\left(  x\right)   &  =\frac{1}%
{2^{N}\pi}\int\psi\left(  x-y\right)  e^{2iyp}V\left(  x+y\right)  \phi\left(
x+y\right)  dy \label{f*V}%
\end{align}
Since $V\left(  x\right)  $ here is arbitrary, there is a simple and direct
application of (\ref{V*f}) and (\ref{f*V}) to write Wigner transforms
themselves as star products of wave functions and their Fourier transforms,
with an intercalated exponential.\ \ Namely%
\begin{align}
f_{\psi\phi}\left(  x,p\right)   &  =\psi\left(  x\right)  \star f_{1\phi
}\left(  x,p\right)  =\psi\left(  x\right)  \star\frac{1}{2^{N}\pi}\int
e^{2iyp}\phi\left(  x+y\right)  dy\nonumber\\
&  =\frac{1}{2^{N-1}}\psi\left(  x\right)  \star\left(  e^{-2ixp}\widehat
{\phi}\left(  -2p\right)  \right) \nonumber\\
&  =\frac{1}{2^{N-1}}\psi\left(  x\right)  \star e^{-2ixp}\star\widehat{\phi
}\left(  -p\right) \label{WFasPsi(x)*Phi(p)}\\
f_{\psi\phi}\left(  x,p\right)   &  =f_{\psi1}\left(  x,p\right)  \star
\phi\left(  x\right)  =\frac{1}{2^{N}\pi}\int\psi\left(  x-y\right)
e^{2iyp}dy\star\phi\left(  x\right) \nonumber\\
&  =\frac{1}{2^{N-1}}\left(  e^{2ixp}\widehat{\psi}\left(  2p\right)  \right)
\star\phi\left(  x\right) \nonumber\\
&  =\frac{1}{2^{N-1}}\widehat{\psi}\left(  p\right)  \star e^{2ixp}\star
\phi\left(  x\right)  \label{WFasPsi(p)*Phi(x)}%
\end{align}
where the momentum space wave functions are defined by%
\begin{equation}
\widehat{\psi}\left(  2p\right)  \equiv\frac{1}{2\pi}\int\psi\left(  w\right)
e^{-2iwp}dw\ ,\ \ \ \widehat{\phi}\left(  -2p\right)  \equiv\frac{1}{2\pi}%
\int\psi\left(  w\right)  e^{2iwp}dw
\end{equation}
Reasoning along these same lines also leads to a compact expression for Wigner
transforms in terms of star products of wave functions with an intercalated
momentum-space delta \cite{Braunss}, either Dirac or Kronecker.%
\begin{equation}
f_{\psi\phi}\left(  x,p\right)  =2^{1-N}\psi\left(  x\right)  \star\delta
_{2p}\star\phi\left(  x\right)
\end{equation}
where $\delta_{2p}=\delta\left(  2p\right)  \equiv\frac{1}{2\pi}\int_{-\infty
}^{+\infty}e^{2isp}ds$ for $x\in\mathbb{R}$ and $\delta_{2p}=\delta
_{2p,0}\equiv\frac{1}{2\pi}\int_{0}^{2\pi}e^{2isp}ds$\ for $x\in\mathbb{S}%
^{1}$.

In the second class, if $G\left(  x,p\right)  =K\left(  p\right)  $ then
integration by parts assuming no surface contributions (IPANS) gives%
\begin{align}
K\left(  p\right)  \star f_{\psi\phi}\left(  x,p\right)   &  =\frac{1}%
{2^{N}\pi}\int e^{2iyp}K\left(  p-\frac{1}{2}i\overrightarrow{\partial_{x}%
}\right)  \psi\left(  x-y\right)  \phi\left(  x+y\right)  dy\nonumber\\
&  =\frac{1}{2^{N}\pi}\int e^{2iyp}K\left(  -\frac{1}{2}i\overleftarrow
{\partial_{y}}-\frac{1}{2}i\overrightarrow{\partial_{x}}\right)  \psi\left(
x-y\right)  \phi\left(  x+y\right)  dy\nonumber\\
&  =\frac{1}{2^{N}\pi}\int e^{2iyp}\phi\left(  x+y\right)  K\left(  -\frac
{1}{2}i\left(  \overrightarrow{\partial_{x}}-\overrightarrow{\partial_{y}%
}\right)  \right)  \psi\left(  x-y\right)  dy\\
f_{\psi\phi}\left(  x,p\right)  \star K\left(  p\right)   &  =\frac{1}%
{2^{N}\pi}\int e^{2iyp}K\left(  p+\frac{1}{2}i\overrightarrow{\partial_{x}%
}\right)  \psi\left(  x-y\right)  \phi\left(  x+y\right)  dy\nonumber\\
&  =\frac{1}{2^{N}\pi}\int e^{2iyp}K\left(  -\frac{1}{2}i\overleftarrow
{\partial_{y}}+\frac{1}{2}i\overrightarrow{\partial_{x}}\right)  \psi\left(
x-y\right)  \phi\left(  x+y\right)  dy\nonumber\\
&  =\frac{1}{2^{N}\pi}\int e^{2iyp}\psi\left(  x-y\right)  K\left(  \frac
{1}{2}i\left(  \overrightarrow{\partial_{x}}+\overrightarrow{\partial_{y}%
}\right)  \right)  \phi\left(  x+y\right)  dy
\end{align}
More generally, some care is needed to unravel the remaining derivatives in
(\ref{JustOneDerivative}) and (\ref{JustOneDerivativeAgain}). \ For general
$G\left(  x,p\right)  $ there will appear an operator matrix element involving
the operator obtained from $G\left(  x,p\right)  $ by the Weyl correspondence.

As an illustration, consider $G\left(  x,p\right)  =xp$. \ Then%
\begin{align*}
xp\star f_{\psi\phi}\left(  x,p\right)   &  =\frac{1}{2^{N}\pi}\int
e^{2iyp}\left(  x-y\right)  \left(  p-\frac{1}{2}i\overrightarrow{\partial
_{x}}\right)  \psi\left(  x-y\right)  \phi\left(  x+y\right)  dy\\
&  =\frac{1}{2^{N}\pi}\int\left(  x-y\right)  \left(  -\frac{1}{2}%
i\partial_{y}\left(  e^{2iyp}\right)  -\frac{1}{2}e^{2iyp}i\overrightarrow
{\partial_{x}}\right)  \psi\left(  x-y\right)  \phi\left(  x+y\right)  dy\\
&  =\frac{1}{2^{N}\pi}\int e^{2iyp}\left(  -\frac{1}{2}i+\left(  x-y\right)
\frac{1}{2}i\left(  \overrightarrow{\partial_{y}}-\overrightarrow{\partial
_{x}}\right)  \right)  \psi\left(  x-y\right)  \phi\left(  x+y\right)  dy
\end{align*}
upon IPANS. \ We recognize part of this as the expected operator matrix
element $\left(  x-y\right)  \frac{1}{2}i\left(  \overrightarrow{\partial_{y}%
}-\overrightarrow{\partial_{x}}\right)  \psi\left(  x-y\right)  =\left\langle
x-y\right\vert \mathsf{xp}\left\vert \psi\right\rangle $. \ The other term in
the integrand, $-\frac{1}{2}i$, is the price we must pay because on the LHS we
took just the simple product, $xp$, and not the star product, $x\star
p=xp+\frac{1}{2}i$. \ Had we taken the later instead, we would have%
\begin{align}
x\star p\star f_{\psi\phi}\left(  x,p\right)   &  =\frac{1}{2^{N}\pi}\int
e^{2iyp}\left(  \left(  x-y\right)  \frac{1}{2}i\left(  \overrightarrow
{\partial_{y}}-\overrightarrow{\partial_{x}}\right)  \right)  \psi\left(
x-y\right)  \phi\left(  x+y\right)  dy\nonumber\\
&  =\frac{1}{2^{N}\pi}\int e^{2iyp}\left\langle x-y\right\vert \mathsf{xp}%
\left\vert \psi\right\rangle ~\phi\left(  x+y\right)  ~dy
\end{align}
Actually, this illustrates the general situation. \ If the WF is acted on by a
\textquotedblleft star function\textquotedblright\ $G_{\star}\left(
x,p\right)  $ of $x$ and $p$, by which terminology we mean $G_{\star}\left(
x,p\right)  $ is a sum of multinomials of star products of $x$ and $p$, then
the result involves an operator matrix element involving an operator function
$\mathsf{G}\left(  \mathsf{x,p}\right)  $ with the operators arranged \emph{in
the same order} as the various $x$ and $p$ factors were arranged in the
multinomials of the star function. \ Indeed, that operator is just the Weyl
correspondent of the star function.%
\begin{equation}
G_{\star}\left(  x,p\right)  \star f_{\psi\phi}\left(  x,p\right)  =\frac
{1}{2^{N}\pi}\int e^{2iyp}\left\langle x-y\right\vert \mathsf{G}\left(
\mathsf{x,p}\right)  \left\vert \psi\right\rangle ~\phi\left(  x+y\right)  ~dy
\end{equation}
A similar simple operator correspondence applies to acting on the right with
$G_{\star}\left(  x,p\right)  $, \emph{provided} we complex conjugate the
second function.%
\begin{equation}
f_{\psi\overline{\phi}}\left(  x,p\right)  \star G_{\star}\left(  x,p\right)
=\frac{1}{2^{N}\pi}\int e^{2iyp}\psi\left(  x-y\right)  ~\left\langle
\phi\right\vert \mathsf{G}\left(  \mathsf{x,p}\right)  \left\vert
x+y\right\rangle ~dy
\end{equation}
%

%TCIMACRO{\TeXButton{TeX}{\section
%*{Appendix C.  Conventions and star product compositions of bilinear Wigner transforms}
%\addcontentsline{toc}{section}%
%{Appendix C.  Conventions and star product compositions of bilinear Wigner transforms}%
%}}%
%BeginExpansion
\section
*{Appendix C.  Conventions and star product compositions of bilinear Wigner transforms}
\addcontentsline{toc}{section}%
{Appendix C.  Conventions and star product compositions of bilinear Wigner transforms}%
%EndExpansion

For functions $\psi$ and $\phi$ either on $\mathbb{R}$ or on $\mathbb{S}^{1}$,
we define the Wigner transform $f_{\psi\phi}$ as%
\begin{equation}
f_{\psi\phi}\left(  x,p\right)  \equiv\frac{1}{2^{N}\pi}\int_{\mathbb{R}\text{
or }\mathbb{S}^{1}}\psi\left(  x-y\right)  \phi\left(  x+y\right)  e^{2iyp}dy
\end{equation}
where $N$ is chosen to simplify the normalization of two such functions:
\ $N=0$ for $\mathbb{R}$; $N=1$ for $\mathbb{S}^{1}$. \ With these choices%
\begin{equation}
\psi\left(  x\right)  \phi\left(  x\right)  =\left\{
\begin{array}
[c]{cc}%
\int_{-\infty}^{+\infty}f_{\psi\phi}\left(  x,p\right)  dp & \text{ \ \ for
\ \ }x\in\mathbb{R}\\
\sum_{2p\in\mathbb{Z}}f_{\psi\phi}\left(  x,p\right)  & \text{ \ \ for
\ \ }x\in\mathbb{S}^{1}%
\end{array}
\right.
\end{equation}
Note that for position coordinates on the circle, $x,y\in\left[
0,2\pi\right]  $, the momentum sum is over all \emph{semi-integer} $p$ so that
the complete periodic Dirac delta is produced, $\sum_{2p\in\mathbb{Z}}%
e^{2iyp}=2\pi\delta\left(  y\right)  =2\pi\delta\left(  y+2\pi\right)  $, such
that $\int_{0}^{2\pi}\delta\left(  y\right)  dy=\int_{-\epsilon}%
^{2\pi-\epsilon}\delta\left(  y\right)  dy=1$ for all $\epsilon$. \ This is
important since in principle the contributing Fourier mode numbers of $\psi
$\ and $\phi$\ could differ by either even or odd integers.

With these choices for $c$\ we also have a uniform appearance for the $\star$
composition of two such functions%
\begin{equation}
f_{\psi\phi}\left(  x,p\right)  \star f_{\eta\chi}\left(  x,p\right)  =\left(
\phi,\eta\right)  ~f_{\psi\chi}\left(  x,p\right)  \label{StarOfTwoWFs}%
\end{equation}
for both cases. \ On the RHS we have used the notation%
\begin{equation}
\left(  \phi,\eta\right)  =\frac{1}{2\pi}\int_{\mathbb{R}\text{ or }%
\mathbb{S}^{1}}\phi\left(  x\right)  \eta\left(  x\right)  dx
\end{equation}
That is to say this metric is local, $K\left(  x,y\right)  =\frac{1}{2\pi
}\delta\left(  x-y\right)  $ in the language of biorthogonal systems as used
in the main text, Eqn(\ref{ScalarProduct}).

For $\mathbb{R}$\ the result (\ref{StarOfTwoWFs}) is obtained as in the
Overview of \cite{ZFC}. \ For $\mathbb{S}^{1}$ the result is obtained with a
modicum of novelty as follows. \ Applying the star product directly to each of
the integral representations for $f_{\psi\phi}$ and $f_{\eta\chi}$, and
carrying out the requisite variable shifts on the integrands, we obtain%
\begin{gather}
f_{\psi\phi}\left(  x,p\right)  \star f_{\eta\chi}\left(  x,p\right)
=\frac{1}{2^{N}\pi}\int\psi\left(  x-y_{1}\right)  \phi\left(  x+y_{1}\right)
e^{2iy_{1}p}e^{y_{1}\overrightarrow{\partial_{x}}}~\frac{1}{2^{N}\pi}\int
e^{-\overleftarrow{\partial_{x}}y_{2}}e^{2iy_{2}p}\eta\left(  x-y_{2}\right)
\chi\left(  x+y_{2}\right)  dy_{2}\nonumber\\
=\frac{1}{\left(  2^{N}\pi\right)  ^{2}}\iint\psi\left(  x-y_{1}-y_{2}\right)
\phi\left(  x+y_{1}-y_{2}\right)  \eta\left(  x+y_{1}-y_{2}\right)
\chi\left(  x+y_{1}+y_{2}\right)  e^{2i\left(  y_{1}+y_{2}\right)  p}%
dy_{1}dy_{2}\nonumber\\
=\frac{1}{\left(  2^{N}\pi\right)  ^{2}}\int\psi\left(  x-y_{1}-y_{2}\right)
\chi\left(  x+y_{1}+y_{2}\right)  e^{2i\left(  y_{1}+y_{2}\right)  p}d\left(
y_{1}+y_{2}\right)  ~\frac{1}{2}\int\phi\left(  x+y_{1}-y_{2}\right)
\eta\left(  x+y_{1}-y_{2}\right)  d\left(  y_{1}-y_{2}\right)
\end{gather}
Now we only need worry about the regions for the two integrations. \ When the
original $y_{1}$\ and $y_{2}$\ are integrated over the whole real line, then
so are $y_{1}+y_{2}$\ and $y_{1}-y_{2}$, and the result (\ref{StarOfTwoWFs}%
)\ follows. \ When the original $y_{1}$\ and $y_{2}$\ are integrated over the
circle, it may seem that the $y_{1}-y_{2}$\ region of integration depends on
$y_{1}+y_{2}$. \ However, this is actually not the case for periodic
integrands. \ The original coupled integrals $\int_{0}^{2\pi}dy_{1}\int
_{0}^{2\pi}dy_{2}\cdots$ may be split into uncoupled integrals $\int_{2\pi
}^{4\pi}d\left(  y_{1}+y_{2}\right)  \cdots\times\frac{1}{2}\int_{-2\pi}%
^{2\pi}d\left(  y_{1}-y_{2}\right)  \cdots$ upon making use of this
periodicity. \
%TCIMACRO{\TeXButton{TeX}{\begin{multicols}{2}}}%
%BeginExpansion
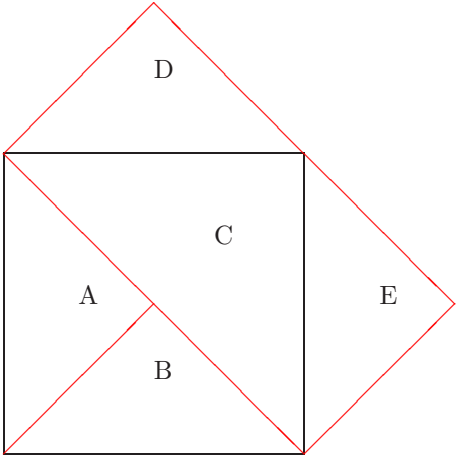
\begin{multicols}{2}%
%EndExpansion
%

%TCIMACRO{\TeXButton{Parallelogram}{{\setlength{\unitlength}{1cm}
%\begin{picture}
%(0,6.0)
%\put(1,2){A}
%\put(2,1){B}
%\put(2.8,2.8){C}
%\put(2,5){D}
%\put(5,2){E}
%\put(0,0){\line(1,0){4}}
%\put(0,0){\line(0,1){4}}
%\put(4,0){\line(0,1){4}}
%\put(0,4){\line(1,0){4}}
%\color{red}
%\put(0,0){\line(1,1){2}}
%\put(0,4){\line(1,-1){4}}
%\put(4,0){\line(1,1){2}}
%\put(4,4){\line(1,-1){2}}
%\put(0,4){\line(1,1){2}}
%\put(4,4){\line(-1,1){2}}
%\color{black}
%\end{picture}
%}}}%
%BeginExpansion
{\setlength{\unitlength}{1cm}
\begin{picture}
(0,6.0)
\put(1,2){A}
\put(2,1){B}
\put(2.8,2.8){C}
\put(2,5){D}
\put(5,2){E}
\put(0,0){\line(1,0){4}}
\put(0,0){\line(0,1){4}}
\put(4,0){\line(0,1){4}}
\put(0,4){\line(1,0){4}}
\color{red}
\put(0,0){\line(1,1){2}}
\put(0,4){\line(1,-1){4}}
\put(4,0){\line(1,1){2}}
\put(4,4){\line(1,-1){2}}
\put(0,4){\line(1,1){2}}
\put(4,4){\line(-1,1){2}}
\color{black}
\end{picture}
}%
%EndExpansion

\noindent The original $2\pi\times2\pi$ square representing the region of
integration on the $y_{1}\times y_{2}$ plane may be partitioned by diagonal
segments into three right-triangular regions (A, B, and C in the Figure).
\ For $2\pi$-periodic integrands the result of the double integration is
unchanged if the left-most triangle (A) is translated $2\pi$ to the right (to
become triangle E), and the lower triangle (B) is translated up by $2\pi$ (to
become triangle D), to form a $2\pi\times4\pi$ rectangle (C+D+E) on the
$\left(  y_{1}+y_{2}\right)  \times\left(  y_{1}-y_{2}\right)  $ integration
plane.\footnote{The scale of the final rectangle in the Figure needs to be
increased by $\sqrt{2}$ to correctly display the relative area of the initial
and final regions on the $y_{1}\times y_{2}$ and $\left(  y_{1}+y_{2}\right)
\times\left(  y_{1}-y_{2}\right)  $ planes, as is evident from the Jacobian
$\left\vert \partial\left(  y_{1}+y_{2},y_{1}-y_{2}\right)  /\partial\left(
y_{1},y_{2}\right)  \right\vert =2$.} Then periodicity is again exploited to
evaluate $\frac{1}{2}\int_{-2\pi}^{2\pi}d\left(  y_{1}-y_{2}\right)
\phi\left(  x+y_{1}-y_{2}\right)  ~\eta\left(  x+y_{1}-y_{2}\right)
=2\pi\left(  \phi,\eta\right)  $ and hence to obtain the final result
(\ref{StarOfTwoWFs}). \ This result is used extensively in the next Appendix.%
%TCIMACRO{\TeXButton{TeX}{\end{multicols}}}%
%BeginExpansion
\end{multicols}%
%EndExpansion
%

%TCIMACRO{\TeXButton{TeX}{\section
%*{Appendix D.  Non-diagonal WFs, conventional and hybrid}
%\addcontentsline{toc}{section}%
%{Appendix D.  Non-diagonal WFs, conventional and hybrid}}}%
%BeginExpansion
\section*{Appendix D.  Non-diagonal WFs, conventional and hybrid}
\addcontentsline{toc}{section}%
{Appendix D.  Non-diagonal WFs, conventional and hybrid}%
%EndExpansion

Consider any biorthogonal system involving countable pairs of functions and
their duals, $\left\{  \psi_{n},\chi_{n}\right\}  $, which has an equivalent
hermitian system involving corresponding functions, $\left\{  \phi
_{n},\overline{\phi_{n}}\right\}  $, orthonormal in the usual sense. \ That
is, in the notation of the previous Appendix,%
\begin{equation}
\left(  \chi_{k},\psi_{n}\right)  =\delta_{k,n}=\left(  \overline{\phi_{k}%
},\phi_{n}\right)
\end{equation}
In phase space we define the following conventional and hybrid Wigner
transforms.
\begin{align}
e_{k,n}\left(  x,p\right)   &  \equiv\frac{1}{2\pi}\int_{0}^{2\pi}\phi
_{k}\left(  x-y\right)  ~\overline{\phi_{n}}\left(  x+y\right)  ~e^{2iyp}dy\\
f_{k,n}\left(  x,p\right)   &  \equiv\frac{1}{2\pi}\int_{0}^{2\pi}\psi
_{k}\left(  x-y\right)  ~\overline{\psi_{n}}\left(  x+y\right)  ~e^{2iyp}dy\\
\widetilde{f_{k,n}}\left(  x,p\right)   &  \equiv\frac{1}{2\pi}\int_{0}^{2\pi
}\overline{\chi_{k}}\left(  x-y\right)  ~\chi_{n}\left(  x+y\right)
~e^{2iyp}dy
\end{align}%
\begin{align}
g_{k,n}\left(  x,p\right)   &  \equiv\frac{1}{2\pi}\int_{0}^{2\pi}\psi
_{k}\left(  x-y\right)  ~\chi_{n}\left(  x+y\right)  ~e^{2iyp}dy\\
h_{k,n}\left(  x,p\right)   &  \equiv\frac{1}{2\pi}\int_{0}^{2\pi}\psi
_{k}\left(  x-y\right)  ~\overline{\phi_{n}}\left(  x+y\right)  ~e^{2iyp}dy\\
\widetilde{h_{k,n}}\left(  x,p\right)   &  \equiv\frac{1}{2\pi}\int_{0}^{2\pi
}\phi_{k}\left(  x-y\right)  ~\chi_{n}\left(  x+y\right)  ~e^{2iyp}dy
\end{align}
Clearly, in the notation of the main text, $e_{n}\left(  x,p\right)  \equiv
e_{n,n}\left(  x,p\right)  $, etc. \ %

%TCIMACRO{\TeXButton{TeX}{\begin{multicols}{2}}}%
%BeginExpansion
\begin{multicols}{2}%
%EndExpansion
From the result in the previous Appendix, (\ref{StarOfTwoWFs}), we immediately
obtain the following star products.%
\begin{align}
e_{j,k}\star e_{l,m}  &  =e_{j,m}~\delta_{k,l}\\
f_{j,k}\star\widetilde{f_{l,m}}  &  =g_{j,m}~\delta_{k,l}\\
\widetilde{f_{j,k}}\star f_{l,m}  &  =\overline{g_{m,j}}~\delta_{k,l}\\
f_{j,k}\star\overline{g_{m,l}}  &  =f_{j,m}~\delta_{k,l}\\
g_{j,k}\star f_{l,m}  &  =f_{j,m}~\delta_{k,l}\\
g_{j,k}\star g_{l,m}  &  =g_{j,m}~\delta_{k,l}\\
\overline{g_{l,m}}\ \star\ \widetilde{f_{k,j}}  &  =f_{m,j}~\delta_{k,l}%
\end{align}
None of these mix the biorthogonal $\left\{  \psi_{k},\chi_{n}\right\}
$\ system with the hermitian $\left\{  \phi_{k},\overline{\phi_{n}}\right\}  $
system. \ But those mixed products are also given immediately by
(\ref{StarOfTwoWFs}).%
\begin{align}
g_{j,k}\star h_{l,m}  &  =h_{j,m}~\delta_{k,l}\\
h_{j,k}\star e_{l,m}  &  =h_{j,m}~\delta_{k,l}\\
e_{j,k}\star\widetilde{h_{l,m}}  &  =\widetilde{h_{j,m}}~\delta_{k,l}\\
h_{j,k}\star\widetilde{h_{l,m}}  &  =g_{j,m}~\delta_{k,l}\\
f_{j,k}\star\overline{\widetilde{h_{m,l}}}  &  =h_{j,m}~\delta_{k,l}\\
\overline{\widetilde{h_{k,j}}}\star\widetilde{h_{l,m}}  &  =\widetilde
{f_{j,m}}~\delta_{k,l}\\
\widetilde{h_{j,k}}\star f_{l,m}  &  =\overline{h_{m,j}}~\delta_{k,l}\\
\widetilde{h_{j,k}}\star g_{l,m}  &  =\widetilde{h_{j,m}}~\delta_{k,l}\\
\widetilde{h_{j,k}}\star h_{l,m}  &  =e_{j,m}~\delta_{k,l}%
\end{align}
There are a few other obvious relations that follow from complex conjugations,
using $\overline{e_{k,n}}=e_{n,k}$,\ $\overline{f_{k,n}}=f_{n,k}$,
and\ $\overline{\widetilde{f_{k,n}}}=\widetilde{f_{n,k}}$, but in writing
these it should be kept in mind that the hybrid conjugates, $\overline{g}$,
$\overline{h}$, and $\overline{\widetilde{h}}$, are in general different
functions, independent of $g$, $h$, and $\widetilde{h}$.

Moreover, there are other star products, such as $g_{j,k}\star\widetilde
{f_{l,m}}$ or $f_{j,k}\star g_{l,m}$, which are \emph{not} proportional to
$\delta_{k,l}$\ as simple consequences of the biorthonormality of $\left\{
\psi_{k},\chi_{n}\right\}  $\ or of $\left\{  \phi_{k},\overline{\phi_{n}%
}\right\}  $. \ These involve the scalar products $\left(  \overline{\chi_{k}%
},\chi_{l}\right)  $, $\left(  \overline{\psi_{k}},\psi_{l}\right)  $,
$\left(  \overline{\phi_{k}},\psi_{n}\right)  $, and $\left(  \phi_{k}%
,\chi_{n}\right)  $, which are model dependent and not single Kronecker
deltas. \ Thus their evaluation depends on specific details of the system.
\ Additional model dependent structure exists, such as the form of specific
$\overline{\psi}$s as infinite series of $\chi$s, and vice versa, or the form
of specific $\phi$s as infinite series of $\psi$s, and vice versa, which
permit re-expressing some star products as infinite sums of others. \ These
re-expressions may amount to relations already known to hold in general, as
above, but often they are new results and highly model dependent. \ 

In summary, the full algebra of Wigner transform star products for a
biorthogonal system is quite complicated and varies from one model to another.
\ But the subset of products that we have explicitly shown are valid for any
combined $\left\{  \left\{  \psi_{k},\chi_{n}\right\}  ,\left\{  \phi
_{k},\overline{\phi_{n}}\right\}  \right\}  $\ system.%
%TCIMACRO{\TeXButton{TeX}{\end{multicols}}}%
%BeginExpansion
\end{multicols}%
%EndExpansion
%

%TCIMACRO{\TeXButton{TeX}{\section
%*{Appendix E.  A brief overview of biorthogonal systems and density operators}
%\addcontentsline{toc}{section}%
%{Appendix E.  A brief overview of biorthogonal systems and density operators}%
%}}%
%BeginExpansion
\section
*{Appendix E.  A brief overview of biorthogonal systems and density operators}
\addcontentsline{toc}{section}%
{Appendix E.  A brief overview of biorthogonal systems and density operators}%
%EndExpansion

As noted in the introduction, for general biorthogonal systems the Hilbert
space structure is at first sight very different than that for hermitian
Hamiltonian systems inasmuch as the dual wave functions are usually not just
the complex conjugates of the wave functions. \ Still, we may keep most of the
compact Dirac notation for a state, $\left\vert \psi\right\rangle $, and its
dual, $\widetilde{\left\langle \psi\right\vert }$, provided we just allow
$\widetilde{\left\langle \psi\right\vert }\left.  x\right\rangle
=\overline{\left\langle x\right.  \widetilde{\left\vert \psi\right\rangle }%
}\neq\overline{\left\langle x\right.  \left\vert \psi\right\rangle }$. \ 

As an alternative notation, we will often write for the dual wave function
$\widetilde{\left\langle \psi\right\vert }\left.  x\right\rangle =\chi\left(
x\right)  $ (no overbar on $\chi$!) thereby reducing the clutter of \ symbol
decorations. \ So then, orthonormality for a discretely indexed biorthogonal
set of states $\left\{  \left\vert \psi_{k}\right\rangle \right\}  $ and their
duals $\left\{  \widetilde{\left\langle \psi_{n}\right\vert }\right\}
$\ (i.e. a countable basis) reads $\widetilde{\left\langle \psi_{n}\right\vert
}\left.  \psi_{k}\right\rangle =\delta_{n,k}=\int dx\ \chi_{n}\left(
x\right)  \psi_{k}\left(  x\right)  $. \
%TCIMACRO{\TeXButton{TeX}{\begin{multicols}{2}}}%
%BeginExpansion
\begin{multicols}{2}%
%EndExpansion

Let us now go through some purely formal manipulations to get a quick overview
of the abstract constructions that are possible for biorthogonal systems. \ If
we are dealing with a discrete biorthonormal basis, $\left\{  \left\vert
\psi_{k}\right\rangle ,\widetilde{\left\langle \psi_{n}\right\vert }\right\}
$, then through the usual conjugation of wave functions we can always
\textquotedblleft double-up\textquotedblright\ the states and their
duals\footnote{Keep in mind that the total number of independent Hilbert space
states ($L^{2}$ functions) may not have actually changed by this doubling-up
procedure, e.g. as is certainly the case if the original wave functions and
their duals were real functions. \ But this doesn't affect the formal
constructions to follow.} to include $\left\{  \widetilde{\left\vert \psi
_{n}\right\rangle },\left\langle \psi_{k}\right\vert \right\}  $ and hence
write the obvious relation $\widetilde{\left\vert \psi_{n}\right\rangle }%
=\sum_{k}\widetilde{\left\vert \psi_{k}\right\rangle }~\delta_{k,n}=\sum
_{k}\widetilde{\left\vert \psi_{k}\right\rangle }\widetilde{\left\langle
\psi_{k}\right\vert }\left.  \psi_{n}\right\rangle $. \ That is to say
\begin{equation}
\widetilde{\left\vert \psi_{n}\right\rangle }=R\left\vert \psi_{n}%
\right\rangle
\end{equation}
where the operator $R$ is just the formal hermitian sum%
\begin{equation}
R=\sum_{k}\widetilde{\left\vert \psi_{k}\right\rangle }\widetilde{\left\langle
\psi_{k}\right\vert }=R^{\dag}%
\end{equation}
This is a positive operator, certainly, but without further assumptions about
the completeness of the $\widetilde{\left\vert \psi_{k}\right\rangle } $, $R$
may well annihilate some states, so it is not necessarily positive definite.
We also have the equally obvious relation%
\begin{equation}
\widetilde{\left\langle \psi_{n}\right\vert }=\left\langle \psi_{n}\right\vert
R
\end{equation}
Thus the biorthonormality of the system may also be written as%
\begin{equation}
\delta_{n,k}=\left\langle \psi_{n}\right\vert R\left\vert \psi_{k}%
\right\rangle
\end{equation}

The operator $R$\ plays the role of a metric in the span of $\left\{
\left\vert \psi_{n}\right\rangle \right\}  $. \ Given the requisite
convergence of the sums, for any linear combination of the basis states
\begin{equation}
\left\vert \psi\right\rangle =\sum_{k}c_{k}\left\vert \psi_{k}\right\rangle
\end{equation}
we have a manifestly positive definite norm%
\begin{equation}
\sum_{k}\left\vert c_{k}\right\vert ^{2}=\left\langle \psi\right\vert
R\left\vert \psi\right\rangle
\end{equation}
Equivalently, we may write
\begin{equation}
\sum_{k}\left\vert c_{k}\right\vert ^{2}=\widetilde{\left\langle
\psi\right\vert }\left.  \psi\right\rangle
\end{equation}
where the dual corresponding to $\left\vert \psi\right\rangle $ \emph{in the
given basis} is just%
\begin{equation}
\widetilde{\left\langle \psi\right\vert }=\left\langle \psi\right\vert
R=\sum_{n}\overline{c_{n}}\widetilde{\left\langle \psi_{n}\right\vert }%
\end{equation}
as well as the equation dual to this%
\begin{equation}
\widetilde{\left\vert \psi\right\rangle }=R\left\vert \psi\right\rangle
=\sum_{k}c_{k}\widetilde{\left\vert \psi_{k}\right\rangle }%
\end{equation}
Now, we emphasize that it is quite possible, and often the case for
non-hermitian systems, that $R\neq1$. \ Moreover, it is also often the case
that $\delta_{k,n}\neq\widetilde{\left\langle \psi_{k}\right.  }%
|\widetilde{\left.  \psi_{n}\right\rangle }\equiv\int dx\ \chi_{k}\left(
x\right)  \overline{\chi_{n}\left(  x\right)  }$ and $\delta_{k,n}%
\neq\left\langle \psi_{k}\right.  \left\vert \psi_{n}\right\rangle \equiv\int
dx\ \overline{\psi_{k}\left(  x\right)  }\psi_{n}\left(  x\right)  $. \ 

For emphasis and clarity, let us restate the previous nine relations in terms
of wave functions $\psi\left(  x\right)  \equiv\left\langle x\right.
\left\vert \psi\right\rangle $, their dual functions $\widetilde{\left\langle
\psi\right\vert }\left.  x\right\rangle \equiv\chi\left(  x\right)  $, and the
integral kernel in the position basis%
\begin{equation}
R\left(  x,y\right)  \equiv\left\langle x\right\vert R\left\vert
y\right\rangle
\end{equation}
We have
\begin{equation}
\overline{\chi_{n}\left(  x\right)  }=\int R\left(  x,y\right)  \psi
_{n}\left(  y\right)  dy
\end{equation}%
\begin{equation}
R\left(  x,y\right)  =\sum_{k}\overline{\chi_{k}\left(  x\right)  }\chi
_{k}\left(  y\right)  =\overline{R\left(  y,x\right)  }%
\end{equation}%
\begin{equation}
\chi_{n}\left(  x\right)  =\int\overline{\psi_{n}\left(  y\right)  }R\left(
y,x\right)  dy
\end{equation}%
\begin{equation}
\delta_{n,k}=\int\overline{\psi_{n}\left(  x\right)  }R\left(  x,y\right)
\psi_{k}\left(  y\right)  dxdy
\end{equation}%
\begin{equation}
\psi\left(  x\right)  =\sum_{k}c_{k}\psi_{k}\left(  x\right)
\end{equation}%
\begin{align}
\sum_{k}\left\vert c_{k}\right\vert ^{2}  &  =\int\overline{\psi\left(
y\right)  }R\left(  y,x\right)  \psi\left(  x\right)  dxdy\\
&  =\int\chi\left(  x\right)  \psi\left(  x\right)  dx
\end{align}%
\begin{align}
\chi\left(  x\right)   &  =\int\overline{\psi\left(  y\right)  }R\left(
y,x\right)  dy\nonumber\\
&  =\sum_{n}\overline{c_{n}}\chi_{n}\left(  x\right)
\end{align}
and finally%
\begin{align}
\overline{\chi\left(  x\right)  }  &  =\int R\left(  x,y\right)  \psi\left(
y\right)  dy\nonumber\\
&  =\sum_{n}c_{n}\overline{\chi_{n}\left(  x\right)  }%
\end{align}

The role of the states and their duals is interchanged upon constructing the
dual metric
\begin{equation}
\widetilde{R}=\sum_{k}\left\vert \psi_{k}\right\rangle \left\langle \psi
_{k}\right\vert =\widetilde{R}^{\dag}%
\end{equation}
In terms of this operator we have%
\begin{align}
\left\langle \psi\right\vert  &  =\widetilde{\left\langle \psi\right\vert
}\widetilde{R}=\sum_{n}\overline{c_{n}}\left\langle \psi_{n}\right\vert \\
\left\vert \psi\right\rangle  &  =\widetilde{R}\widetilde{\left\vert
\psi\right\rangle }=\sum_{k}c_{k}\left\vert \psi_{k}\right\rangle
\end{align}
so $\widetilde{R}$\ is effectively the inverse to $R$, and vice versa, at
least on the appropriate subspaces. \ Correspondingly, we have the formal
relations%
\begin{align}
\widetilde{R}R  &  =\sum_{k}\left\vert \psi_{k}\right\rangle \widetilde
{\left\langle \psi_{k}\right\vert }\\
R\widetilde{R}  &  =\sum_{k}\widetilde{\left\vert \psi_{k}\right\rangle
}\left\langle \psi_{k}\right\vert \\
\widetilde{R}R  &  =\widetilde{R}R\widetilde{R}R\ ,\ \ \ R\widetilde
{R}=R\widetilde{R}R\widetilde{R}%
\end{align}
The first of these acts as the unit operator on the span of $\left\{
\left\vert \psi_{k}\right\rangle \right\}  $\ while the second acts as the
unit on the span of $\left\{  \widetilde{\left\vert \psi_{k}\right\rangle
}\right\}  $. \ So $\widetilde{R}R$ and $R\widetilde{R}$ are at least
projection operators onto those respective spaces, but they are not
necessarily $1$ acting on all functions. \ The effect of $\widetilde{R}R$
acting on the span of $\left\{  \widetilde{\left\vert \psi_{k}\right\rangle
}\right\}  $ is not clearly discernible, nor is $R\widetilde{R}$ acting on the
span of $\left\{  \left\vert \psi_{k}\right\rangle \right\}  $.

For many PT symmetric systems, the duals are proportional to the wave
functions. \ When these are chosen to be PT eigenstates, as is usually the
case, then in a broad class of situations $\chi_{n}\left(  x\right)  =\pm
\psi_{n}\left(  x\right)  $. \ Again, we stress the absence of complex
conjugation. \ However, there are some interesting situations where the dual
functions are not so simply related to the wave functions \cite{CM,CMS}.
\ This is particularly the case when dealing with eigenfunctions at so-called
\textquotedblleft spectral singularities.\textquotedblright\ \ These
exceptional cases seem to illustrate more general features of non-hermitian
systems, so we will focus on them in this discussion. \ 

Perhaps the most interesting feature of these cases is that the wave functions
dual to energy eigenfunctions do not also obey homogeneous equations when
acted upon by the Hamiltonian. \ Rather, in Dirac notation, the structure is
like this:%
\begin{equation}
H\left\vert E\right\rangle =E\left\vert E\right\rangle \ ,\ \ \ \widetilde
{\left\langle E\right\vert }H=E\widetilde{\left\langle E\right\vert
}+\left\langle I_{E}\right\vert
\end{equation}
The inhomogeneity $\left\langle I_{E}\right\vert $\ varies with $E$, usually,
but for the situation of interest to us here, it is orthogonal to \emph{all}
the energy eigenstates, $\left\langle I_{E}\right.  \left\vert E^{\prime
}\right\rangle =0$. \ For the PT symmetric theories of interest, the energy
eigenvalues are real.

It is evident at this point that we may have some freedom in our construction
of density operators, given that we may have $\widetilde{\left\vert \psi
_{k}\right\rangle }\neq\left\vert \psi_{k}\right\rangle $, etc. \ This is so.
\ There are four choices in general. \ For pure states in the various spans of
the basis vectors, as specified above, but otherwise arbitrary, we have
\begin{align}
\rho &  =\left\vert \psi\right\rangle \left\langle \psi\right\vert =\rho
^{\dag}\\
\widetilde{\rho}  &  =\widetilde{\left\vert \psi\right\rangle }\widetilde
{\left\langle \psi\right\vert }=\widetilde{\rho}^{\dag}%
\end{align}
as well as the less symmetrical%
\begin{align}
\left(  \symbol{126}\rho\right)   &  =\widetilde{\left\vert \psi\right\rangle
}\left\langle \psi\right\vert =\left(  \rho\symbol{126}\right)  ^{\dag}\\
\left(  \rho\symbol{126}\right)   &  =\left\vert \psi\right\rangle
\widetilde{\left\langle \psi\right\vert }=\left(  \symbol{126}\rho\right)
^{\dag}%
\end{align}
These are all interrelated by the $R$ and $\widetilde{R}$ operators.%
\begin{align}
\left(  \rho\symbol{126}\right)   &  =\rho R=\widetilde{R}\widetilde{\rho}\\
\left(  \symbol{126}\rho\right)   &  =R\rho=\widetilde{\rho}\widetilde{R}\\
\rho &  =\widetilde{R}\widetilde{\rho}\widetilde{R}=\left(  \rho
\symbol{126}\right)  \widetilde{R}=\widetilde{R}\left(  \symbol{126}%
\rho\right) \\
\widetilde{\rho}  &  =R\rho R=\left(  \symbol{126}\rho\right)  R=R\left(
\rho\symbol{126}\right)
\end{align}
When $\widetilde{\left\langle \psi\right\vert }\left.  \psi\right\rangle =1$,
the various density operators also obey several variants of the standard pure
state condition.%
\begin{align}
\rho\widetilde{\rho}  &  =\left(  \rho\symbol{126}\right) \\
\widetilde{\rho}\rho &  =\left(  \symbol{126}\rho\right) \\
\left(  \rho\symbol{126}\right)  \left(  \rho\symbol{126}\right)   &  =\left(
\rho\symbol{126}\right) \\
\left(  \symbol{126}\rho\right)  \left(  \symbol{126}\rho\right)   &  =\left(
\symbol{126}\rho\right) \\
\widetilde{\rho}\left(  \rho\symbol{126}\right)   &  =\widetilde{\rho}\\
\left(  \symbol{126}\rho\right)  \widetilde{\rho}  &  =\widetilde{\rho}\\
\left(  \rho\symbol{126}\right)  \rho &  =\rho\\
\rho\left(  \symbol{126}\rho\right)   &  =\rho
\end{align}
However, we note that $\rho^{2}\neq\rho$ and $\widetilde{\rho}^{2}%
\neq\widetilde{\rho}$, in general, but rather
\begin{equation}
\rho R\rho=\rho\ ,\ \ \ \widetilde{\rho}\widetilde{R}\widetilde{\rho
}=\widetilde{\rho}%
\end{equation}
as implicitly stated in the previous set of equations.

The results of the previous Appendix can be recast into the density operator
language. \ A full equivalence with the results there can be achieved provided
there is another set of states and their duals, $\left\{  \left\vert \phi
_{k}\right\rangle ,\left\langle \phi_{n}\right\vert \right\}  $, in one-to-one
correspondence with $\left\{  \left\vert \psi_{k}\right\rangle ,\widetilde
{\left\langle \psi_{n}\right\vert }\right\}  $, where the other set is
equipped with a trivial metric (i.e. $R=1$), as would be the case for a
biorthogonal system which admits a similarity transformation to an hermitian
Hamiltonian. \ The full equivalence is then given by%
\begin{align}
e_{k,n}  &  \sim\left\vert \phi_{k}\right\rangle \left\langle \phi
_{n}\right\vert \\
f_{k,n}  &  \sim\left\vert \psi_{k}\right\rangle \left\langle \psi
_{n}\right\vert \\
\widetilde{f_{k,n}}  &  \sim\widetilde{\left\vert \psi_{k}\right\rangle
}\widetilde{\left\langle \psi_{n}\right\vert }\\
g_{k,n}  &  \sim\left\vert \psi_{k}\right\rangle \widetilde{\left\langle
\psi_{n}\right\vert }\\
h_{k,n}  &  \sim\left\vert \psi_{k}\right\rangle \left\langle \phi
_{n}\right\vert \\
\widetilde{h_{k,n}}  &  \sim\left\vert \phi_{k}\right\rangle \widetilde
{\left\langle \psi_{n}\right\vert }%
\end{align}
The star products of the previous Appendix correspond to the operator products
of these dyadics, in an obvious way.

When the metric $R$ is positive definite and hence invertible, the other set
of states could be realized through its hermitian square root $S$, with%
\begin{equation}
R=S^{2}%
\end{equation}
In this case we may take
\begin{align}
\left\vert \phi_{k}\right\rangle  &  =S\left\vert \psi_{k}\right\rangle \\
\widetilde{\left\langle \psi_{n}\right\vert }  &  =\left\langle \phi
_{n}\right\vert S\\
\left\vert \psi_{k}\right\rangle  &  =S^{-1}\left\vert \phi_{k}\right\rangle
\\
\left\langle \phi_{n}\right\vert  &  =\widetilde{\left\langle \psi
_{n}\right\vert }S^{-1}%
\end{align}
If $R$ is positive but not definite, only the first two of these are
guaranteed valid, but then the role of $S^{-1}$ to give the latter two
relations may be effectively played on the appropriate subspace by the
\textquotedblleft other\textquotedblright\ square root, $\widetilde{S}$, where%
\begin{align}
\widetilde{R}  &  =\widetilde{S}^{2}\\
\left\vert \psi_{k}\right\rangle  &  =\widetilde{S}\left\vert \phi
_{k}\right\rangle \\
\left\langle \phi_{n}\right\vert  &  =\widetilde{\left\langle \psi
_{n}\right\vert }\widetilde{S}%
\end{align}%
%TCIMACRO{\TeXButton{TeX}{\end{multicols}}}%
%BeginExpansion
\end{multicols}%
%EndExpansion
%

%TCIMACRO{\TeXButton{TeX}{\section
%*{Appendix F.  Operator expressions from Weyl transforms}
%\addcontentsline{toc}{section}%
%{Appendix F.  Operator expressions from Weyl transforms}}}%
%BeginExpansion
\section*{Appendix F.  Operator expressions from Weyl transforms}
\addcontentsline{toc}{section}%
{Appendix F.  Operator expressions from Weyl transforms}%
%EndExpansion

\paragraph{Real Liouville theory}

(adapted from \cite{CUZ}) \ Given the factorized phase-space generating
function
\begin{equation}
\mathcal{G}(z;x,p)=K_{ip}\left(  e^{z}\right)  ~\exp\left(  -\frac{1}%
{2}~e^{2x-z}\right)  \label{RealLiouvilleGF}%
\end{equation}
what is the operator corresponding to it? \ According to Weyl's prescription
the associated operator\footnote{The operator corresponding to $O$ will be
distinguished from it by using the same letter but sans serif: \ $\mathsf{O}%
$.} is
\begin{align}
\mathsf{G}(z)  &  =\frac{1}{(2\pi)^{2}}\int d\tau d\sigma dxdp~\mathcal{G}%
(z;x,p)\exp(i\tau(\mathsf{p}-p)+i\sigma(\mathsf{x}-x))\\
&  =\frac{1}{(2\pi)^{2}}\int d\tau d\sigma dxdp~\exp(i\tau\mathsf{p}%
+i\sigma\mathsf{x})\,\exp\left(  -\frac{1}{2}\,e^{2x-z}-i\sigma x\right)
K_{ip}\left(  e^{z}\right)  ~\exp(-i\tau p).\nonumber
\end{align}
The integrals over $x$ and $p$ may be evaluated separately, if the $\sigma$
contour is first shifted slightly above the real axis, $\sigma\rightarrow
\sigma+i\epsilon$, thereby suppressing contributions to the $x$-integral as
$x\rightarrow-\infty$. \ Now $s\equiv\frac{1}{2}e^{2x-z}$ gives
\begin{align}
\int_{-\infty}^{+\infty}dx~\exp\left(  -\frac{1}{2}~e^{2x-z}-i\left(
\sigma+i\epsilon\right)  x\right)   &  =\int_{0}^{\infty}\frac{ds}{2s}\left(
2se^{z}\right)  ^{-i\left(  \sigma+i\epsilon\right)  /2}\exp\left(  -s\right)
\nonumber\\
&  =\frac{1}{2}~e^{-i\left(  z+\ln2\right)  \sigma/2}~\Gamma\left(  -i\left(
\sigma+i\epsilon\right)  /2\right)
\end{align}
But then again%
\begin{equation}
\int_{-\infty}^{\infty}dp~K_{ip}\left(  e^{z}\right)  \exp(-i\tau p)=\frac
{1}{2}\int_{-\infty}^{\infty}dX\,e^{-e^{z}\cosh{X}}~2\pi\delta\left(
X-\tau\right)  =\pi\,e^{-e^{z}\cosh{\tau}}%
\end{equation}
So
\begin{equation}
\mathsf{G}(z)=\frac{1}{8\pi}\int d\tau d\sigma~e^{-i\left(  z+\ln2\right)
\sigma/2}~\Gamma\left(  -i\left(  \sigma+i\epsilon\right)  /2\right)
~e^{-e^{z}\cosh{\tau}}~\exp(i\tau\mathsf{p}+i\sigma\mathsf{x})
\end{equation}
The shifted $\sigma$ contour avoids the pole in $\Gamma$ at the origin.

Re-ordering with all $\mathsf{p}$s to the right (thereby departing from Weyl
ordering but without actually changing the operator) yields $\exp
(i\tau\mathsf{p}+i\sigma\mathsf{x})=\exp(i\sigma\mathsf{x})\exp\left(
i\sigma\tau/2\right)  \exp(i\tau\mathsf{p})$. \ Performing the $\sigma$
integration before the $\tau$ integration permits taking the limit
$\epsilon\rightarrow0$ to obtain
\begin{align}
\mathsf{G}(z)  &  =\frac{1}{8\pi}\int\!d\tau\left(  \int\!d\sigma\Gamma\left(
-i\left(  \sigma+i\epsilon\right)  /2\right)  \exp(i\sigma\mathsf{x}%
+i\sigma\tau/2-i\sigma\left(  z+\ln2\right)  /2)\right)  \,e^{-e^{z}\cosh
{\tau}}\exp(i\tau\mathsf{p})\nonumber\\
&  =\frac{1}{8\pi}\int d\tau\left(  4\pi\exp\left(  -e^{2\mathsf{x}%
+\tau-\left(  z+\ln2\right)  }\right)  \right)  e^{-e^{z}\cosh{\tau}}%
\exp(i\tau\mathsf{p})\nonumber\\
&  =\frac{1}{2}\int d\tau\,e^{-e^{z}\cosh{\tau}}\ \exp\left(  -\frac{1}%
{2}e^{2\mathsf{x}+\tau-z}\right)  \ \exp(i\tau\mathsf{p}) \label{OperatorG}%
\end{align}
This is the operator correspondent of $\mathcal{G}(z;x,p)$. \ In this form it
is straightforward to show that
\begin{equation}
\left[  \mathsf{p}^{2}+e^{2\mathsf{x}},\mathsf{G}(z)\right]  =0 \label{[H,G]}%
\end{equation}
for all values of $z$, which suggests an interpretation of $\mathsf{G}(z)$ as
a propagator and a concomitant interpretation of $z$ as a generalized
\textquotedblleft time.\textquotedblright

The form (\ref{OperatorG}) also leads to a more intuitive Hilbert space
representation. \ Acting to the right of a position eigen-bra, we have
$\left\langle x\right\vert \mathsf{x}=\left\langle x\right\vert x$, while the
subsequent exponential of the momentum operator just translates, $\left\langle
x\right\vert \exp(i\tau\mathsf{p})=\left\langle x+\tau\right\vert $. So the
full right-operation of $\mathsf{G}$ is
\begin{align}
\left\langle x\right\vert \mathsf{G}(z)  &  =\frac{1}{2}\int d\tau
\,\left\langle x+\tau\right\vert \exp\left(  -\frac{1}{2}e^{2x+\tau-z}%
-\frac{1}{2}e^{z+\tau}-\frac{1}{2}e^{z-\tau}\right) \nonumber\\
&  =\frac{1}{2}\int dy\,\left\langle y\right\vert \exp\left(  -\frac{1}%
{2}e^{x+y-z}-\frac{1}{2}e^{z+y-x}-\frac{1}{2}e^{z-y+x}\right)
\end{align}
Inserting $\boldsymbol{1}=\int dx\,\left\vert x\right\rangle \left\langle
x\right\vert $ gives $\mathsf{G}(z)=\int dx\,\left\vert x\right\rangle
\left\langle x\right\vert \mathsf{G}(z)$, and leads to a coordinate space
realization of the operator involving an $x,y$-symmetric kernel, in which form
it is clear that $\mathsf{G}(z)=\mathsf{G}(z)^{\dag}$ for real $z$.%
\begin{equation}
\mathsf{G}(z)=\frac{1}{2}\int dxdy\,\left\vert x\right\rangle \,\exp\left(
-\frac{1}{2}e^{x+y-z}-\frac{1}{2}e^{x-y+z}-\frac{1}{2}e^{-x+y+z}\right)
\ \left\langle y\right\vert \label{OperatorGAsDyadic}%
\end{equation}
All the operator characteristics are now carried by the dyadics $\left\vert
x\right\rangle \cdots\left\langle y\right\vert $. \ The composition law of
this operator parallels its phase-space isomorph.
\begin{equation}
\mathsf{G}(u)\mathsf{G}(v)={\frac{1}{2}}\int dw~\exp\left(  -{\frac{1}{2}%
}\left(  e^{u+v-w}+e^{u-v+w}+e^{-u+v+w}\right)  \right)  ~\mathsf{G}%
(w)=\mathsf{G}(v)\mathsf{G}(u)
\end{equation}

\paragraph{Imaginary Liouville theory}

Let us apply these same methods to the imaginary Liouville case. \ Here, we
can just write down the final answer from what we know about real Liouville
theory, as given above, and then work backwards, a procedure often followed in
\cite{CM}. \ It is easily seen that the following operator similarity
transformation is hermitian, for real $s$, and converts operator $\mathsf{H}$
into operator $\mathsf{H}^{\dag}$.%

\begin{subequations}
\begin{align}
\mathsf{F}\left(  s\right)   &  =\frac{1}{2}\int dxdy\ \left\vert
x\right\rangle \ \exp\left(  \frac{1}{2s}e^{i\left(  -x+y\right)  }-\frac
{s}{2}e^{i\left(  -x-y\right)  }-\frac{s}{2}e^{i\left(  x+y\right)  }\right)
\ \left\langle y\right\vert \label{OperatorFAsDyadic}\\
& \nonumber\\
\mathsf{F}\left(  s\right)  \mathsf{H}  &  =\mathsf{H}^{\dag}\mathsf{F}\left(
s\right) \\
& \nonumber\\
\mathsf{H}  &  =\mathsf{p}^{2}+e^{2i\mathsf{x}}\ ,\ \ \ \mathsf{H}^{\dag
}=\mathsf{p}^{2}+e^{-2i\mathsf{x}}\text{ \ \ where }\mathsf{x,p}\text{ are
operators}%
\end{align}
However, it is not obvious that $\mathsf{F}\left(  s\right)  $\ is positive
definite, even on a restricted range of $\mathsf{p}$\ eigenstates.

Acting on a position eigenbra, $\left\langle x\right\vert $, it can be seen
that $\mathsf{F}$ is given by the parity operator multiplying a single
parametric integral, whose integrand can be written in factorized form as a
product of a spatial translation and an exponential of a momentum translation,
just like the real Liouville case. \ That is to say
\end{subequations}
\begin{equation}
\mathsf{F}\left(  s\right)  =\frac{1}{2}~\mathbb{P}\int d\sigma\ e^{-s\cos
\sigma}\ \exp\left(  \frac{1}{2s}e^{i\left(  2\mathsf{x}+\sigma\right)
}\right)  \ \exp\left(  i\sigma\mathsf{p}\right)  \label{OperatorF}%
\end{equation}
where again $\mathsf{x}$ and $\mathsf{p}$ are position and momentum operators,
and where $\mathbb{P}$ is the usual parity operator: $\left\langle
x\right\vert \mathbb{P}=\left\langle -x\right\vert $, $\mathbb{P}%
\mathsf{x}\mathbb{P}=-\mathsf{x}$, $\mathbb{P}\mathsf{p}\mathbb{P}%
=-\mathsf{p}$. \ The range of the $\sigma$ integration is whatever is needed
for completeness of position eigenstates: \ $\boldsymbol{1}=\int
d\sigma\ \left\vert \sigma\right\rangle \left\langle \sigma\right\vert $.
\ The form (\ref{OperatorF})\ should be compared to (\ref{OperatorG}), with
the identification $s=-e^{z}$. \ 

The presence of $\mathbb{P}$\ in $\mathsf{F}(s)$ has interesting consequences.
\ From (\ref{OperatorF}) it follows for real $s$\ that we can write%
\begin{equation}
\mathsf{F}\left(  s\right)  =\int d\sigma\ e^{-s\cos\sigma}\ \mathsf{D}\left(
s\right)  \ \mathbb{P\ }\mathsf{D}^{\dag}\left(  s\right)
\label{OperatorFasDPDbar}%
\end{equation}
where%
\begin{equation}
\mathsf{D}\left(  s\right)  \equiv\exp\left(  -\frac{1}{2}i\sigma
\mathsf{p}\right)  \exp\left(  \frac{1}{4s}e^{-2i\mathsf{x}}\right)
\ ,\ \ \ \mathsf{D}^{\dag}\left(  s\right)  \equiv\exp\left(  \frac{1}%
{4s}e^{2i\mathsf{x}}\right)  \exp\left(  \frac{1}{2}i\sigma\mathsf{p}\right)
\end{equation}
From the form (\ref{OperatorFasDPDbar}) it would seem that $\mathsf{F}$\ is
\emph{not} positive, although it is manifestly hermitian. \ From the form
(\ref{OperatorF}) we can also check that $\mathsf{F}=\mathsf{F}^{\dag}$ for
real $s$, but not so easily as from (\ref{OperatorFAsDyadic}) or from
(\ref{OperatorFasDPDbar}).

In the course of obtaining (\ref{OperatorF})\ from (\ref{OperatorFAsDyadic}%
)\ it is remarkably easy to see how the parity operator appears. \ No prior
knowledge of its presence is needed. \ However, if we remove $\mathbb{P}$ from
the RHS of (\ref{OperatorF}), we see that%
\begin{equation}
\left[  \mathsf{H},\mathbb{P}\mathsf{F}\left(  s\right)  \right]  =0=\left[
\mathsf{H}^{\dag},\mathsf{F}\left(  s\right)  \mathbb{P}\right]
\end{equation}
which is similar to (\ref{[H,G]}). \ All this again suggests an interpretation
of $\mathbb{P}\mathsf{F}(s)$, or $\mathsf{F}\left(  s\right)  \mathbb{P}$,\ as
a propagator for the system governed by $\mathsf{H}$,\ or $\mathsf{H}^{\dag}$,
as well as an interpretation of $s$ as a generalized time. \ Moreover, from
(\ref{OperatorF})\ the phase-space correspondent of $\mathbb{P}\mathsf{F}%
\left(  s\right)  $ is obviously given by%
\begin{align}
\mathcal{F}(s;x,p)  &  =\frac{1}{2}\int d\sigma\ e^{-s\cos\sigma}\ \exp\left(
\frac{1}{2s}e^{i\left(  2x+\sigma\right)  }\right)  \star\exp\left(  i\sigma
p\right) \\
&  =\frac{1}{2}\int d\sigma\ e^{-s\cos\sigma}\ \exp\left(  i\sigma p\right)
\text{ }\exp\left(  \frac{1}{2s}e^{2ix}\right)
\end{align}
This is nicely factorized into a function of $p$ times a function of $x$.
\ For integer $p$, with $\sigma$ integrated over $\left[  0,2\pi\right]  $, it
is simply%
\begin{equation}
\mathcal{F}(s;x,p\in\mathbb{N})=\pi I_{p}\left(  -s\right)  \exp\left(
\frac{1}{2s}e^{2ix}\right)  \label{ImaginaryLiouvilleGF}%
\end{equation}
This should be compared to (\ref{RealLiouvilleGF}), again with the
identification $s=-e^{z}$.

We leave as exercises for the interested reader to determine the operators
that result from taking the Weyl transforms of the phase-space metrics and/or
the dual metrics.

%

%TCIMACRO{\TeXButton{LaTeX}{\LaTeX{}}}%
%BeginExpansion
\LaTeX{}%
%EndExpansion

\end{document}